\documentclass[twocolumn]{aastex631}
\usepackage{amsmath}
\usepackage{upgreek}
\usepackage{CJK}
\usepackage{graphicx}
\usepackage{subfigure}
\usepackage{enumerate}
\usepackage{amssymb}
\usepackage{multirow}

\makeatletter

\newcommand{\Rmnum}[1]{\expandafter\@slowromancap\romannumeral #1@}
\makeatother

\submitjournal{AAS Journals}

\shorttitle{Extremely Inclined Orbit of S-type Planet $\gamma$ Cep Ab}
\shortauthors{Huang X.M. \& Ji J.}

\begin{document}
\begin{CJK*}{UTF8}{gbsn}

\title{Extremely Inclined Orbit of S-type Planet $\gamma$ Cep Ab Induced by Eccentric Kozai--Lidov Mechanism}
\correspondingauthor{Jianghui Ji}
\email{jijh@pmo.ac.cn}

\author{Xiumin Huang}
\affiliation{CAS Key Laboratory of Planetary Sciences, Purple Mountain Observatory, Chinese Academy of Sciences, Nanjing 210023, China}
\affiliation{School of Astronomy and Space Science, University of Science and Technology of China, Hefei 230026, China}

\author{Jianghui Ji}
\affiliation{CAS Key Laboratory of Planetary Sciences, Purple Mountain Observatory, Chinese Academy of Sciences, Nanjing 210023, China}
\affiliation{School of Astronomy and Space Science, University of Science and Technology of China, Hefei 230026, China}
\affiliation{CAS Center for Excellence in Comparative Planetology, Hefei 230026, China}

\begin{abstract}
$\gamma$ Cep Ab is a typical S-type planet, which occupies a nearly perpendicular planetary orbit relative to the binary. Here we use the Markov Chain Monte Carlo (MCMC) sampler to conduct full N-body fitting and derive self-consistent orbital solutions for this hierarchical system. Then we employ the Eccentric Kozai--Lidov (EKL) mechanism to explain the extremely inclined orbit of S-type planet $\gamma$ Cep Ab. The EKL mechanism plays an essential role in exploring significant oscillations of the mutual inclination $i_{\mathrm{mut}}$ between the planet and the secondary star. We perform qualitative analysis and extensive numerical integrations to investigate the flip conditions and timescales of $\gamma$ Cep Ab's orbit. When the planetary mass is 15 $M_{\mathrm{Jup}}$, the planet can reach $i_{\mathrm{mut}} \sim$ 113$^{\circ}$ with the critical initial conditions of $i_{\mathrm{mut}} < 60^{\circ}$ and $e_1<0.7$. The timescale for the first orbital flip decreases with the increase of the perturbation Hamiltonian. Flipping orbits of $\gamma$ Cep Ab are confirmed to have a large possibility to retain stable based on surfaces of section and the secular stability criterion. Furthermore, we extend the application of EKL to general S-type planetary systems with $a_1/a_2\leq0.1$, where the most intense excitation of $i_{\mathrm{mut}}$ occurs when $a_1/a_2=0.1$ and $e_2 \sim 0.8$, and the variation of planetary mass mainly affect the flip possibility where $e_1\leq 0.3$.
\end{abstract}


\section{Introduction}\label{sec:introduction}
As of today, more than 200 exoplanets are discovered in binary systems, which consist of circumbinary planets (P-type) and satellite-like orbit bodies (S-type) \citep{Schwarz2016}.
P-type planets detected by \textit{Kepler} Space Telescope are largely coplanar with the binary, where the mutual inclination is less than 2.5$^\circ$ \citep{Kostov2014}. The catalogue of exoplanets in binary systems (\url{http://www.univie.ac.at/adg/schwarz/ multiple.html}) reports that only 26 percent of S-type planets are detected with an orbital inclination of $\sim$ 90$^{\circ}$ through transit observations, while over 50 percent of this population are observed by  radial velocity (RV) without determined inclinations. Therefore, the distribution of orbital inclination of planets in binaries plays a significant role in estimating their occurrence rate and understanding the evolution \citep{Armstrong2014,Gong2018}.

\begin{figure}
   \includegraphics[width=\columnwidth,height=7cm]{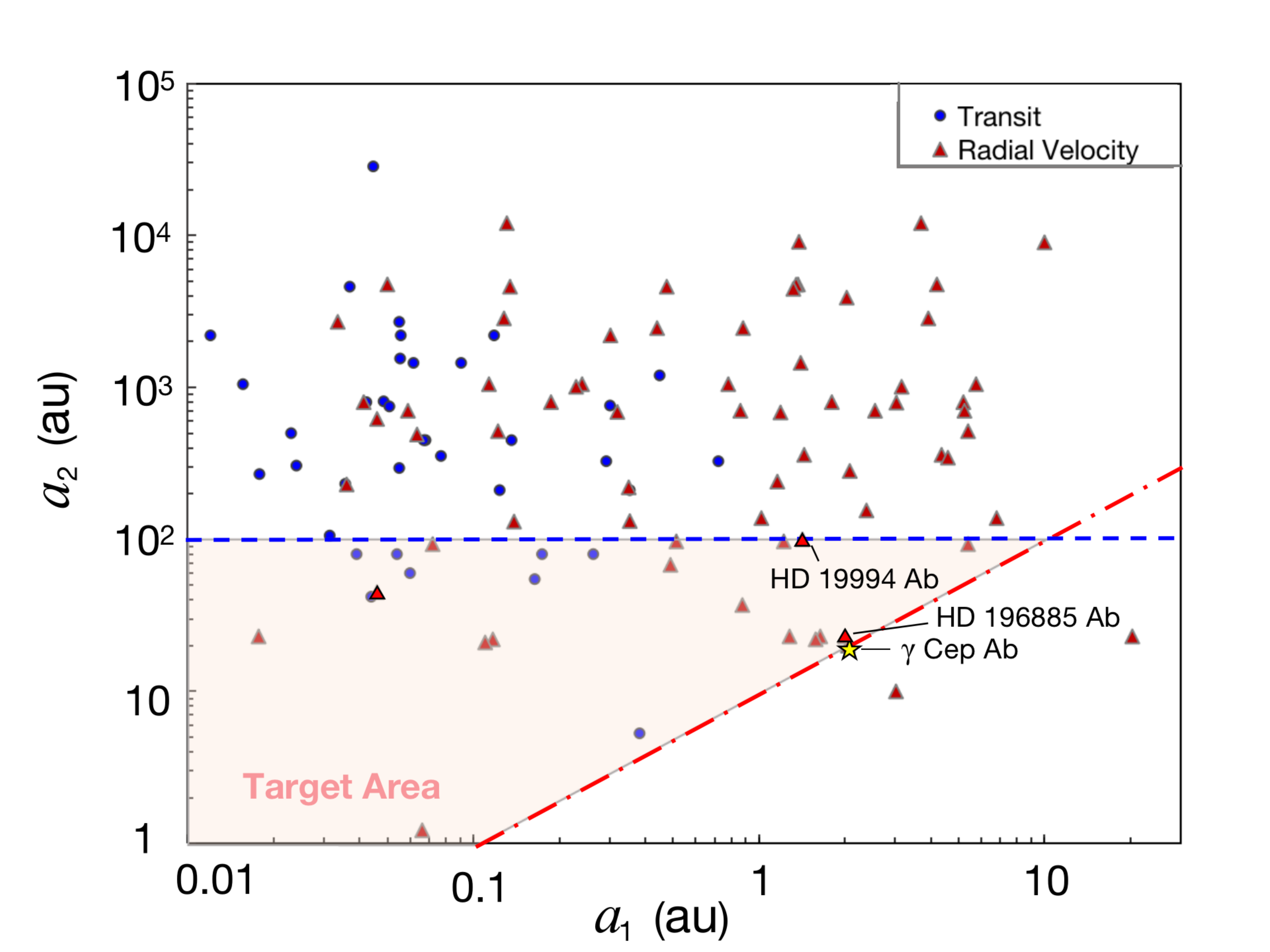}
    \caption{Distribution of semi-major axis of the binaries and  planets in S-type systems. Blue dots represent the S-type planets by transit, while red triangles for those by radial velocity. $\gamma$ Cep Ab is marked by the yellow pentagram.
\label{fig:distributiona}}
\end{figure}

Figure \ref{fig:distributiona} shows the distribution of semi-major axis (SMA) of the planet and the secondary for S-type systems. Here blue dashed line denotes an upper limit of SMA of the secondary, where the perturbation from the secondary is very significant. The red dot dashed line indicates the limit of SMA ratio $a_1/a_2$ = 0.1, where $a_1$ and $a_2$ are, respectively, the SMA of the planet and secondary. In particular, three potentially inclined S-type planets of HD 19994 Ab, HD 196885 Ab and $\gamma$ Cep Ab detected via radial velocity, are labelled out in Figure \ref{fig:distributiona}.

The planet $\gamma$ Cep Ab is one of the best-known S-type planets in  close-binary systems. The radial velocity signal of the planet in $\gamma$ Cep was first measured in 1988 \citep{Campbell1988}, but the observation errors and influence of the secondary star made the planetary detection confusing. \citet{hatzes2003} revealed the planetary companion to $\gamma$ Cep A with high-precision radial velocity measurements spanning from 1981 to 2002. In-situ formation for this planet seems to be less likely due to the truncation model of the proto-planetary disk \citep{Artymowicz1994} and the dynamical instability of planetesimals in the presence of the star companion $\gamma$ Cep B \citep[e.g.][]{Jang-Condell2008,Giuppone2011}. \citet{Marti2012} suggested that the binary orbital configuration of this system was established after the formation of the planet, by scattering scenario of the third fly-by star, then $\gamma$ Cep AB forms the close-binary configuration.

$\gamma$ Cep Ab occupies a nearly perpendicular orbit relative to the binary \citep{reffert2011}, which presents a tremendous challenge to migration scenario in the proto-planetary disk with the angular momentum exchange. However, close binaries may have a remarkable influence on the formation and evolution of S-type planets through dynamical perturbations \citep{Xie2010}. \citet{Andrade-Ines2016} studied in a large parameter space of S-type planets to determine the applicability of the disturbing regime up to the second order, including the orbital stability, the mean-motion resonances, and the short-period oscillations. With a relative high occurrence rate of 50 percent of double stars \citep{Tokovinin1997}, it is very likely and naturally to yield a hierarchical triple system that hosts an S-type planet, which consists of the inner close-binary and the third object in a distant outer orbit.

In the pioneering work of \citet{Kozai1962} and \citet{Lidov1962}, secular dynamics was applied to the hierarchical triple system, indicating the inclined test particle with an inclination of $39^{\circ}\le i \le141^{\circ}$ would lead to periodic oscillations of its inclination and eccentricity, which is now called the Kozai-Lidov mechanism. However, the fundamental theories and phenomenons in the Kozai-Lidov mechanism were investigated and published by \citet{vonZeipel1910} before the middle twentieth century. Afterwards, the Kozai-Lidov mechanism for hierarchical triple systems was extensively investigated \citep{Harrington1968, Lee2003, Naoz2013, Teyssandier2013, Lei2018, Lei2019, Lei2021, Tan2020} under a wide variety of circumstances. \citet{Ford2000} and \citet{Eggleton2001} showed that the Kozai-Lidov mechanism plays an essential part in the secular evolution of hierarchical triple star system and explored the interaction of tidal friction with Kozai cycles in a triple star system. Such scenario was further employed to explain the observed close binaries \citep{Fabrycky2007, Perets2009, Thompson2011, Shappee2013, Naoz2014}.

\citet{Li2014a} systematically explored Kozai--Lidov mechanism and characterized the parameter space that allows large amplitude oscillations in eccentricity and inclination under the test particle limit. As the outer perturber is eccentric and the SMA ratio of the inner orbit and the outer orbit meets $\alpha$ = $a_1 / a_2 \ll 1$, the polynomial in classical outer perturbation equation can be expanded to the third order.  This attributes to the octupole level term, which can give rise to EKL mechanism \citep{Lithwick2011, Naoz2016}. In addition, \citet{Li2014b} confirmed the sufficient initial mutual inclination could produce extremely large eccentricities and flips of the inner orbit, i.e., the orbital inclination transforms between $I_1  <  90^{\circ}$ and  $I_1  > 90^{\circ}$ for the originally prograde or retrograde orbit.

Additionally, for perturbed orbits with an initial low eccentricity and high inclination, orbital flips induced by EKL mechanism is demonstrated to be a kind of resonance, with the libration of a critical angle \citep{Sidorenko2018}. Recently, \citet{Lei2022} systematically studied the EKL mechanism, consisting of the analysis of the orbital flipping and its parameterization. The analytical averaging theory interprets the flip orbits as solutions around polar periodic orbits or a kind of resonant trajectories.

This work aims to explore the inclination evolution and the secular stability of potential inclined S-type planetary systems under the EKL mechanism, particularly for the orbital flip in the $\gamma$ Cep Ab B system. We conduct Markov Chain Monte Carlo (MCMC) search  with the radial velocity data to derive the best-fitting orbital solution of $\gamma$ Cep Ab, suggesting that its orbital plane is nearly perpendicular to that of the secondary star. We conclude that such high mutual inclination can provide substantial evidence of orbital flips with different timescale under specific conditions of the planetary mass, eccentricity and mutual inclination, implying that $\gamma$ Cep Ab may have flipped due to EKL. When the planetary mass is 15 $M_{\mathrm{Jup}}$, $\gamma$ Cep Ab can easily reach the target mutual inclination above 120$^{\circ}$ with the critical initial conditions of $i_{\mathrm{mut}} < 60^{\circ}$ and $e_1<0.7$. Moreover, the flipping cases of $\gamma$ Cep Ab are proved to be stable from  the stability index and Poincar\'{e} surfaces of section. Finally, we explore the parameter space of planetary mass, SMA, and eccentricities to provide theoretical prediction for searching potential inclined S-type planets in close binary systems. When $a_1/a_2$ is fixed, the flip occurs where $e_1$ and $e_2$ are both larger than 0.2 or $e_1 < 0.2$ and $e_2 > 0.3$. The flip likelihood of typical S-type planets is further addressed to search for inclined S-type planets induced by EKL (see Figure \ref{fig:distributiona}).

This work is structured as follows. In Section \ref{sec:orbital solutions}, the published high-precision radial velocity data are employed to derive the orbital solution of $\gamma$ Cep Ab through the full N-body fitting with MCMC sampler. Section \ref{sec:model} describes the Kozai-Lidov and EKL mechanisms. Section \ref{sec:secular_evolution1} provides qualitative analysis of initial conditions and numerical results of the maximum inclination and the orbital flip timescale under various initials of $\gamma$ Cep Ab. The stability of flipping cases is mapped in the planes of ($e_1$, $e_2$) and ($e_1$, $g_1$) as well as the Poincar\'{e} surface of section. Section \ref{sec:secular_evolution2} extends EKL to more general S-type planets to investigate flip conditions. In Section \ref{sec:conclusions}, we summarize the major outcomes.

\section{Orbital solutions of $\gamma$ Cep A$\rm \lowercase {b}$} \label{sec:orbital solutions}
\subsection{N-body fitting of the RV data} \label{subsec:radial velocity fitting}
The $\gamma$ Cep system is known as a close binary at a distance of 13.79 pc \citep{hatzes2003}, which is a candidate target of \textit{CHES} mission \citep{Ji2022}. The primary star $\gamma$ Cep A is a planet-hosting bright star with the spectral type of K1III-IV and a stellar mass of $m_0$ = $1.40 \pm0.12$ $M_{\odot}$ \citep{neuhauser2007}. \citet{neuhauser2007} presented a direct detection of the companion $\gamma$ Cep B, where the parameters of the secondary star are $m_2$ = 0.409 $\pm$ 0.018 $M_{\odot}$, $a_2$ = 20.18 $\pm$ 0.66 au, $I_2$ = 119.3$^{\circ}$, $\Omega_2$ = 18.04 $\pm$ $0.98^{\circ}$ and the orbital period is $P_2$ = 67.5 $\pm$ 1.4 yr. {\citet{reffert2011} conducted the fitting only for the inclination $I_1$ and the ascending node $\Omega_1$, and adopted $P$, $a$, $e$ and $K$  from the best-fitting solution from \citet{Butler2006}. Thus they obtained the best-fitting values of $I_1$ = 5.7$^{\circ}$, $\Omega_1$ = 37.5$^{\circ}$ or $I_1$ = 173.1$^{\circ}$, $\Omega_1$ = 356.1$^{\circ}$ (see Table \ref{tab:rvresults}), where $I_2$ and $\Omega_2$ are constrained by \citet{neuhauser2007}.}

With the most recent spectroscopic observations from the Canada-France-Hawaii Telescope (CFHT) \citep{Walker1992} and the McDonald Observatory Planetary Search (MOPS) program \citep{hatzes2003}, the two-Keplerian solutions of $\gamma$ Cep system were previously given \citep{hatzes2003,torres2007,neuhauser2007}. In these literatures, the uncoupled two Keplerian orbital fitting approach was utilized with the standard non-linear least-squares technique. {However, for the hierarchical system with a massive secondary companion, the significant mutual interaction between the planet-star pair is supposed to be considered under the Jacobi frame to represent the real dynamics \citep{Lee2003}.}

\begin{figure}
        \includegraphics[width=\columnwidth,height=8cm]{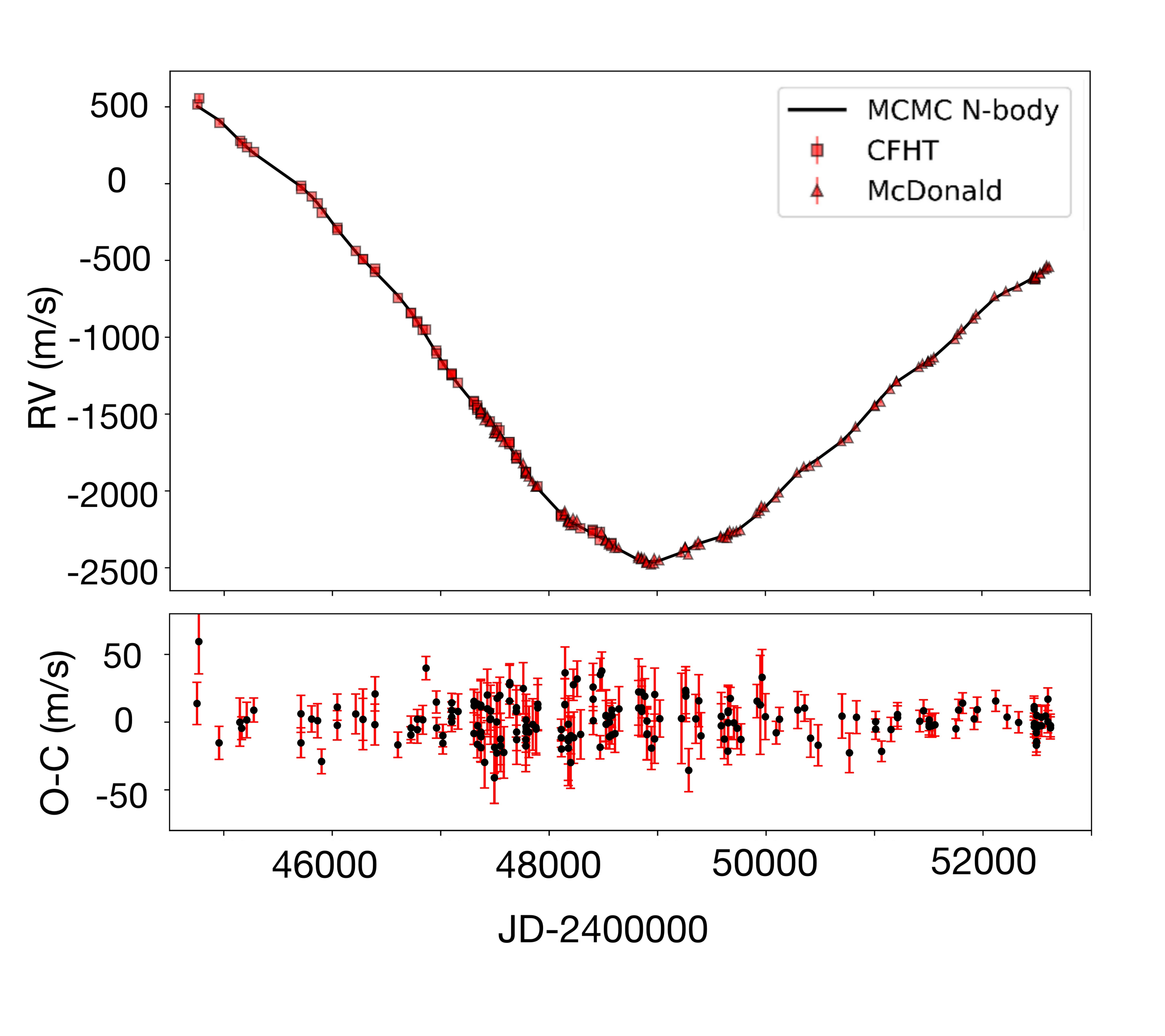}
    \caption{{The best-fitting radial velocity signals of $\gamma$ Cep AbB system. \textit{Upper panel}: red triangles and squares show the  published observations \citep{hatzes2003, Walker1992}, where the black solid line denotes the RV signals of N-body model with mutual perturbation between the companion and planet b. \textit{Lower panel}: O-C  for N-body model. \label{fig:totfit}}}
\end{figure}

\begin{figure}
        \includegraphics[width=\columnwidth,height=6cm]{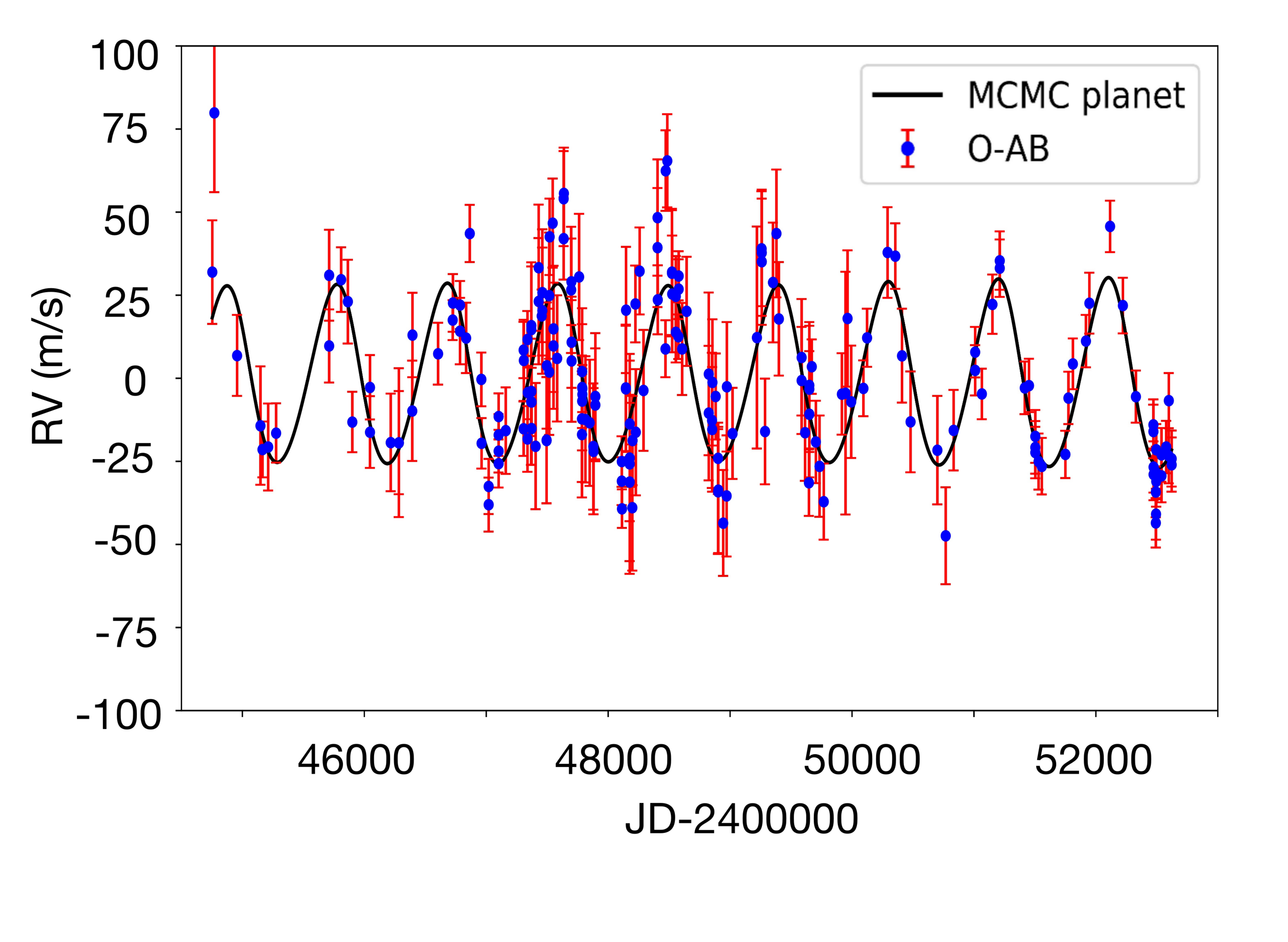}
    \caption{Blue dots with red error bars are the residuals after subtracting calculated radial velocity of $\gamma$ Cep AB from the observations (O-AB). The derived radial velocity signals induced by $\gamma$ Cep Ab from the N-body fitting is presented by the black solid curve, where all the orbital elements change over time.\label{fig:residual}}
\end{figure}

\begin{figure}
        \includegraphics[width=\columnwidth,height=7.5cm]{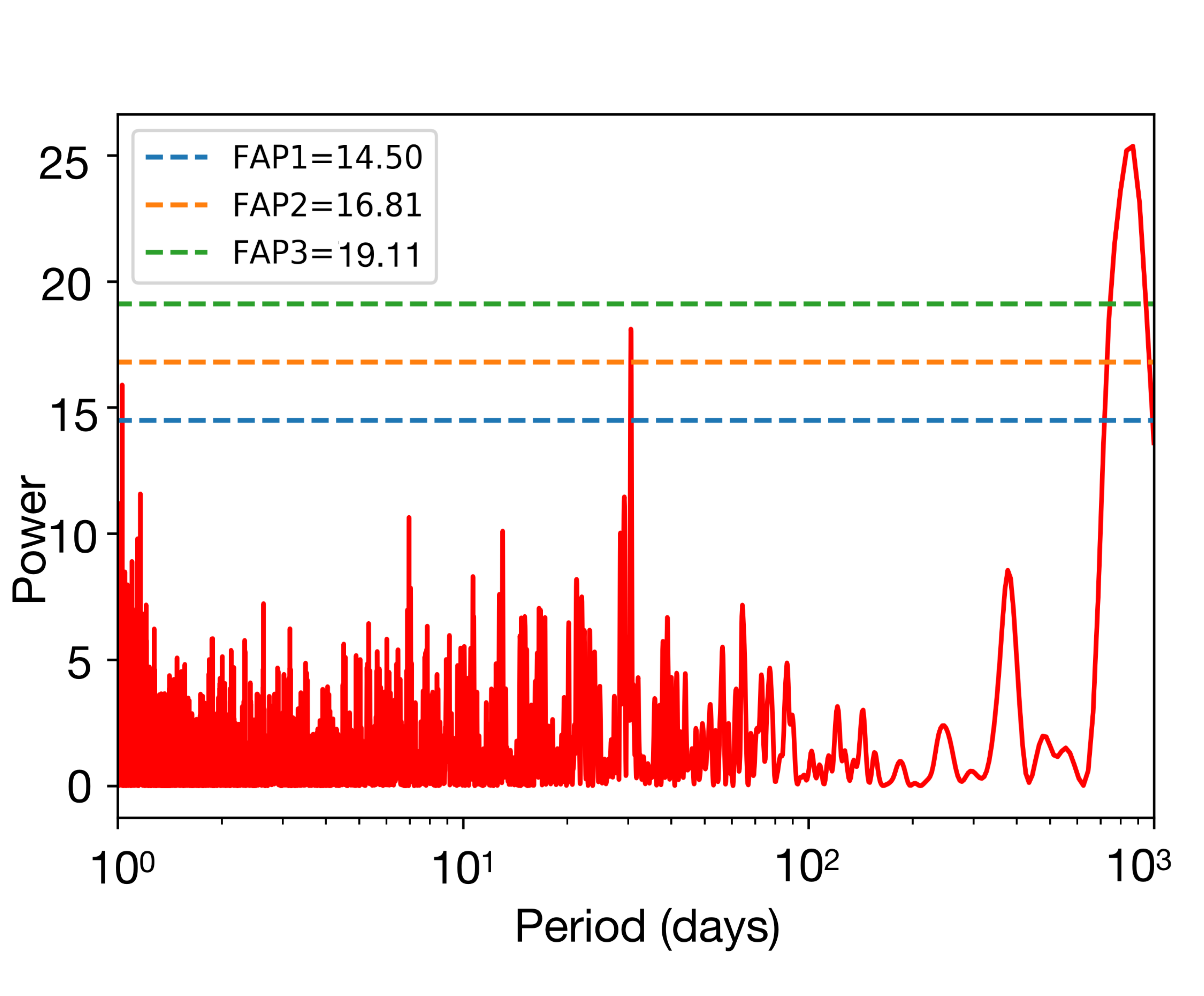}
    \caption{Lomb-Scargle Periodogram analysis of $\gamma$ Cep Ab. The value of power is the maximum power of periodic signals, dashed lines of FAP1, FAP2 and FAP3 represent the  1$\%$, 0.1$\%$, and 0.01$\%$ of FAP respectively. There is a significant periodic signal of roughly 900 days, which means this signal is truly periodic.}
    \label{fig:period}
\end{figure}

\begin{figure*}
\centering
       \includegraphics[width=1.95\columnwidth,height=16cm]{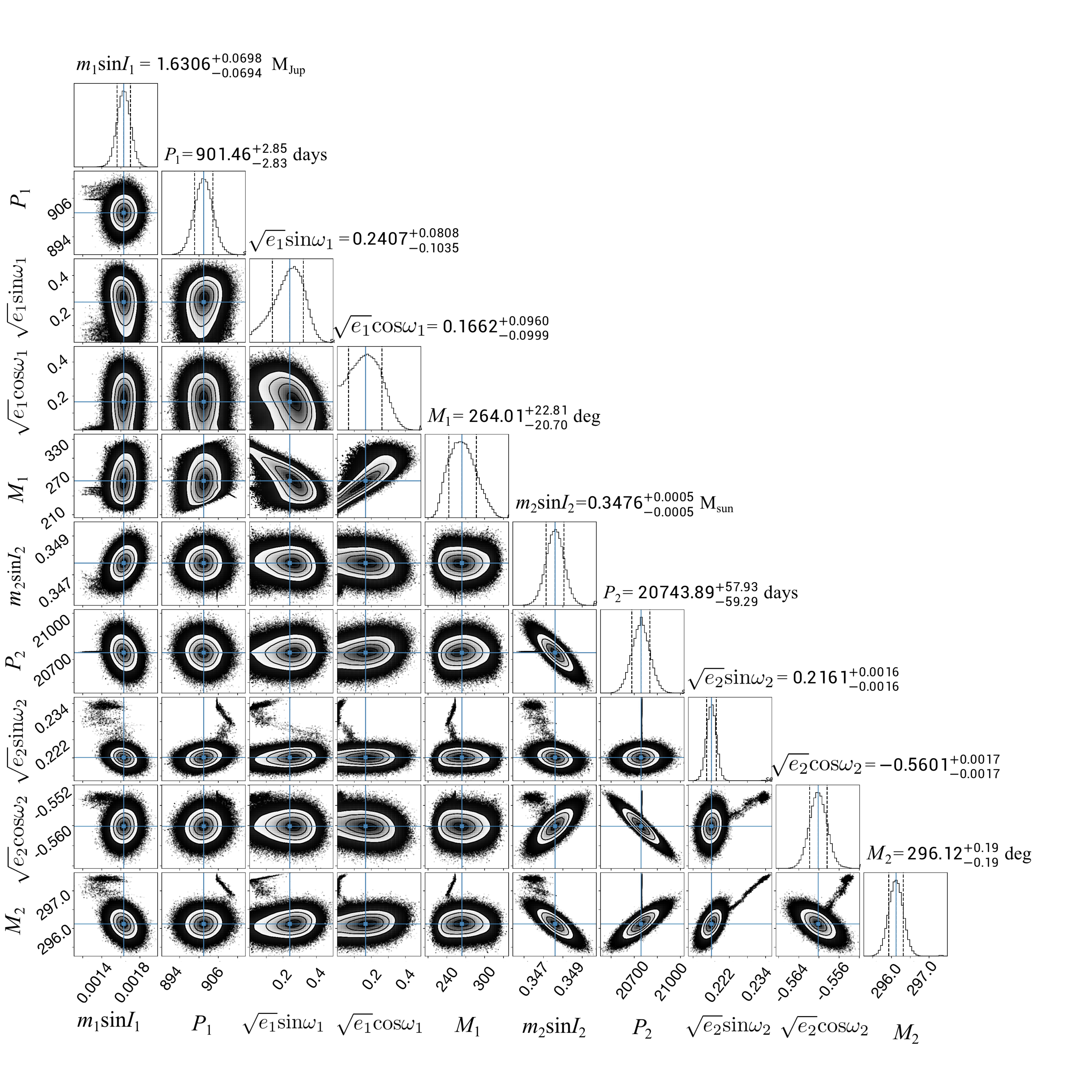} \caption{{The corner diagram of our N-body best-fitting solutions with $\chi^{2}$ = 1.44. Here $m\sin I$, $P$, $\sqrt{e}\sin\omega$, $\sqrt{e}\cos\omega$, $M$ with subscripts 1, 2 denote the minimum mass, orbital period, $\sqrt{e}$ vector and the mean anomaly of $\gamma$ Cep Ab and $\gamma$ Cep AB, respectively}. The figure illustrates the one and two dimensional projections of the posterior probability distributions of  parameters, whose structure directly shows the covariances between  two parameters. The distribution histograms of each parameter are listed on the diagonal.\label{fig:triangle}}
\end{figure*}

{Here we solve orbits of the planet and the star companion with the MCMC ensemble sampler \textit{emcee} \citep{Foreman-Mackey2013} in the full N-body model \citep{Ford2006,Nelson2016}.  Fourteen parameters of  $\{m_1\mathrm{sin}I_1$, $P_1$, $\sqrt{e_1}\mathrm{sin}\omega_1$, $\sqrt{e_1}\mathrm{cos}\omega_1$, $M_{1}$, $m_2\mathrm{sin}I_2$, $P_2$, $\sqrt{e_2}\mathrm{sin}\omega_2$, $\sqrt{e_2}\mathrm{cos}\omega_2$, $M_{2}\}$ plus RV offsets $\{RV_{0,1}$, $RV_{0,2}$, $RV_{0,3}$, $RV_{0,4}\}$ of four time series (CFHT, MOPS $\mathrm{\Rmnum1}$, MOPS $\mathrm{\Rmnum2}$, MOPS $\mathrm{\Rmnum3}$) are adopted for fitting at the first observation epoch (HJD-2444754.129) in the Jacobi reference frame \citep{Lee2003}. $\sqrt{e}\mathrm{sin}\omega$ and $\sqrt{e}\mathrm{cos}\omega$ are more efficient than $e$ and $\omega$ for low-eccentricity planets, and can avoid the situation of multiple solutions  \citep{Ford2006}.}

{Then we utilize the N-body integrator \textit{IAS15} \citep{Rein2015}, which stands for integrator with an adaptive step-size control, to compute the perturbed orbits of the planet and the secondary star in each step of \textit{emcee} sampler. Initial orbital inclinations and ascending nodes are assumed to be known \citep{reffert2011}: $I_1$ = 5.7$^{\circ}$, $\Omega_1$ = 37.5$^{\circ}$ or $I_1$ = 173.1$^{\circ}$, $\Omega_1$ = 356.1$^{\circ}$. Other initials include the planetary mass $m_1$ and the secondary mass $m_2$, the semi-major axis $a_{1,2}$, the eccentricity $e_{1,2}$ and the argument of periastron $\omega_{1,2}$ can be derived from the resultant fitting parameters. The integration precision is given to be $10^{-15}$ and the minimum time step is set to be 0.001 day. Theoretical RV signals of observation epochs are calculated from the Jacobian orbital elements, with RV semi-amplitude $K_{1,2}$ defined in \citet{Lee2003}:
 \begin{equation}
 \begin{aligned}
\label{equ:RV}
&K_{1}=\left(\frac{2 \pi G}{P_{1}}\right)^{1 / 3} \frac{m_{1} \sin I_{1}}{\left(m_{0}+m_{1}\right)^{2 / 3}} \frac{1}{\sqrt{1-e_{1}^{2}}}, \\
&K_{2}=\left(\frac{2 \pi G}{P_{2}}\right)^{1 / 3} \frac{m_{2} \sin I_{2}}{\left(m_{0}+m_{1}+m_{2}\right)^{2 / 3}} \frac{1}{\sqrt{1-e_{2}^{2}}},
 \end{aligned}
\end{equation}
where $P_{1,2}$, $I_{1,2}$, $e_{1,2}$ are time-variable orbital elements at each observation epoch.
 }

{To derive the N-body best-fitting solutions for this system, we run 100,000 steps for each of 28 walkers in the 14-dimensional parameter spaces. The mean acceptance fraction of sampler walkers is 0.36, which is suggested to be between 0.2 and 0.5 to produce representative samples \citep{Foreman-Mackey2013}, and the autocorrelation time for each fitting parameter ranges from 200 to 500 steps, which is smaller than our sampling steps to ensure the convergence of MCMC chains. Here the autocorrelation time represents the number of steps required for the chain to produce an independent sample \citep{Foreman-Mackey2013}.}

{Finally, we report two set of solutions of the minimum mass $m\sin I$, the RV semi-amplitude \textit{K}, the semi-major axis $a$, the eccentricity \textit{e}, the argument of periastron $\omega$ and the epoch of periastron passage $T_{0}$ in Table \ref{tab:rvresults} with $\chi^{2}$ = 1.48 for $I_1$ = 5.7$^{\circ}$ and  $\chi^{2}$ = 1.44 for $I_1$ = 173.1$^{\circ}$, respectively. In Figure \ref{fig:totfit} and \ref{fig:residual}, we plot the best-fitting RV signals from N-body solutions (\textit{black solid line}) with the measurements versus the epoch with $\chi^{2}$ = 1.44, with respect to $RV_{0,1}$ = $1290.0 ^{+18.5}_{-17.6}$ ms$^{-1}$, $RV_{0,2}$ =  $2035.2 ^{+20.3}_{-18.8} $ ms$^{-1}$, $RV_{0,3}$ =  $2231.1 ^{+19.2}_{-16.8}$ ms$^{-1}$ and $RV_{0,4}$ = $865.0 ^{+22.4}_{-19.4} $ ms$^{-1}$, respectively.}

By subtracting the theoretical radial velocity induced by the secondary, Figure \ref{fig:period} presents the Lomb-Scargle Periodogram analysis of the planet b, where the strongest signal reveals an orbital period of roughly 900 days. Figure \ref{fig:triangle} shows the posterior distributions of $\{m_1\mathrm{sin}I_1$, $P_1$, $\sqrt{e_1}\mathrm{sin}\omega_1$, $\sqrt{e_1}\mathrm{cos}\omega_1$, $M_{1}$, $m_2\mathrm{sin}I_2$, $P_2$, $\sqrt{e_2}\mathrm{sin}\omega_2$, $\sqrt{e_2}\mathrm{cos}\omega_2$, $M_{2}$$\}$, we report the median (50th percentile) of the posterior distribution as the best-fitting value, 1-$\sigma$ uncertainties are provided using the 16th and 84th percentiles.

\subsection{Estimation of the planetary mass}\label{subsec:inclinationfit}
To derive more reliable orbital solutions, high-precision radial velocity  and astrometric measurements are required to utilize in the fitting procedure at the same time.  As well-known, Gaia Data Release 2 (DR2) \citep{Gaia2018} was first accessible in 2018, and the science team recently announced the Gaia Early Data Release 3 (EDR3) \citep{Gaia2021}. For most sources in the Gaia catalogues, the sequential astrometric data are not yet available and will be released with Gaia Data Release 4. Thus the improvement of the planetary mass from Gaia will be expected in the future.

Table \ref{tab:rvresults} summarizes our derived orbital parameters of $\gamma$ Cep system. The minimum planetary mass is fitted to {be $1.7420^{+0.0733}_{-0.0743}$ $M_{\mathrm{Jup}}$ or $1.6306^{+0.0698}_{-0.0694}$ $M_{\mathrm{Jup}}$}. Note that the real mass of planet b is significantly dependent on the accuracy of radial velocity observations and the dynamical integrations, here we simply estimate the mass range of $\gamma$ Cep Ab. When the observed inclination of the planet is $I_1$ $\in$ [3.8$^{\circ}$, 20.8$^{\circ}$] \citep{reffert2011}, the estimated planetary mass $m_1$ $\in$ [5.0, 26.6] $M_{\mathrm{Jup}}$, while for $I_1$ $\in$ [166.6$^{\circ}$, 174.8$^{\circ}$] \citep{reffert2011}, it can be derived that $m_1$ $\in$ [7.1, 26.2] $M_{\mathrm{Jup}}$. In addition, based on the planetary mass limit of 16.9 $M_{\mathrm{Jup}}$ \citep{torres2007}, here we assume $m_1$ $\in$ [5, 16.9] $M_{\mathrm{Jup}}$.

To determine the current observed mutual inclination $i_{\rm mut}$ in $\gamma$ Cep system, we use the law of cosines for angles of a spherical triangle \citep{Gellert1977}:
\begin{equation}
\label{equ:itot}
\cos i_{\mathrm{mut}}=\cos I_1 \cos I_2 + \sin I_1 \sin I_2 \cos (\Omega_1 - \Omega_2),
\end{equation}
to derive $i_{\mathrm{mut}} = 113.9^{\circ}$ for $I_1$=5.7$^{\circ}$. Thus, there naturally arises a question - what kind of the dynamical process can trigger such high mutual inclinations of two bodies, whether the EKL mechanism can play a role in the secular evolution of $\gamma$ Cep Ab?  In the following, we describe the EKL mechanism and explore the secular evolution of $\gamma$ Cep Ab under EKL.

\begin{deluxetable*}{clccccc} \label{tab:rvresults}
\linespread{1.2}
\tablecaption{Orbital solutions of $\gamma$ Cep system}
\tablehead{
\multirow{2}*{Object} &\multirow{2}*{Parameters} & \multicolumn{2}{c}{This Work}& \multirow{2}*{\citet{hatzes2003}} &  \multirow{2}*{\citet{torres2007}} &  \multirow{2}*{\citet{neuhauser2007}}\\
\cline{3-4}
 & &$\chi^{2}$=1.48& $\chi^{2}$=1.44&&&
}
\startdata
\hline
& $K_{1}$ (m s$^{-1}$)& $28.08^{+1.45}_{-1.23}$ &$26.40^{+1.41}_{-1.18}$&$27.5 \pm 1.5$& $27.1 \pm 1.5$ &$27.0 \pm 1.5$\\
& $P_{1}$ (days) &$905.64\pm 2.83$& $901.46 ^{+2.85}_{-2.83}$&$905.574 \pm 3.08$ &$902.8 \pm 3.5$ & $902.9 \pm 3.5$\\
&$a_1$ (AU)& $2.1459\pm0.0048$&$2.1376^{+0.0048}_{-0.0047}$ & $2.13\pm0.05$& $1.94\pm0.06$& $2.044\pm0.057$\\
&$e_1$& $0.0724^{+0.0879}_{-0.0575}$ & $0.0856^{+0.0866}_{-0.0624}$&$0.12 \pm 0.05$& $0.113 \pm 0.058$ & $0.115\pm0.058$\\
$\gamma$ Cep Ab&$\omega_{1}$ (deg)& $48.47^{+4.56}_{-1.81}$ &$55.37^{+8.81}_{-4.57}$&$49.6 \pm25.6$& $63.0\pm27.0$&$63.0\pm27.0$ \\
&$T_{0,1}$ (HJD-2400000)& $53140.16^{+34.15}_{-38.31}$ &$53107.63^{+25.47}_{-30.90}$& $53121.9\pm66.9$ & $53146.0\pm72.0$ & $53146.0\pm71.0$\\
&$m_1 \sin I_1 \left(M_{\mathrm{Jup}}\right)$ & $1.7420^{+0.0733}_{-0.0743}$ & $1.6306^{+0.0698}_{-0.0694}$&$1.70 \pm 0.40$& $1.43 \pm 0.13$ &$1.60 \pm 0.13$\\
&$\Omega_1$ (deg)& 37.5$^{*}$ &356.1$^{*}$ &...&...&...\\
&$I_1$ (deg)& 5.7$^{*}$& 173.1$^{*}$&...& $>4.9$ &...\\
\hline
&$K_{2}$ (m s$^{-1}$) & $1821.70^{+0.95}_{-1.01}$&$1699.94^{+3.45}_{-3.19}$& $1820.0\pm49.0$&$1925.0\pm14.0$&$1932.0\pm14.0$\\
&$P_2$  (days)& $20754.66^{+59.01}_{-57.09}$&$20731.68^{+57.18}_{-59.53}$& $20750.658 \pm1568.6$&$24392\pm522$&$24654.375\pm511$\\
&$a_2$ (AU)& $18.6421^{+0.0392}_{-0.0381}$&$18.6217^{+0.0381}_{-0.0389}$ & $18.5 \pm 1.1$& $19.02\pm0.64$& $20.18\pm0.66$\\
&$e_2$ & $0.3603\pm0.0012 $& $0.3605\pm0.0026$&$0.3610\pm0.0230$& $0.4085 \pm 0.0065$&$0.4112\pm0.0063$\\
$\gamma$ Cep AB&$\omega_{2}$ (deg)& $158.92 \pm 0.20$& $158.90 \pm 0.20$&$158.76\pm1.20$&$160.96 \pm0.40$&$161.01\pm0.40$\\
&$T_{0,2}$ (HJD-2400000)& $48435.22^{+0.80}_{-0.37}$& $48435.04^{+0.46}_{-0.88}$& $48429.03\pm27.0$ & $48479.0\pm12.0$ & $48444.8\pm11.16$\\
&$m_2 \left(M_{\odot}\right)$ &$0.3991\pm0.0005$& $0.3986\pm0.0005$&... &$0.362 \pm0.022$& $0.409\pm0.018$\\
&$\Omega_2$ (deg) &$18.04\pm0.98^{**}$ & $18.04\pm0.98^{**}$&... &$13.0\pm2.4$&$18.04\pm0.98$\\
&$I_2$ (deg)&$119.3\pm1.0^{**}$&$119.3\pm1.0^{**}$ &...& $118.1\pm1.2$ &$119.3\pm1.0$\\
\enddata
 * The astrometric data fitting results adopted from \citet{reffert2011}.\\
**Orbital solutions adopted from imaging observations in \citet{neuhauser2007}.
 \\
\end{deluxetable*}

\section{Dynamical Model}\label{sec:model}
\subsection{Kozai--Lidov mechanism}\label{subsec:Lidov-Kozai}
In the secular evolution of triple body systems, the perturbation arising from the third body acts on a time scale much longer than its orbital period. One typical origin of secular resonance stems from the perturbation potential from adjacent orbits. When the mass of the perturbed body is negligibly small in comparison to those bodies in a hierarchical system, the test particle approximation comes into play. At this time, if the perturbation of a circular outer orbit works, the Hamiltonian of this system can be expanded to the quadrupole level, known as Kozai--Lidov mechanism as aforementioned.

The Kozai--Lidov mechanism addresses that the inclination and eccentricity of the inner test particle oscillate over secular evolution in a hierarchical system, where the test body and the primary star are surrounded by a distant companion \citep{vonZeipel1910, Kozai1962, Lidov1962}. With the secular approximation, the inner and the outer orbits only exchange angular momentum, thus the semi-major axes of orbits do not change. When the SMA ratio $\alpha_r$ = $a_1$/$a_2$ is a small parameter, and the perturbation term of the complete Hamiltonian can be expanded as a power series in $\alpha_r$ \citep{Naoz2016}:
\begin{equation}
\begin{aligned}
\label{equ:Hal}
\mathcal{H}&=\frac{k^{2} m_{0} m_{1}}{2 a_{1}}+\frac{k^{2} m_{2}\left(m_{0}+m_{1}\right)}{2 a_{2}} \\
&+\frac{k^{2}}{a_{2}} \sum_{j=2}^{\infty} \alpha^{j}_r M_{j}\left(\frac{r_{1}}{a_{1}}\right)^{j}\left(\frac{a_{2}}{r_{2}}\right)^{j+1}
P_{j}(\cos \Phi)
\end{aligned}
\end{equation}

\begin{equation}
\begin{aligned}
M_{j}&=m_{0} m_{1} m_{2} \frac{m_{0}^{j-1}-\left(-m_{1}\right)^{j-1}}{\left(m_{0}+m_{1}\right)^{j}} ,
\end{aligned}
\end{equation}
where $m_0$ is the mass of the primary, $m_{1}$ and $m_{2}$ are masses of the inner and the outer body. $k^2$ is the gravitational constant (with the mass unit of $M_{\odot}$ and the length unit of au), $r_{1}$ is the distance between $m_0$ and $m_1$, ${r}_2$ is the distance between the center of mass of the inner binary and $m_{2}$. $P_{\mathrm{j}}$ is the  Legendre polynomial, $\Phi$ is the angle between vectors $\mathbf{r}_{1}$ and $\mathbf{r}_2$ (the subscript $j=1,2$ represents the inner and the outer orbit, respectively).

\begin{figure}
        \includegraphics[width=0.85\columnwidth,height=6cm]{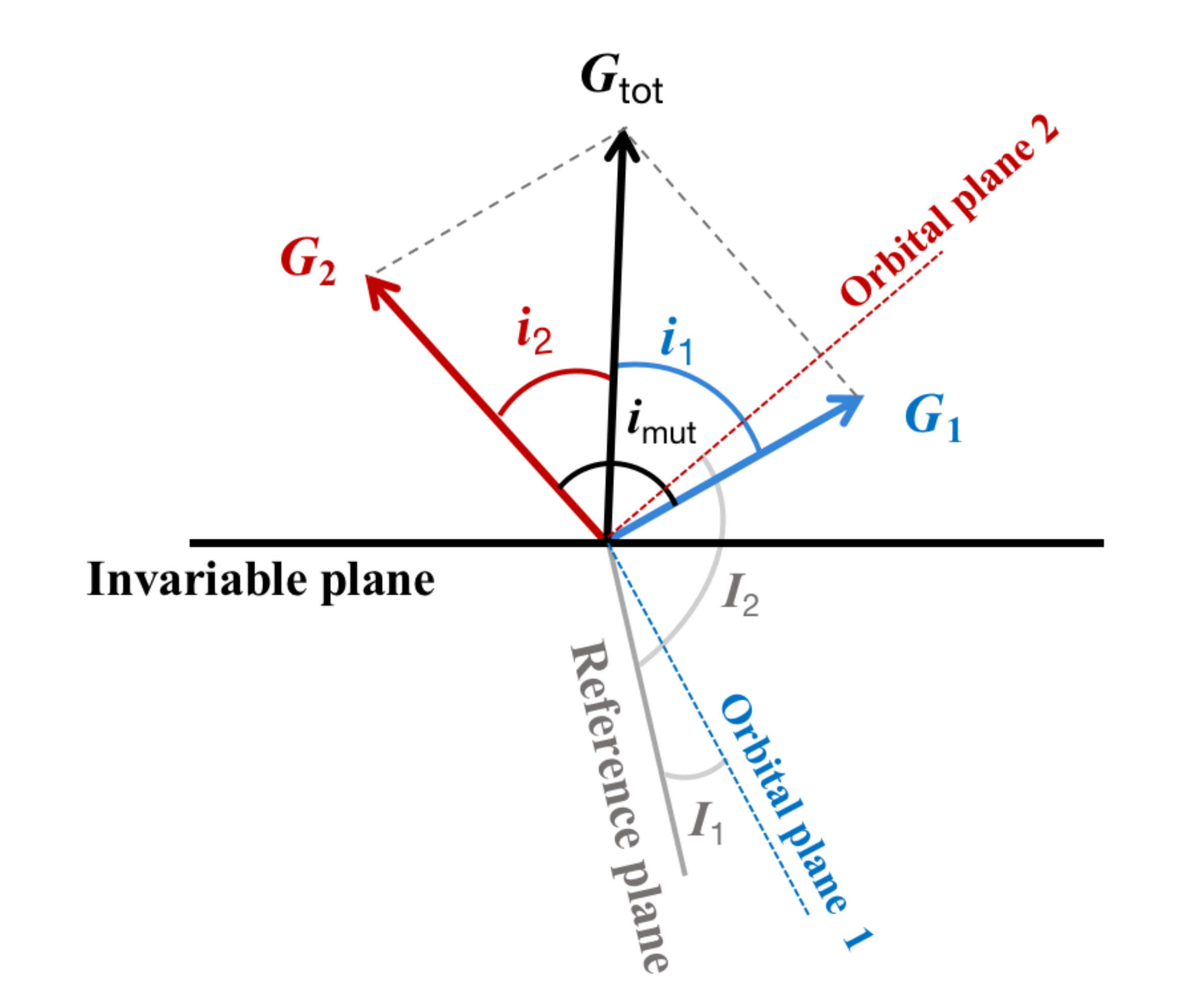}
    \caption{Geometry of the invariable plane and the reference plane system. $G_{\mathrm{1}}$ and $G_{\mathrm{2}}$ are angular momentum vectors of the inner and the outer orbits. $G_{\mathrm{tot}}$ is the total angular momentum vector of the system, the invariable plane is perpendicular to $G_{\mathrm{tot}}$. $i_{1,2}$ is the angle between the orbital angular momentum vector and $\boldsymbol{G}_{\mathrm{tot}}$, where $i_{\mathrm{mut}}$ is the mutual inclination between the inner and outer orbit. $I_{1,2}$ represents the observed inclination of the inner and outer orbit, which is the angle between the orbital plane and the reference plane (the sight plane). \label{fig:invariableplane}}
\end{figure}

The invariable plane reference frame is imported here to define three important Delaunay's elements \citep{Valtonen2006}: $l$, $g$, $h$ and their conjugate momenta $L$, $G$, $H$, where $l$, $g$, and $h$ are the notations of the mean anomaly $M$, the argument of periastron $\omega$ and the longitude of ascending node $\Omega$, respectively. As shown in Figure \ref{fig:invariableplane}, $\boldsymbol{G}_{\mathrm{tot}}$ is the total angular momentum vector of the system, which is conserved over the secular evolution.

Three conjugate momenta $L$, $G$, $H$ are expressed as \citep{Naoz2013}:
\begin{equation}
\label{equ:LGH}
\begin{aligned}
&L_{1}=\frac{m_{0} m_{1}}{m_{0}+m_{1}} \sqrt{k^{2}\left(m_{0}+m_{1}\right) a_{1}} ,\\
&L_{2}=\frac{m_{2}\left(m_{0}+m_{1}\right)}{m_{0}+m_{1}+m_{2}} \sqrt{k^{2}\left(m_{0}+m_{1}+m_{2}\right) a_{2}} ,\\
&G_{1}=L_{1} \sqrt{1-e_{1}^{2}}, \quad G_{2}=L_{2} \sqrt{1-e_{2}^{2}} ,\\
&H_{1}=G_{1} \cos i_{1}, \quad H_{2}=G_{2} \cos i_{2} ,\\
&G_{\mathrm{tot}}=H_{1}+H_{2} ,
\end{aligned}
\end{equation}
where $L$ is only determined by constant parameters, including the masses $m_0$, $m_1$, and $m_2$, the semi-major axis $a_1$, $a_2$, and the gravitational constant $k^2$. Thus $L$ is a constant for a specific system in the evolution, while $G$ and $H$ are time-varying. $G$ represents the magnitude of each orbit's angular momentum and $H$ is the component of $G$ along the z-axis.

According to the geometric relations and the assumption of $h_1-h_2=\pi$, the mutual inclination between the inner and outer orbit $i_{\mathrm{mut}}$ could be derived as \citep{Naoz2013}:
\begin{equation}
\label{equ:cositot}
\cos i_{\mathrm{mut}}=\cos (i_1+ i_2)=\frac{G_{\mathrm{tot}}^{2}-G_{1}^{2}-G_{2}^{2}}{2 G_{1} G_{2}},
\end{equation}

Generally, the equations of motion can be expressed by canonical relations of three conjugate momenta $L$, $G$, $H$ and three Delaunay's elements $l$, $g$, $h$.
As the mean anomaly can be eliminated under the double-averaged secular approximation, and $h_1$ and $h_2$ in equations of motion have been removed by the relation $h_1-h_2=\pi$. The time evolution for $\omega$, $e$ and $i$ can be easily derived from the reduced canonical relations \citep{Naoz2016}:
\begin{equation}
\label{equ:canonical}
\begin{aligned}
\frac{\mathrm{d} G_{j}}{\mathrm{~d} t}=\frac{\partial \mathcal{H}}{\partial g_{j}}, \quad \frac{\mathrm{d} g_{j}}{\mathrm{~d} t}=-\frac{\partial \mathcal{H}}{\partial G_{j}}
\end{aligned}
\end{equation}
where $j$ = 1, 2. The original non-planar three body can be reduced to 2-degree of freedom (DOF) dynamical system.

In \citet{Naoz2013}, $i_2$ is set to be 0, thus $i_{\mathrm{mut}}$ is equal to $i_1$. In this work, we treat the orbital plane of the secondary as the invariable plane, where $i_{\mathrm{mut}}$ = $i_1$. In the following sections, we redefine the orbital flip of the planet b as the variation of $i_{\mathrm{mut}}$ around $90^{\circ}$, instead of the real observed orbital inclination.

\subsection{The EKL mechanism in non-restricted triple systems}\label{subsec:eklmechanism}
In $\gamma$ Cep AbB system, $e_2$ is close to 0.36 and the estimated maximum mass of the planet  is near the deuterium-burning limit. Thus $\gamma$ Cep AbB is a non-restricted triple system with an eccentric outer orbit, where the octupole level terms in the Hamiltonian can become important, and the eccentricities of two orbits are coupled and oscillate simultaneously over secular evolution. \citet{Naoz2013} called it the EKL mechanism in non-restricted triple systems and derived the complete Hamiltonian of the system including the octupole term in addition to the quadrupole term:
\begin{equation}
\label{equ:octupoleH}
\begin{aligned}
\mathcal{H}_{3}(\Delta h \rightarrow \pi)=& \mathcal{H}_{\mathrm{quad}}+\mathcal{H}_{\mathrm{oct}}\\
=& C_{2}[(2+3 e_{1}^{2})(3 \cos ^{2}i_{\mathrm{mut}}-1)\\
&+15 e_{1}^{2} \sin ^{2} i_{\mathrm{mut}} \cos (2 g_{1})] \\
&+C_{3} e_{1} e_{2}[A \cos \phi+10 \cos i_{\mathrm{mut}} \sin ^{2} i_{\mathrm{mut}}\\
&\times(1-e_{1}^{2}) \sin g_{1} \sin g_{2}],
\end{aligned}
\end{equation}
where,
\begin{equation}
C_{2}=-\frac{k^{4}}{16} \frac{\left(m_{0}+m_{1}\right)^{7}}{\left(m_{0}+m_{1}+m_{2}\right)^{3}} \frac{m_{2}^{7}}{\left(m_{0} m_{1}\right)^{3}} \frac{L_{1}^{4}}{L_{2}^{3} G_{2}^{3}}
\end{equation}
\begin{equation}
 \epsilon_{M}=\left(\frac{m_{0}-m_{1}}{m_{0}+m_{1}}\right)\left(\frac{a_{1}}{a_{2}}\right) \frac{e_{2}}{1-e_{2}^{2}},
\end{equation}
\begin{equation}
  C_{3}=\frac{15}{4}\epsilon_{M}C_{2},
  \end{equation}
\begin{equation}
\begin{split}
&A=4+3 e_{1}^{2}-\frac{5}{2} B \sin i_{\mathrm{mut}}^{2}, \\
&B=2+5 e_{1}^{2}-7 e_{1}^{2} \cos \left(2 g_{1}\right),
\end{split}
\end{equation}
\begin{equation}
\cos \phi=-\cos g_{1} \cos g_{2}-\cos i_{\mathrm{mut}} \sin g_{1} \sin g_{2}.
\end{equation}
The secular perturbation theory in this specific case is called the EKL mechanism as previously mentioned. The time evolution for $\omega$, $e$, and $i_{\mathrm{mut}}$ can be derived through Equation (\ref{equ:octupoleH}) as well. Here, $\epsilon_{M}$ can be used as an indicator to characterize the strength of the octupole level effect in the system's perturbation potential.

\section{Secular evolution of $\gamma$ Cep A$\rm \lowercase {b}$}\label{sec:secular_evolution1}
As we described in Section \ref{sec:introduction}, the EKL mechanism plays a crucial role in the secular evolution of the celestial bodies. Here, in $\gamma$ Cep Ab B system, the inner planet b is assumed to be a sub-stellar object with a remarkable mass \citep{reffert2011}, thus the test particle approximation is not applicable. With $a_1$ = 2.1376 au and $a_2$ = 18.6217 au, the SMA ratio of planet b and the secondary $\alpha_r \sim 0.1$, thus the perturbation term of the complete Hamiltonian can be expanded as a power series in $\alpha_r $. As the eccentricity of the outer orbit $e_2 \sim 0.4$, the EKL mechanism can play a significant part in exploring the secular evolution of $\gamma$ Cep Ab B system. Here we will investigate the mutual inclination oscillations of $\gamma$ Cep Ab B system to examine whether the planet b can achieve an extremely inclined orbit through EKL.

In this Section, we are mainly concerned on the flip conditions and timescale for the first flip of $i_1$, which is equal to $i_{\mathrm{mut}}$. The stability of the rolling-over orbits is further explored. To investigate the relative global dynamics of the system, we show several kinds of representative planes of initial conditions. In Section \ref{subsec:inclination1}, the ($e_1$, $i_{\mathrm{mut}}$) plane is first used to search the flip conditions. In Section \ref{subsec:stability2}, to obtain the global, qualitative structure of the averaged 2-DOF Hamiltonian system, we import the representative plane of ($e_1$, $e_2$)  \citep{Michtchenko2004} and the phase space of ($e_1$, $g_1$)  \citep{Tan2020}. Then we apply the ($\sqrt{1-e_{1}^{2}}$, $\sqrt{1-e_{2}^{2}}$) plane as the representative plane to study the relative global dynamical features of $\gamma$ Cep system, and employ ($e_1$, $g_1$) plane to theoretically define quasi-periodic and circulating orbits.

\subsection{Orbital flip conditions}\label{subsec:inclination1}
The investigation of the amplitude of inclination oscillation reveals that $\gamma$ Cep Ab could evolve into extremely inclined orbit.  However, one may have difficulty in accurately predicting theoretical correlation between the extent of orbital inclination excitation and initial conditions. \citet{Katz2011} defined a complex function of analytical critical flip condition by averaging over the quadrupole level effect.

We provide the planetary mass $m_1$ $\in$ [5, 16.9] $M_{\mathrm{Jup}}$ in Section \ref{subsec:inclinationfit}. Here we assume the mass of $\gamma$ Cep Ab to be $m_1$ = $\{5, 9, 11, 15\}$ $M_{\mathrm{Jup}}$, then calculate the ranges of perturbation Hamiltonian $\mathcal{H}$ and the total angular momentum $G_{\mathrm{tot}}$ for each value of $m_1$ with known parameters $\{m_0, m_2, a_1, a_2\}$ and $e_1$ $\in$ [0, 1], $e_2$ $\in$ [0.35, 0.45], $i_{\mathrm{mut}}$ $\in$ [0$^{\circ}$, 180$^{\circ}$]. Furthermore, we set $g_1$ = 0$^{\circ}$ and $g_2$ = 0$^{\circ}$ in the initial conditions, since $g_1$ and $g_2$ can always go through either 0 or 180$^{\circ}$ over  secular evolution.

\begin{figure*}
\begin{center}
\subfigure[]{\includegraphics[width=1\columnwidth,height=7cm]{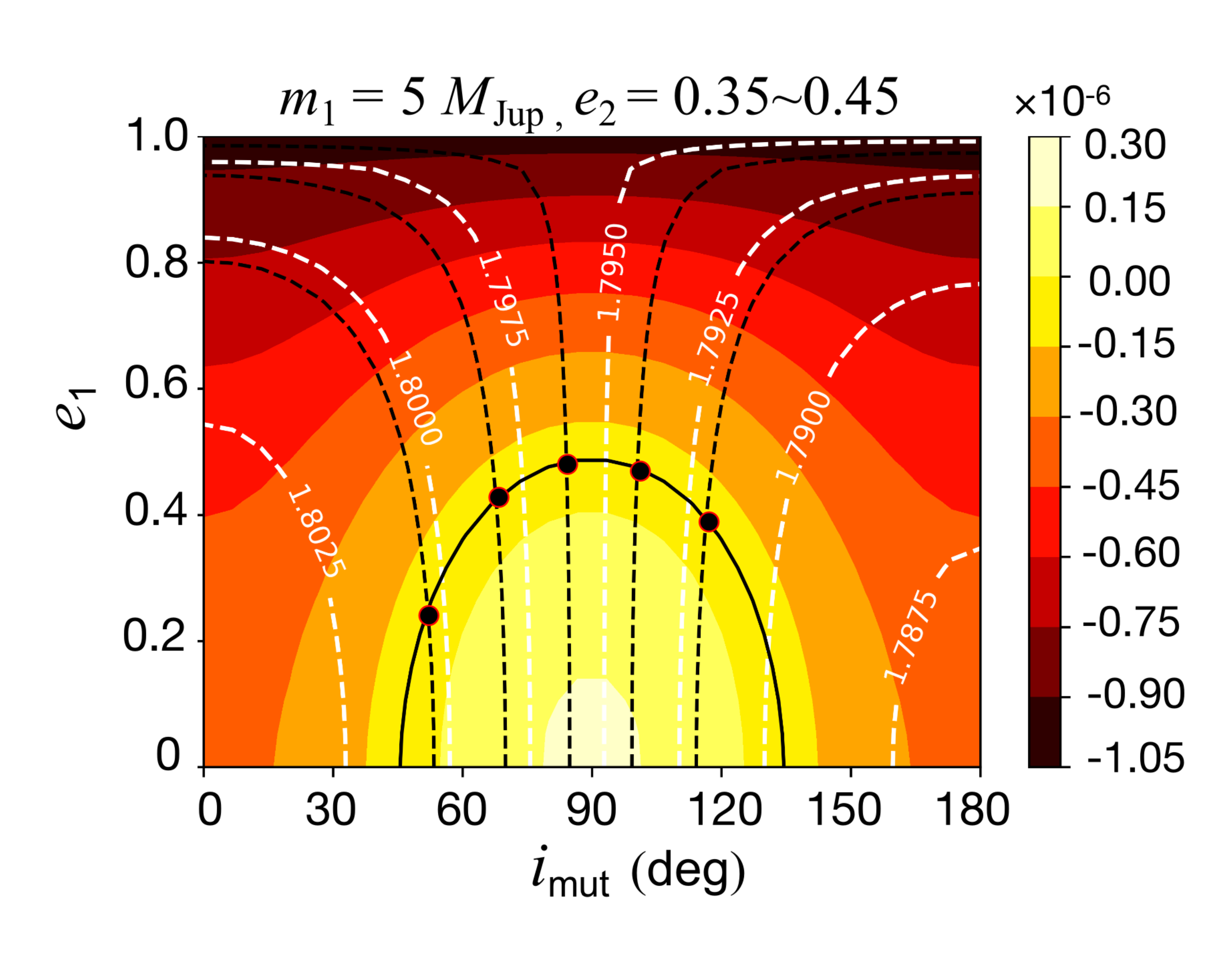}}
\subfigure[]{\includegraphics[width=1\columnwidth,height=7cm]{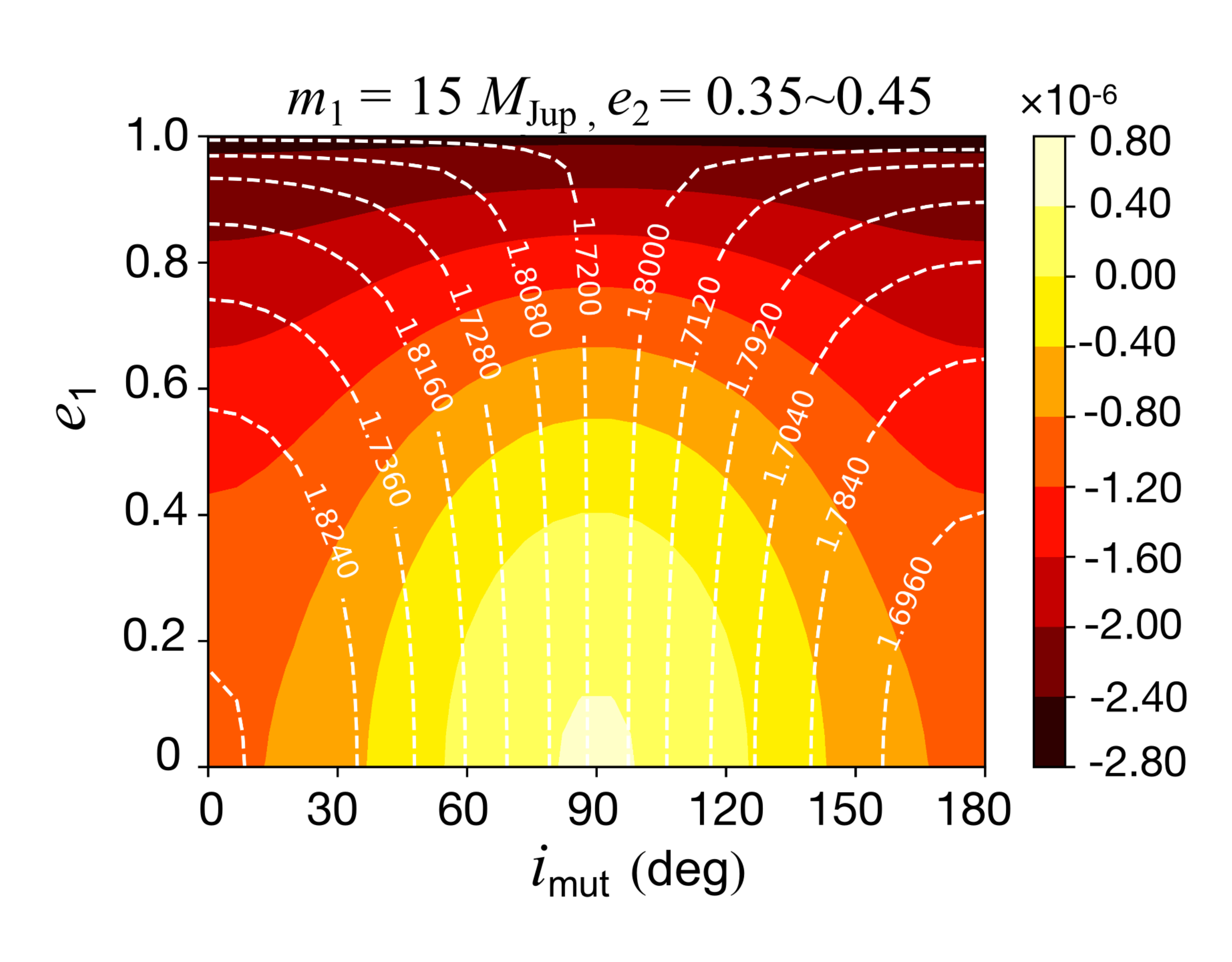}}
\caption{The contour maps of the total angular momentum $G_{\mathrm{tot}}$ and the perturbation Hamiltonian $\mathcal{H}$ for $m_1 $= 5 and 15 $M_{\mathrm{Jup}}$ in the initial conditions plane of ($e_1$, $i_{\mathrm{mut}}$). The white dashed lines represent the constant values of $G_{\mathrm{tot}}$ for $e_2$ $\in$ [0.35, 0.45], the colour bars on the right indicate different values of the perturbation Hamiltonian. In panel (a), when  $e_2$ selects 5 different values, there will be 5 level curves of $G_{\mathrm{tot}}$. We also mark out the initial conditions for $G_{\mathrm{tot}}$ = 1.75318 (the black dashed lines) and $\mathcal{H}$ = $-8.06 \times 10^{-8}$ (the solid line) with black dots, corresponding to data points with red error bars in Figure \ref{fig:imax1} (a).}
    \label{fig:H_G_contour}
 \end{center}
\end{figure*}

The contour maps of the total angular momentum $G_{\mathrm{tot}}$ and the perturbation Hamiltonian $\mathcal{H}$ are simultaneously plotted in the ($e_1$, $i_{\mathrm{mut}}$) plane. In Figure \ref{fig:H_G_contour}, we show two examples of $m_1$ = $\{5, 15\}$ $M_{\mathrm{Jup}}$ with $e_2$ $\in$ [0.35, 0.45], which is around the current observation and will not vary significantly over the evolution. For other planetary masses, the structures of the $\mathcal{H}$-$G_{\mathrm{tot}}$ contour maps have similar features with different ranges of $G_{\mathrm{tot}}$ and $\mathcal{H}$. When $m_1 $= 5 $M_{\mathrm{Jup}}$, $G_{\mathrm{tot}}$ $\in$ [1.7050, 1.8025], $\mathcal{H}$ $\in$ [$-1.05\times 10^{-6}$, $3.0\times 10^{-7}$], when $m_1 $= 9 $M_{\mathrm{Jup}}$, $G_{\mathrm{tot}}$ $\in$ [1.6994, 1.8137], $\mathcal{H}$ $\in$ [$-1.75\times 10^{-6}$, $5.0\times 10^{-7}$]. When $m_1 $= 11 $M_{\mathrm{Jup}}$, $G_{\mathrm{tot}}$ $\in$ [1.6975, 1.8187], $\mathcal{H}$ $\in$ [$-2.10\times 10^{-6}$, $6.0\times 10^{-7}$], when $m_1 $= 15 $M_{\mathrm{Jup}}$, $G_{\mathrm{tot}}$ $\in$ [1.696, 1.824] and $\mathcal{H}$ $\in$ [$-2.8\times 10^{-6}$, $8.0\times 10^{-7}$]. Cross points of $G_{\mathrm{tot}}$ and $\mathcal{H}$ contours indicate all the possible initial conditions of $i_{\mathrm{mut}}$ and $e_1$ over the secular evolution. The range of values of $G_{\mathrm{tot}}$ and $\mathcal{H}$ in Figure \ref{fig:H_G_contour} will change with variational planetary masses, while the structure of contours is similar.

To further derive the evolution results of these general initial conditions, we uniformly choose the specific values of $\mathcal{H}$ and $G_{\mathrm{tot}}$ between the upper and lower limit in Figure \ref{fig:H_G_contour}. For a given $\mathcal{H}$, based on the conserved Hamiltonian and total angular momentum shown in Equation (\ref{equ:LGH}) and Equation (\ref{equ:octupoleH}) of the octupole perturbation theory, we choose at least 40 cases of initial $e_1$ and $i_{\mathrm{mut}}$. Considering the conservation of energy and total angular momentum, we further explore the extreme value of the orbital inclination and orbital stability under these selected initials in detail.

In the non-test particle approximation under the classical Kozai--Lidov mechanism, the eccentricity and the inclination oscillate regularly in a well-defined timescale $t_{\mathrm{quad}}$ \citep{Antognini2015}:
\begin{equation}
\label{equ:t2}
t_{\mathrm{quad}} \sim \frac{16}{15} \frac{a_{2}^{3}\left(1-e_{2}^{2}\right)^{3 / 2} \sqrt{m_{0}+m_{1}}}{a_{1}^{3 / 2} m_{2} k}.
\end{equation}
This relationship was derived under the consideration of the equation of motion of $\omega$, by integrating between the maximum and minimum eccentricities. Here $t_{\mathrm{quad}}$ can be applied to estimate the timescale in the EKL scenario.

According to values of $m_0$, $m_1$, $m_2$, $e_2$, $a_1$, and $a_2$, the quadrupole period $t_{\mathrm{quad}}$ of $\gamma$ Cep system is estimated to be $\sim$ 1000 yr, which is consistent with our numerical simulation results. Here we investigate secular evolution of $\gamma$ Cep Ab by considering a diverse planetary mass and perform the simulation for 100 Myr ($\sim 10^5~ t_{\mathrm{quad}}$) using RKF7(8) integrator. The observed orbital inclinations $I_1$ and $I_2$ are required to calculate the constant total angular momentum and Hamiltonian. Hereafter, $i_1$ and $i_2$ denote the angles of the orbital plane relative to the invariable plane. However, the true orbital inclinations of the inner and outer orbits over the evolution are not well known.

\begin{figure*}
\begin{center}
\subfigure[]{\includegraphics[width=1\columnwidth,height=7cm]{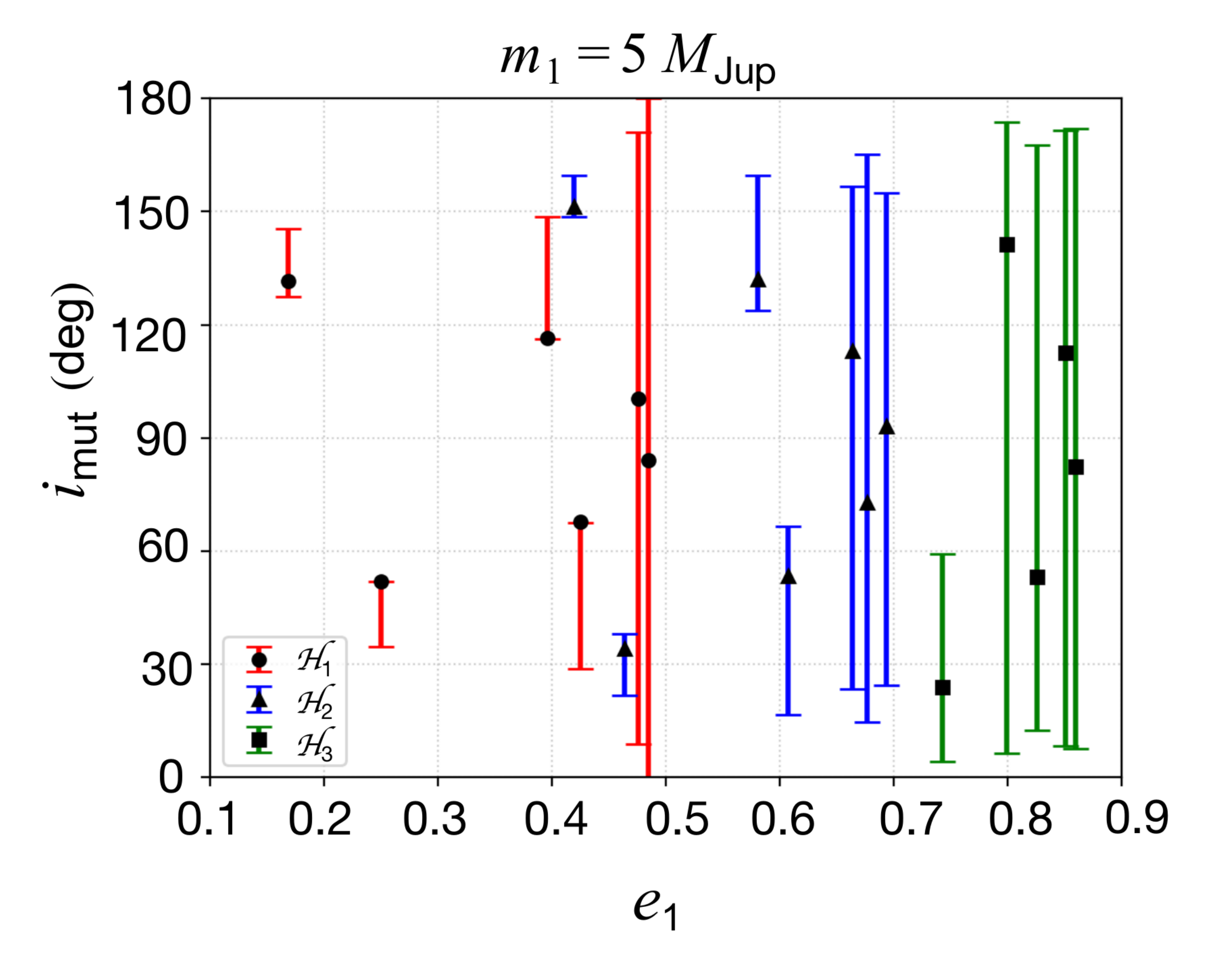}}
\subfigure[]{\includegraphics[width=1\columnwidth,height=7cm]{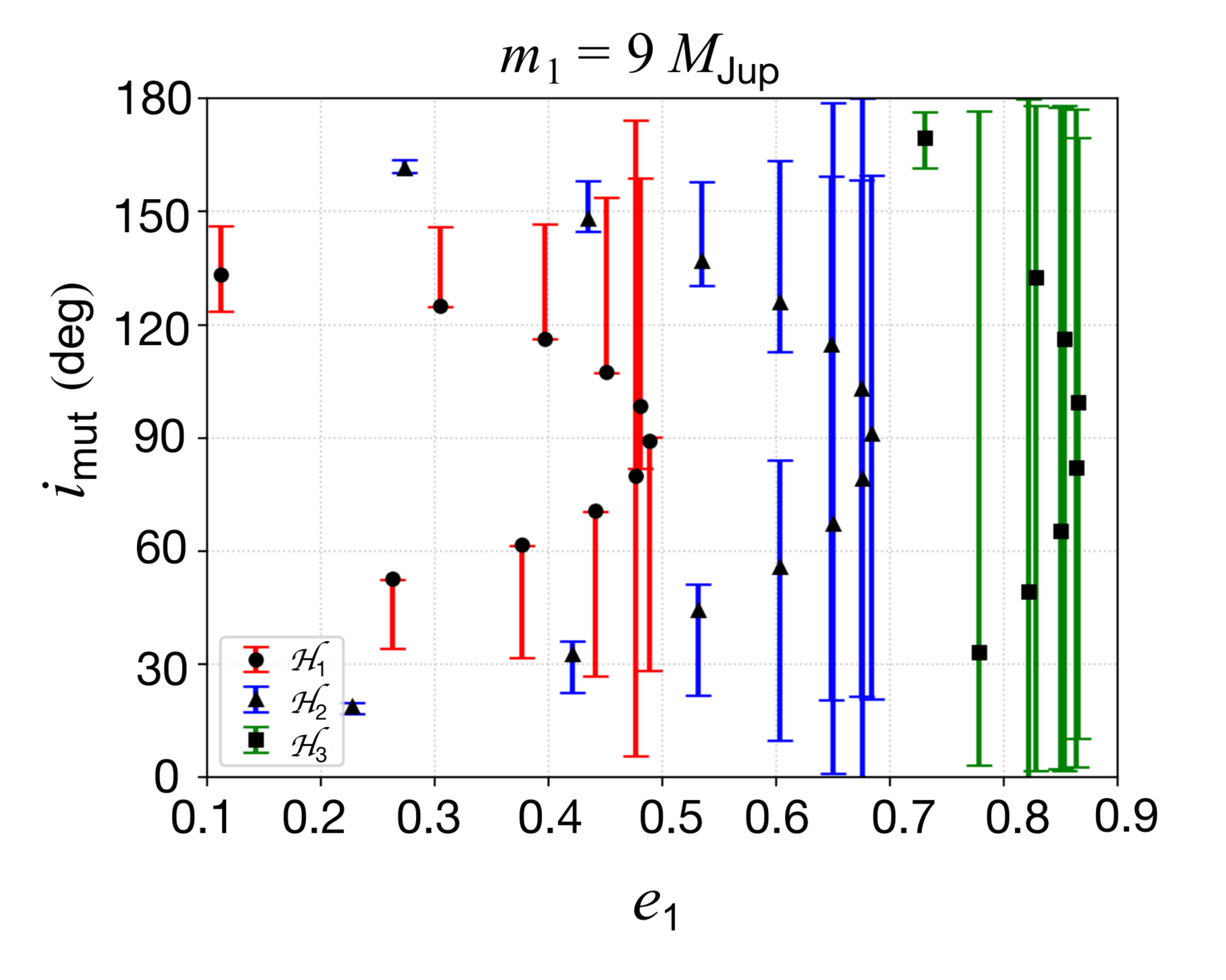}}\\
\subfigure[]{\includegraphics[width=1\columnwidth,height=7cm]{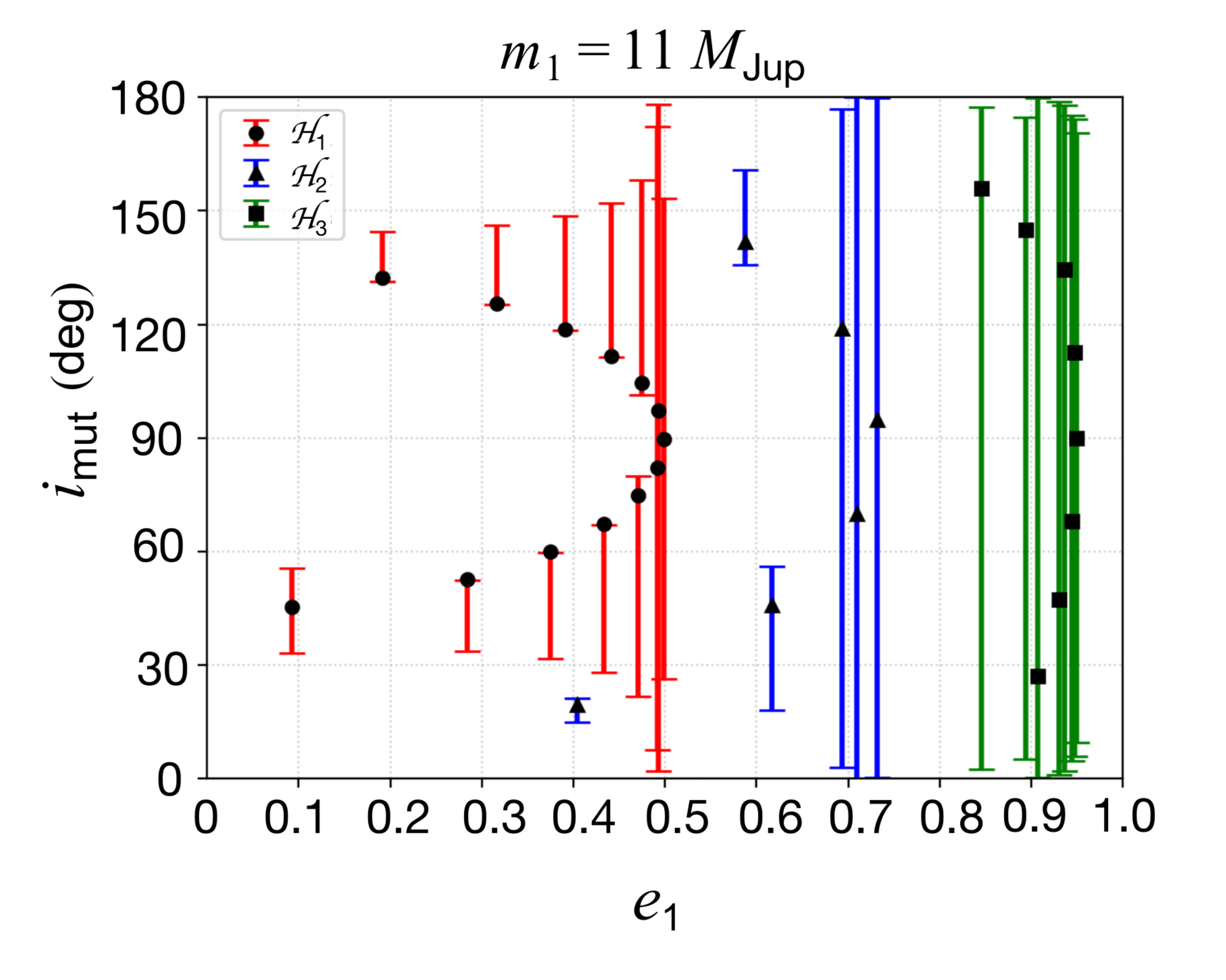}}
\subfigure[]{\includegraphics[width=1\columnwidth,height=7cm]{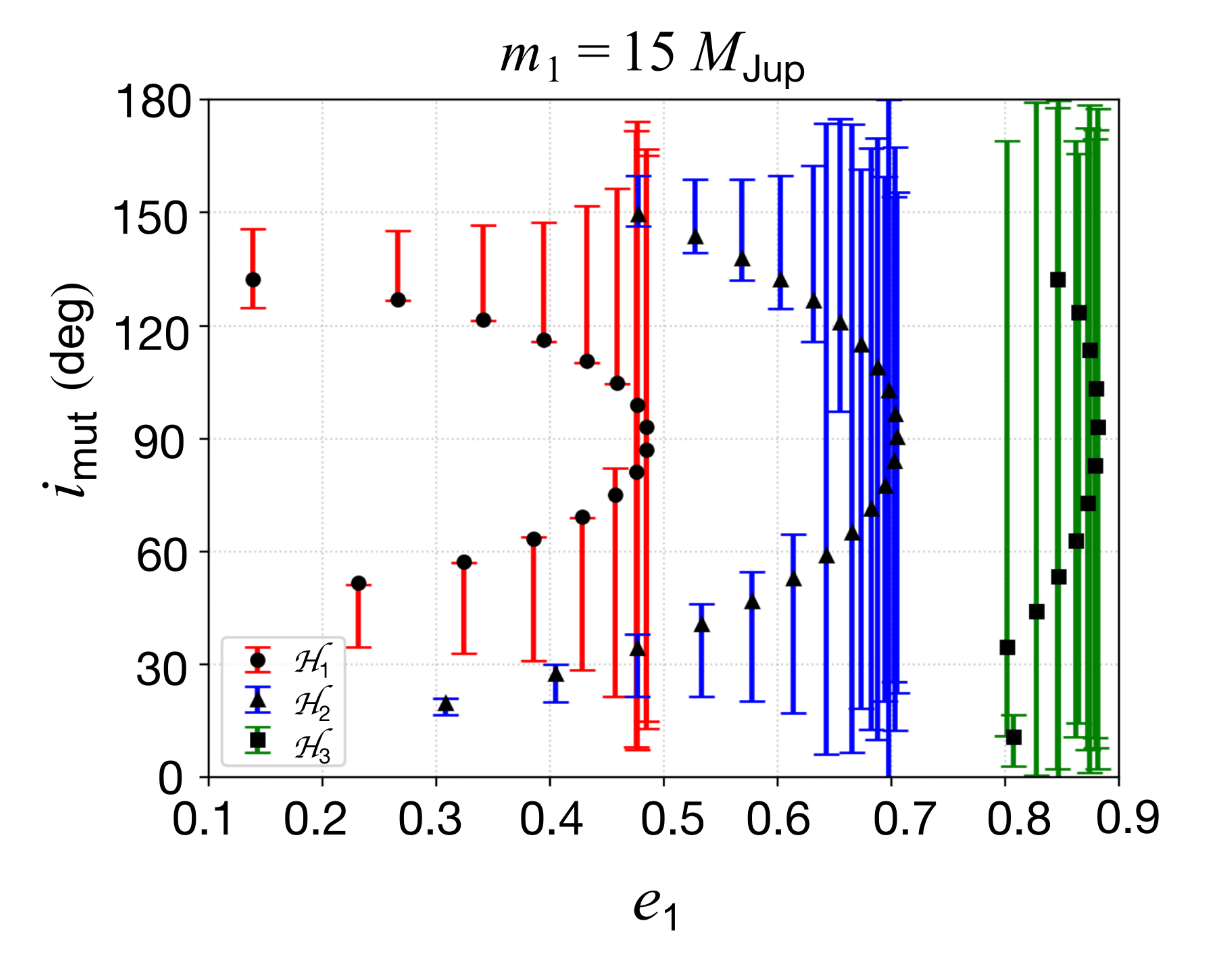}}
\caption{Ranges of $i_{\mathrm{mut}}$ for $m_1 $= 5 , 9, 11, 15 $M_{\mathrm{Jup}}$ over the timescale of 100 Myr. Black dots represents the position of initial parameters $\{e_{1,0}, i_{\mathrm{mut,0}}\}$. The upper limit and the lower limit of the "error bar" represent the maximum and the minimum values of the mutual orbital inclination $i_{\mathrm{mut}}$, respectively.  Different colors correspond to specific values of the perturbation Hamiltonian.}
    \label{fig:imax1}
 \end{center}
\end{figure*}

The panels (a) -- (d) of Figure \ref{fig:imax1} each plots three sets of typical outcomes with {$\mathcal{H}_1$, $\mathcal{H}_2$ and $\mathcal{H}_3$}. If $m_1$ = 5 $M_{\mathrm{Jup}}$, both the prograde and retrograde orbits will flip when $e_1 < 0.5 $ for $\mathcal{H}_1$ = $-8.06 \times 10^{-8}$,  $e_1 > 0.65$ for $\mathcal{H}_2$ = $-4.95 \times 10^{-7}$,  and  $e_1 > 0.8$ for $\mathcal{H}_3$ = $-6.08 \times 10^{-7}$. We note that most intense orbital flips take place when the initial $i_{\mathrm{mut}}$ is close to 90$^{\circ}$.

From the estimated $m_1 \sin I_1$  in Table \ref{tab:rvresults} and the law of cosines for the angles of a spherical triangle in Equation (\ref{equ:cositot}), if $m_1$ = 5 $M_{\mathrm{Jup}}$ and the observed inclination of planet b is roughly 20$^{\circ}$, the mutual inclination $i_{\mathrm{mut}}$ $\sim$ 100$^{\circ}$ as $I_2$ = 119.3$^{\circ}$. As a consequence, it is clear that under the selected values of $\mathcal{H}_1$, $\mathcal{H}_2$, $\mathcal{H}_3$, such high mutual inclination could always be achieved for original prograde orbits. In fact, as shown in Figure \ref{fig:imax1}, the evolution of mutual inclination of prograde and retrograde orbits is approximately symmetric about 90$^{\circ}$, which can also been obtained from Equation (\ref{equ:octupoleH}).

Similarly, if $m_1$ = 9 $M_{\mathrm{Jup}}$, the observed inclination of planet b is nearly 11$^{\circ}$, thus the mutual inclination $i_{\mathrm{mut}}$ is estimated to be 108$^{\circ}$. When $\mathcal{H}_1$ = $-1.45\times10^{-7}$, $e_1$ $\sim$ 0.5, $\mathcal{H}_2$ = $-6.35 \times 10^{-7}$, $e_1 \sim 0.65 $ and $\mathcal{H}_3$ = $-1.22 \times 10^{-6}$, $e_1 \sim 0.8$, the amplitude of inclination for orbital flip could be raised up to about 180$^{\circ}$.

When $m_1$ is equal to 11 $M_{\mathrm{Jup}}$, the calculated inclination of $\gamma$ Cep Ab is about 9$^{\circ}$ and the target mutual inclination $i_{\mathrm{mut}}$ $\sim$ 110$^{\circ}$ can be achieved when $\mathcal{H}_1$ = $-1.98\times10^{-7}$ and $e_1$ $\sim$ 0.5, $\mathcal{H}_2$ = $-8.56 \times10^{-7}$ with $e_1$ $\sim$ 0.7, and $\mathcal{H}_3$ = $-1.7 \times 10^{-6}$ with $e_1$ $\sim$ 0.85. Thus the critical value of initial eccentricity for orbital flips of $\mathcal{H}_2$ in Figure \ref{fig:imax1} (c) is larger than that of Figure \ref{fig:imax1} (a), (b) and (d).

When $m_1$ is 15 $M_{\mathrm{Jup}}$, the observed inclination of $\gamma$ Cep Ab is about 6$^{\circ}$ and the critical conditions for the target mutual inclination $i_{\mathrm{mut}}$ $\sim$ 113$^{\circ}$ are: $\mathcal{H}_1$ = $-2.4\times10^{-7}$ with $e_1$ $\sim$ 0.5, $\mathcal{H}_2$ = $-1.44 \times10^{-6}$ with $e_1$ $\sim$ 0.65, and $\mathcal{H}_3$ = $-2.03 \times 10^{-6}$, $e_1$ $\sim$ 0.8. In Figure \ref{fig:imax1} (d), for $\mathcal{H}_2$ = $-1.44 \times10^{-6}$, orbital flip occurs when the initial $i_{\mathrm{mut}} < $ 60$^{\circ}$. Thus it is easier for $\gamma$ Cep Ab to reach the target mutual inclination with $m_1$ = 15 $M_{\mathrm{Jup}}$, $e_1 < 0.7$, and the critical initial $i_{\mathrm{mut}}$ lower than $60^{\circ}$.

Above all, we conclude that the initial conditions for orbital flips under investigation for $\gamma$ Cep system are $e_1 > 0.5$ and $i_{\mathrm{mut}}$ $\in$ [60$^{\circ}$, 120$^{\circ}$], and various planetary masses simply affect the critical eccentricity when $e_1 > 0.6$, with little influence on the  maximum mutual inclination. The distribution tendency of flipping conditions with various initial inclination and eccentricity is similar to that of \citet{Lei2022}, while we perform the variation of flipping conditions under different $m_1$, which affects the octupole-level factor $\epsilon_M$.

\subsection{The orbital flip timescale}\label{subsec:stability1}
To in-depth understand the rolling-over orbits reported in Section \ref{subsec:inclination1}, we will further explore the orbital flip timescale over secular evolution, which may rely on the initial conditions, e.g., $m_1$, $e_1$ and $i_{\mathrm{mut}}$. The duration of flip is critical for the observation possibility of potential extremely inclined S-type planets which are transforming between prograde and retrograde orbits.
An approximate analytical timescale for the first flip for the non-chaotic orbit when $m_{1} \rightarrow 0$ was given by \citet{Katz2011} and \citet{Antognini2015}:
\begin{equation}
\label{equ:tflip}
t \sim \frac{128}{15 \pi} \frac{a_{2}^{3}}{a_{1}^{3 / 2}} \frac{\sqrt{m_{0}}}{k m_{2}} \sqrt{\frac{10}{\epsilon}}\left(1-e_{2}\right)^{3 / 2},
\end{equation}
where,
\begin{equation}
\label{equ:tflip}
 \epsilon \equiv \frac{e_{2}}{1-e_{2}^{2}} \frac{a_{1}}{a_{2}}.
\end{equation}
This theoretical flip timescale is effective when the initial conditions meet $e_{1} \rightarrow 0,~ \omega_{1} \rightarrow 0, \text { and } i_{\mathrm{mut}} \rightarrow 90^{\circ}$. Note that the flip timescale for the circular test particle approximation is mainly determined by $e_2$ when the object masses and the orbital SMA are fixed. Nevertheless, in the case of high-inclination oscillation, the timescale for the first flip is difficult to be quantitatively assessed because this evolution is likely to be chaotic.
\begin{figure*}
\begin{center}
\subfigure[]{\includegraphics[width=1\columnwidth,height=7cm]{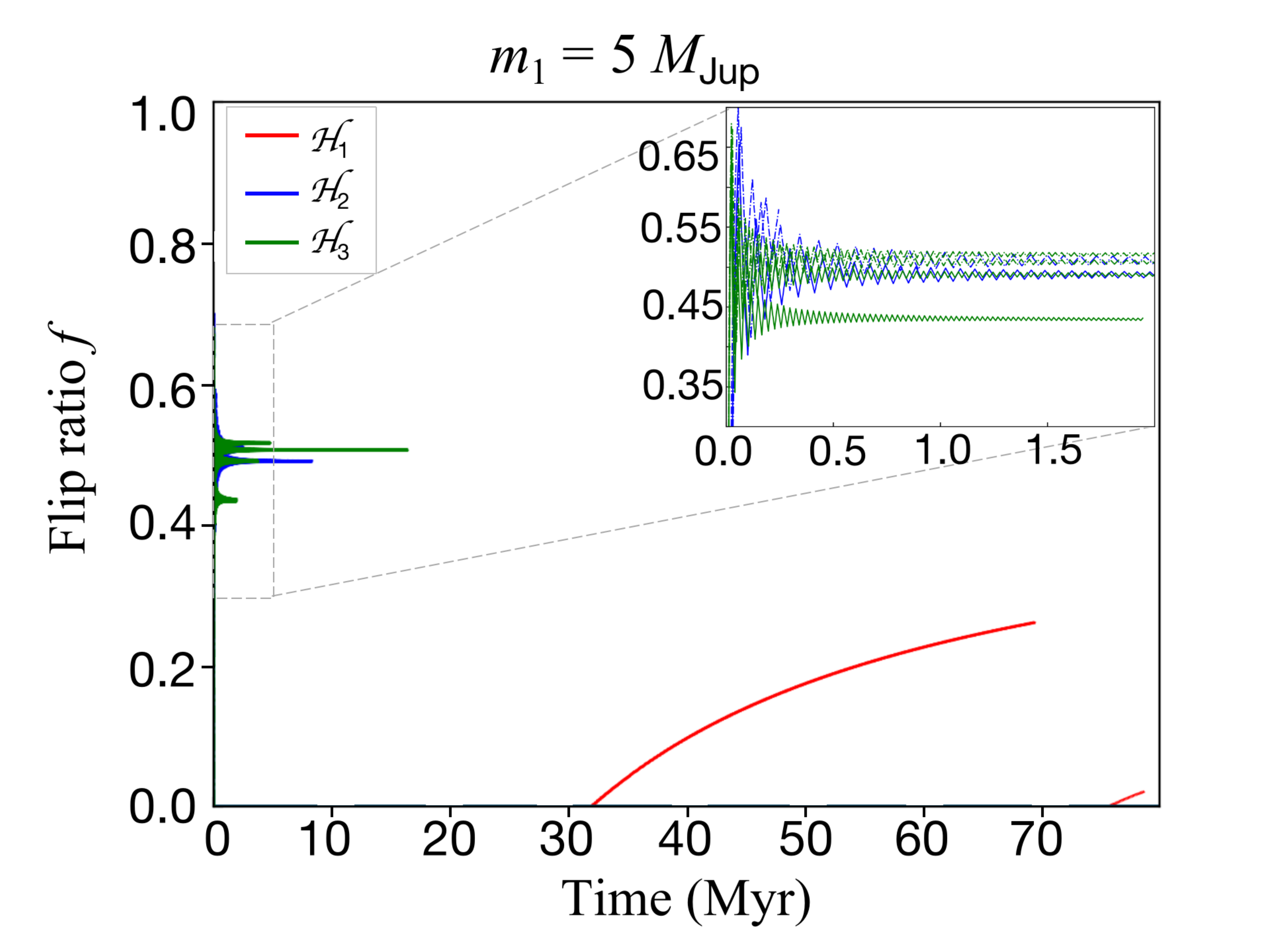}}
\subfigure[]{\includegraphics[width=1\columnwidth,height=7cm]{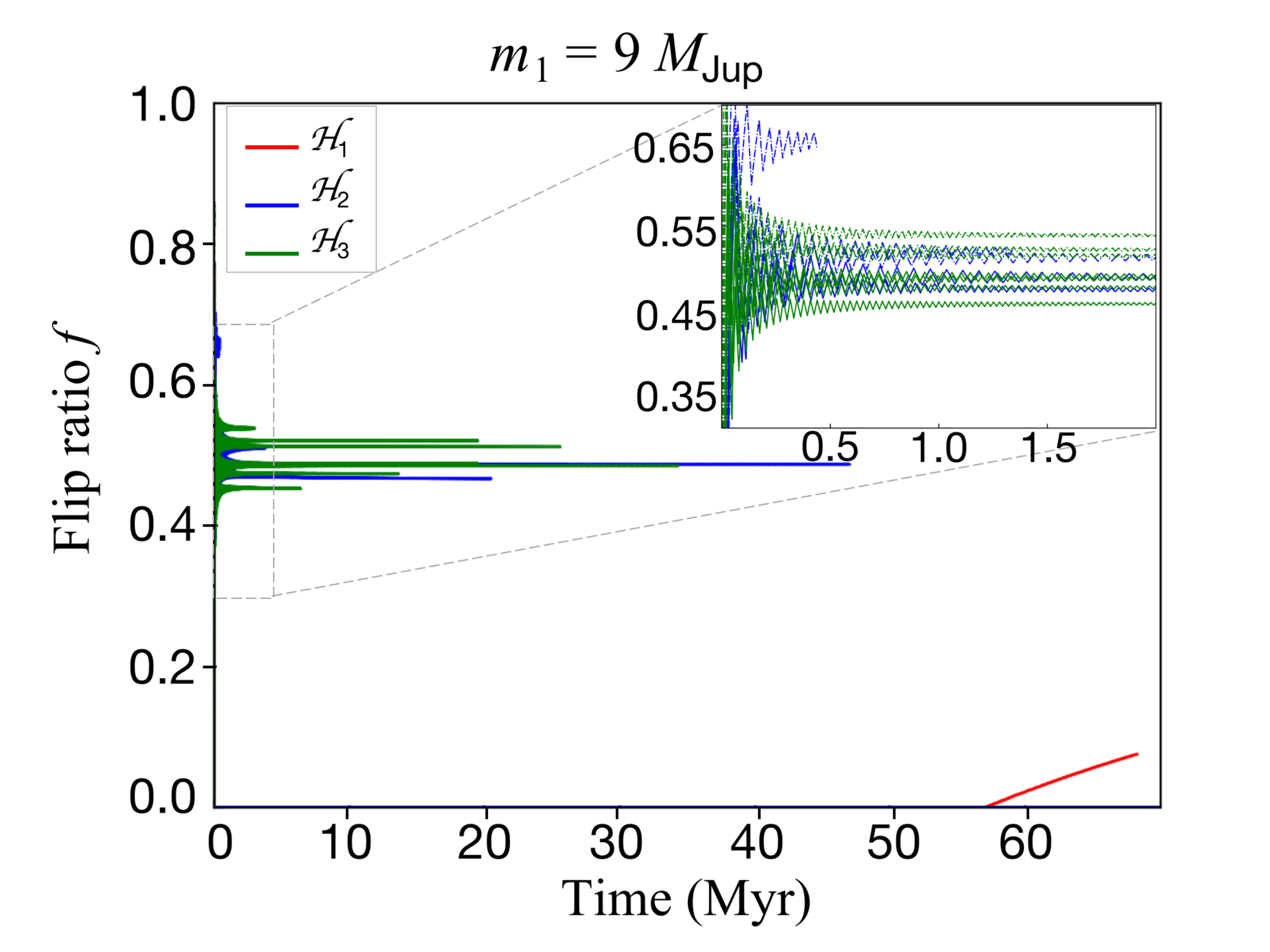}}\\
\subfigure[]{\includegraphics[width=1\columnwidth,height=7cm]{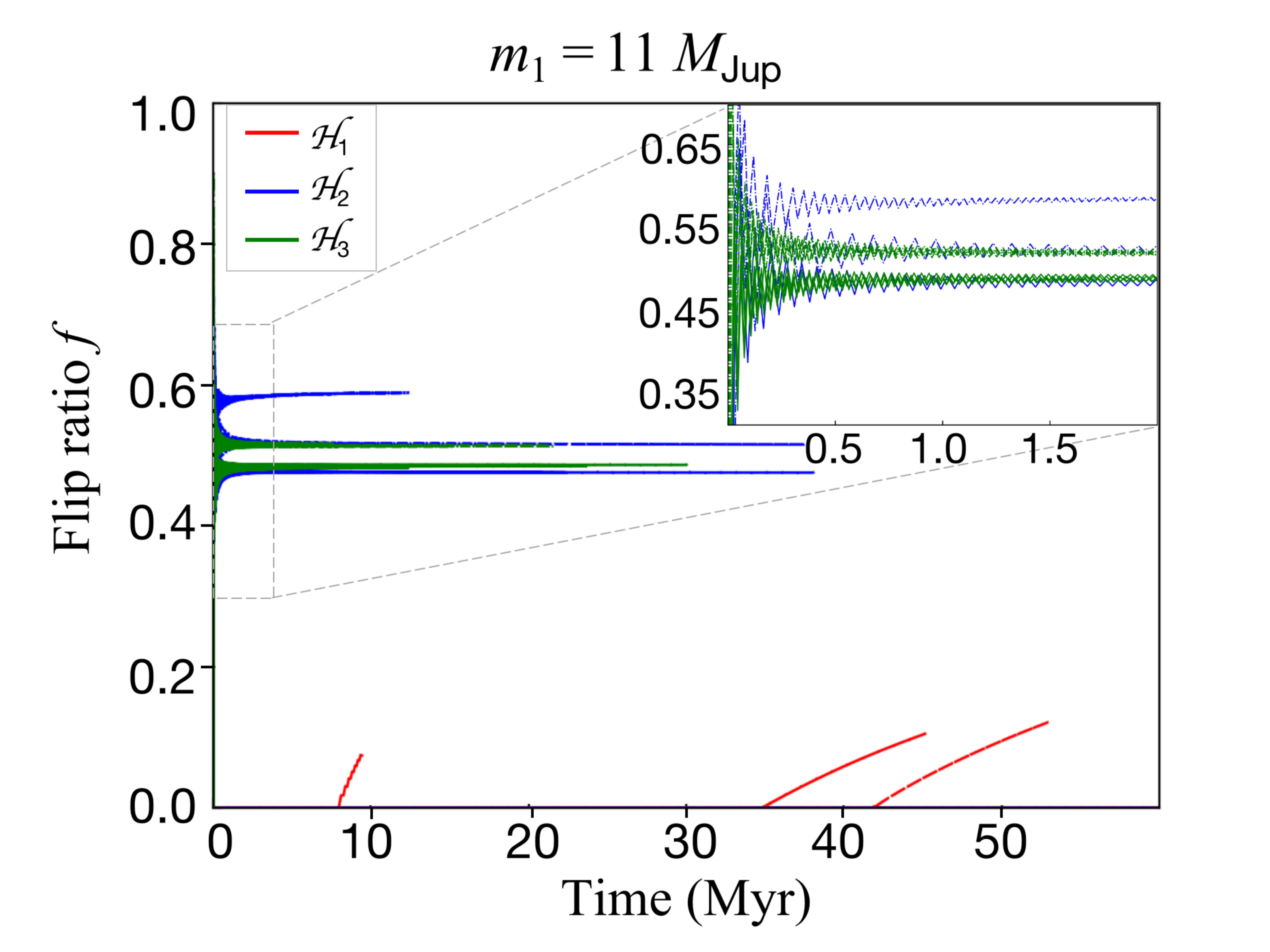}}
\subfigure[]{\includegraphics[width=1\columnwidth,height=7cm]{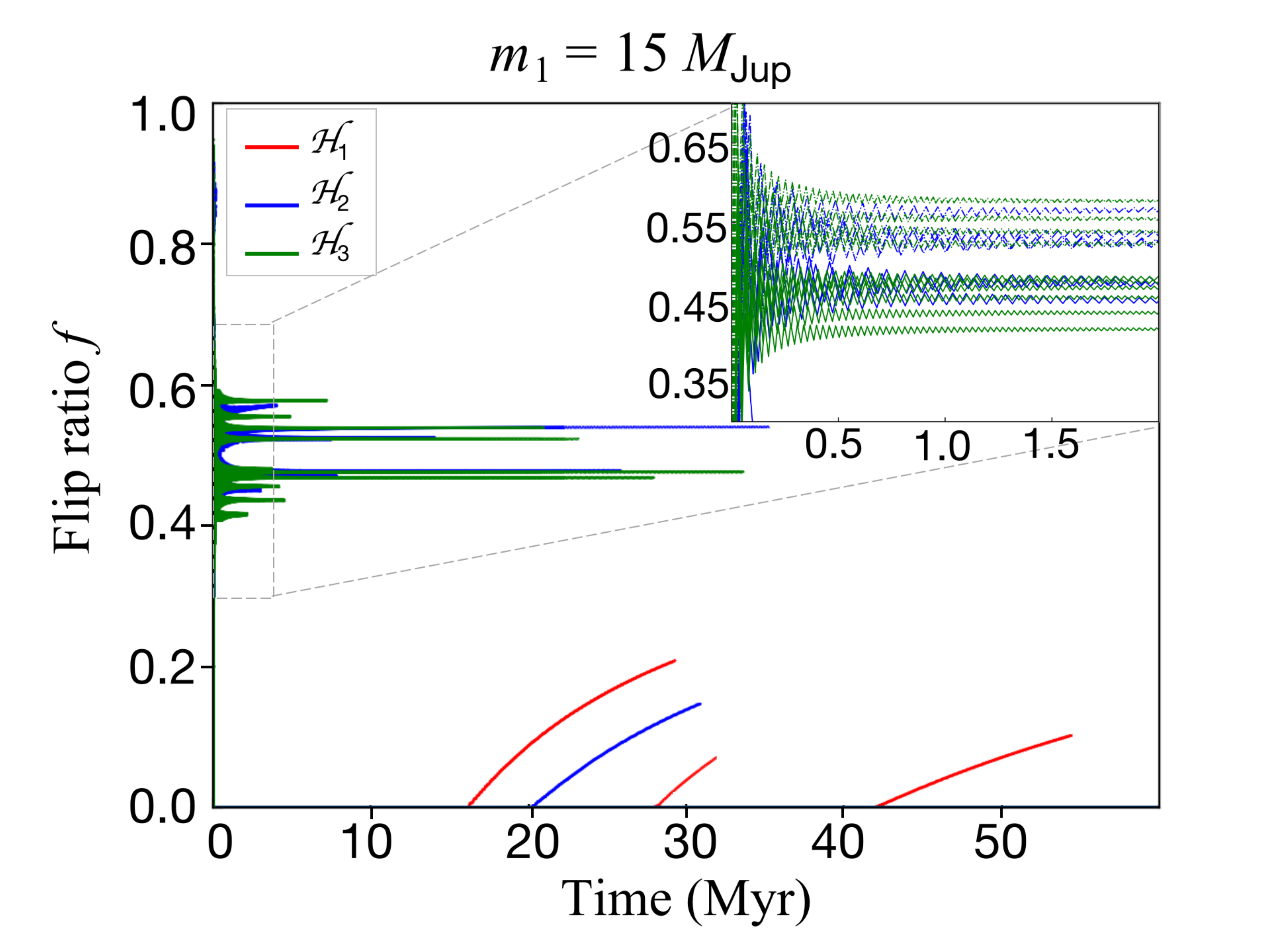}}
\caption{Evolution of the flip ratio for $m_1 $= 5 , 9, 11, 15 $M_{\mathrm{Jup}}$ over the timescale of 100 Myr. Same as Figure \ref{fig:imax1}, $\mathcal{H}_1$, $\mathcal{H}_2$, $\mathcal{H}_3$ correspond to the selected perturbation Hamiltonian. In each panel, the evolution before 2 Myr is given as a sub-figure. In these sub-figures, dotted dashed lines with the equilibrium value $f>0.5$ represent prograde orbits transformed from retrograde orbits, while solid lines with $f<0.5$ are transformed from prograde orbits to retrograde orbits.}
    \label{fig:ft12}
    \end{center}
\end{figure*}
In order to investigate the dependence of orbital flip possibility on the integration timescale, we adopt the definition of the flip ratio $f = t_{\mathrm{flip}} / t_{\mathrm{total}}$ \citep{Teyssandier2013} to describe the timescale for the first flip and observation possibility of orbital flip process over secular evolution, where $t_{\mathrm{flip}}$ is the duration of orbital flip process, and $t_{\mathrm{total}}$ represents the total evolution timescale. We extract cases in Section \ref{subsec:inclination1} in which orbital flips occur in Figure \ref{fig:ft12}. For these rolling-over orbits, the integrations stop when $e_1$ is getting larger than 0.9999 and the planet will fall into the Roche limit of the primary star.

In Figure \ref{fig:ft12} (a), we observe that when $\mathcal{H}_1$ = $-8.06\times10^{-8}$, the orbits turn over after 30 Myr and spend more than 80 Myr to arrive at the equilibrium  $f$. However, the flips under $\mathcal{H}_2$ and $\mathcal{H}_3$ have occurred from the very beginning of evolution, and reach the equilibrium value before 1 Myr. This phenomenon also shows up in Figure \ref{fig:ft12} (b) and Figure \ref{fig:ft12} (c), indicating that the flip timescale decreases with an increase of $\mathcal{H}$ under most circumstances. The oscillation timescale of $f$ over the flip procedure decreases with the rise of the perturbation Hamiltonian, which can be seen from the blue and green curves in four sub-figures.

In Figure \ref{fig:ft12} (d), there is a special flipping case for $\mathcal{H}_2$ when $m_1$ = 15 $M_{\mathrm{Jup}}$. This rolling-over orbit has initial conditions of $e_1$ = 0.64 and $i_{\mathrm{mut}}$ = 58.9$^{\circ}$ in Figure \ref{fig:imax1} (d). In comparison to other cases under $\mathcal{H}_2$ for different masses of planet, it can be concluded that relative low initial $i_{\mathrm{mut}}$ below 60$^{\circ}$ leads to relative large timescale for the first orbital flip and more time to get to the equilibrium value of the flip ratio.

Moreover, we investigate the equilibrium value $f$ for flip cases for $\mathcal{H}_2$ and $\mathcal{H}_3$. We find that the final $f$ of dotted dashed lines all locate above 0.5, while those of solid lines are entirely lower than 0.5. Thus the time duration of flip process for those original retrograde orbits is larger. The maximum $f$ occurs in Figure \ref{fig:ft12} (b) with $f \sim 0.65$ when $m_1$ = 9 $M_{\mathrm{Jup}}$, $\mathcal{H}_2$ = $-6.35\times10^{-7}$, whereas the minimum $f$ occurs in Figure \ref{fig:ft12} (d) with $f \sim 0.4$ when $m_1$ = 15 $M_{\mathrm{Jup}}$, $\mathcal{H}_3$ = $-2.03\times10^{-6}$. Thus the equilibrium value of flip ratio under EKL mechanism is related to the planetary mass, initial eccentricities and mutual inclinations. When the equilibrium value of $f$ is more close to 0.5, the observation possibility of the retrograde orbit transforming from the prograde is higher.

\subsection{Stability of flipping orbits}\label{subsec:stability2}
For the planetary system with given orbital parameters, the periodic orbits and the stability are identified by the representative plane of ($e_1$, $e_2$), level curves in the ($e_1$, $g_1$) plane, the Poincar\'{e} surface of section and the long-term stability criterion. These methods are applied to our investigation of the stability of specific S-type planets in binary systems, while the perturbative treatment and the invariant manifolds characterize the stability with a fixed Hamiltonian \citep{Lei2022}.

To derive global and comprehensive view of the system dynamics, we first attempt to construct the parametrical analysis of the secular model. We first show the representative plane of ($e_1$, $e_2$)  by \citet{Michtchenko2004}, which was then followed by the studies in secular dynamics and resonances \citep{Libert2006, Libert2007, Libert2008, Henrard2008}. This approach may rely on the secular averaging of the full Hamiltonian and could be applied to construct global phase portraits of the systematic variables, thereby detecting resonances and discerning periodic orbits in the N-body dynamics.

\begin{figure}
        \includegraphics[width=\columnwidth,height=7cm]{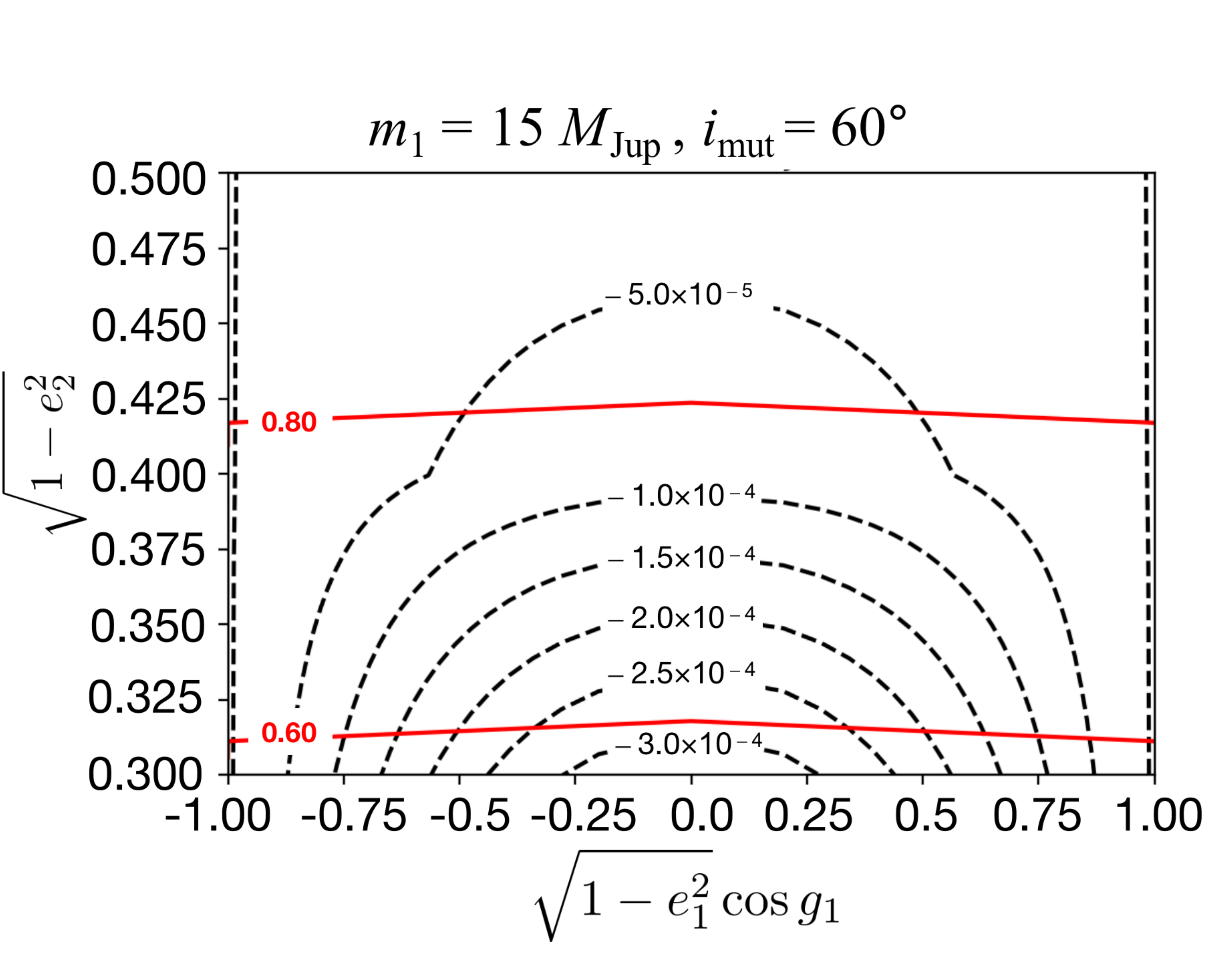}
    \caption{One of the contour map of the total angular moment and the perturbation Hamiltonian in the representative plane of ($\sqrt{1-e_{1}^{2}}$, $\sqrt{1-e_{2}^{2}}$). The black dashed line indicate the level curves of Hamiltonian and the red solid line is the conserved $G_{\mathrm{tot}}$.\label{fig:Gcontour1}}
\end{figure}

According to \citet{Michtchenko2004}, the secular motion of planetary systems is mainly decided by the global quantities of total energy and the Angular Momentum Deficit (\textit{AMD}). The phase structures are determined by two constants: $a_1$/$a_2$ and $m_1$/$m_2$.  The description of the secular behaviour of this system is derived by the distribution of  initial values of $e_1$, $e_2$, $g_1$, $g_2$ and $i_{\mathrm{mut}}$. To simplify the model, we fix $m_1$ = 15 $M_{\mathrm{Jup}}$ and $i_{\mathrm{mut}}$ = 60$^\circ$ in Figure \ref{fig:Gcontour1},  $\sqrt{1-e_{1}^{2}}$ cos$g_1$ $>$ 0 with $g_1$ = 0$^{\circ}$, while $\sqrt{1-e_{1}^{2}}$ cos$g_1$ $<$ 0 with $g_1$ =180$^{\circ}$, and $g_1$ can always go through 0$^{\circ}$ and 180$^{\circ}$ over the evolution. As seen from Figure \ref{fig:Gcontour1}, we may come to conclusion  that the periodic orbits occur when $G_{\mathrm{tot}}$ and $\mathcal{H}$ have two cross points.

\begin{figure*}
\begin{center}
\subfigure[]{\includegraphics[width=1\columnwidth,height=6.5cm]{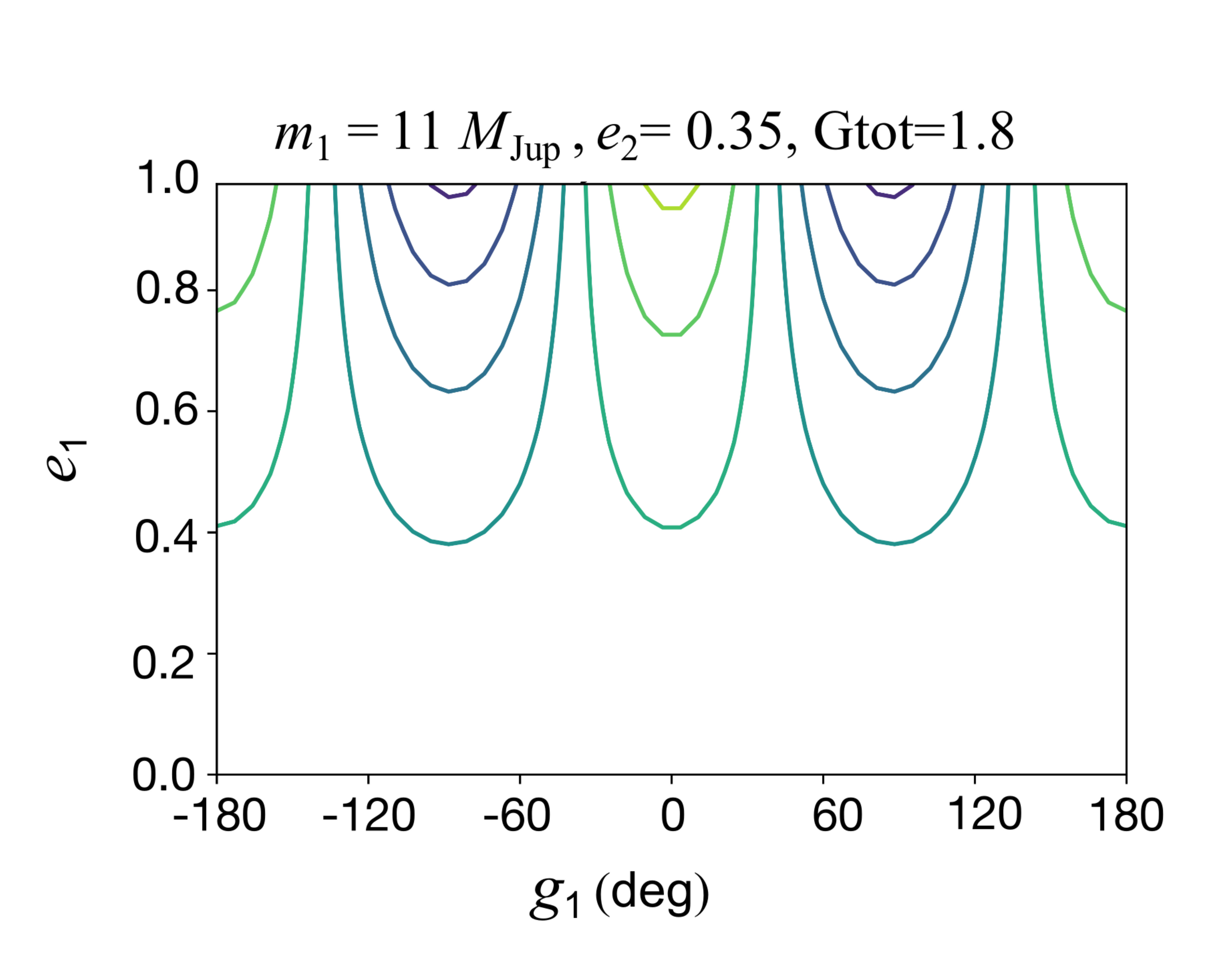}}
\subfigure[]{\includegraphics[width=1\columnwidth,height=6.5cm]{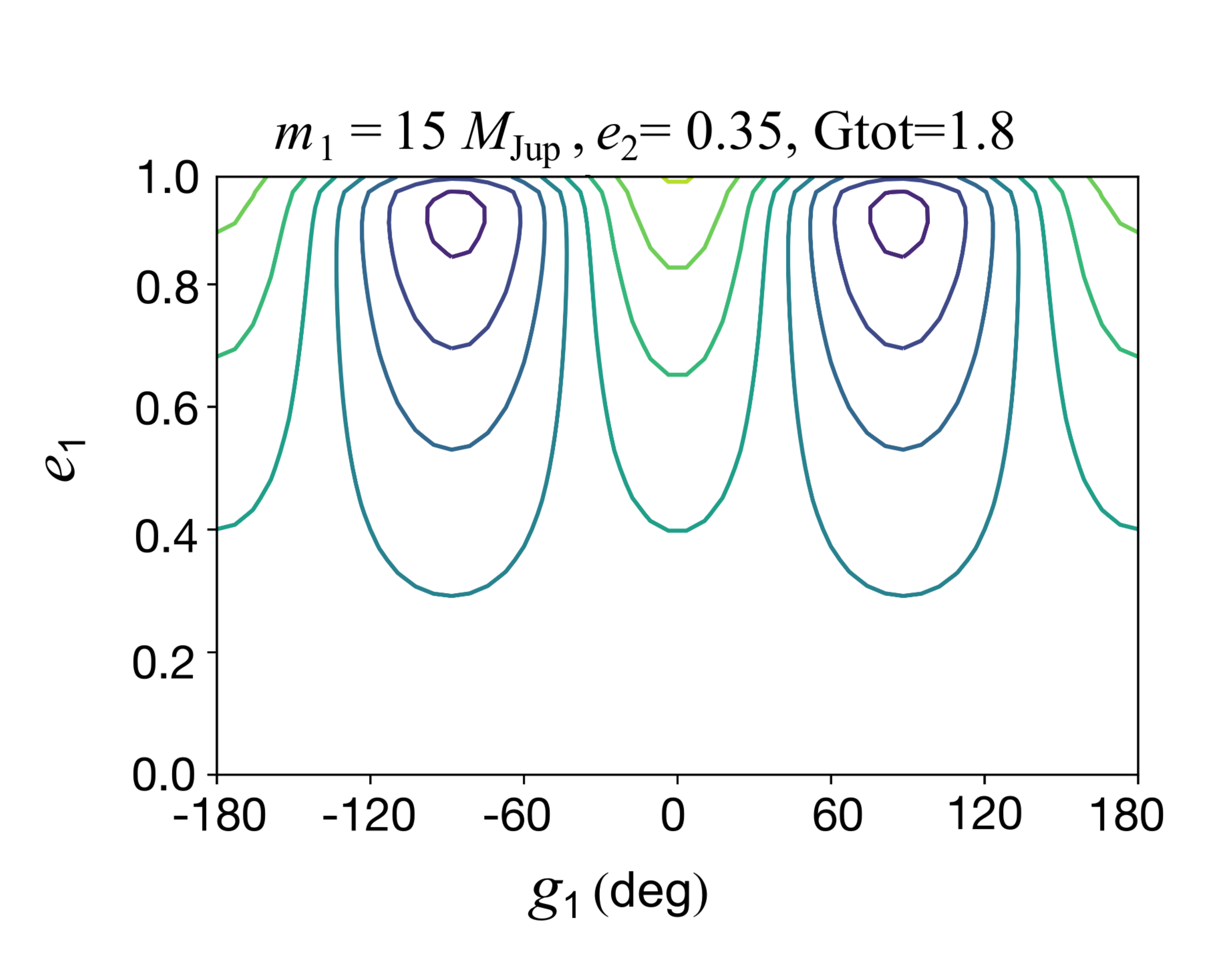}}
\caption{Two examples of Hamiltonian level curves with $m_1$ = 11 $M_{\mathrm{Jup}}$ and 15 $M_{\mathrm{Jup}}$ in the ($e_1$, $g_1$) plane, where $e_2$ and $G_{\mathrm{tot}}$ are fixed. Blue circles and green curves are resonant and oscillating orbits with the equilibrium point of 90$^{\circ}$ and 180$^{\circ}$ respectively.}
    \label{fig:Hcontour1}
    \end{center}
\end{figure*}

\begin{figure*}
\begin{center}
\subfigure[]{\includegraphics[width=0.4\textwidth]{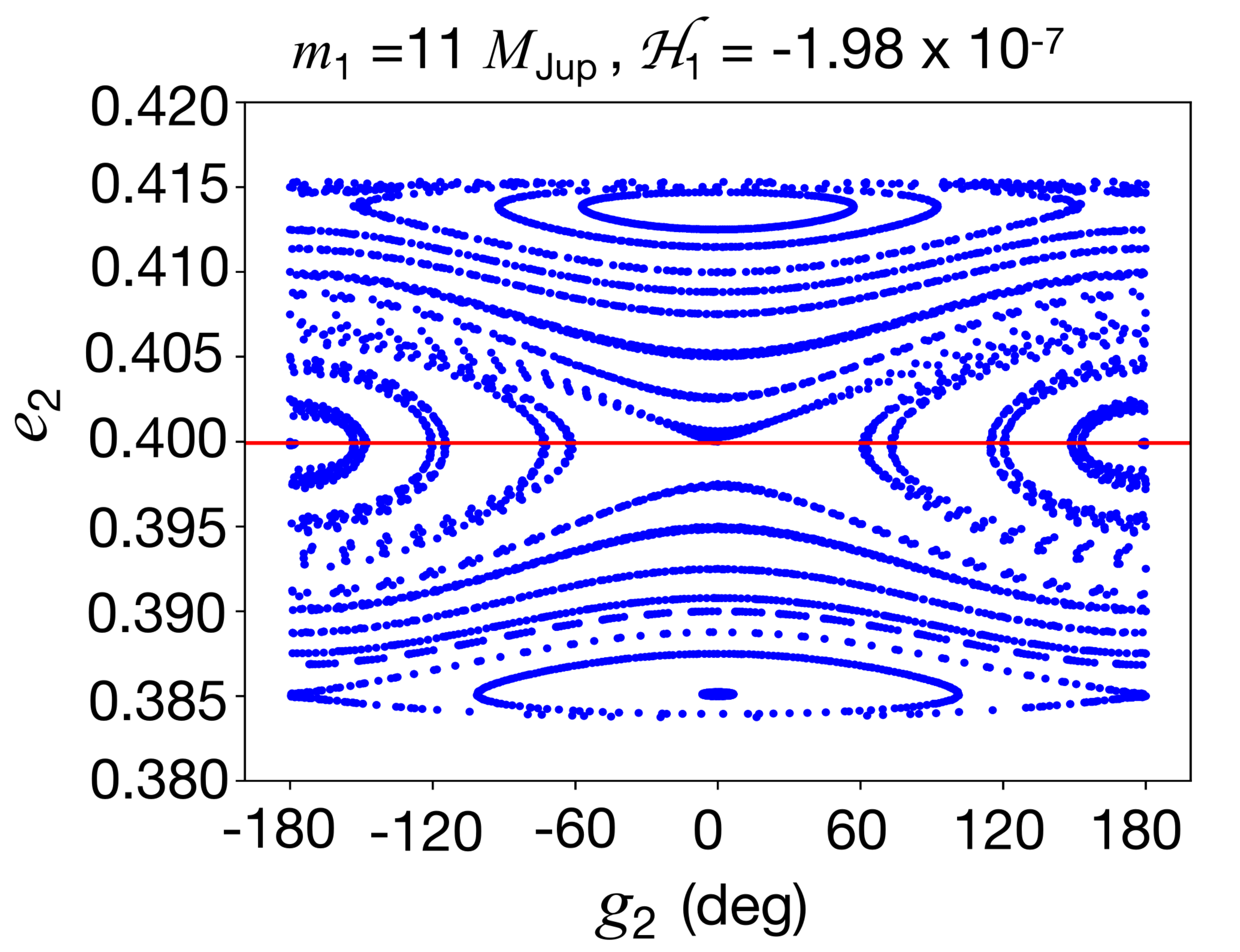}}
\subfigure[]{\includegraphics[width=0.4\textwidth]{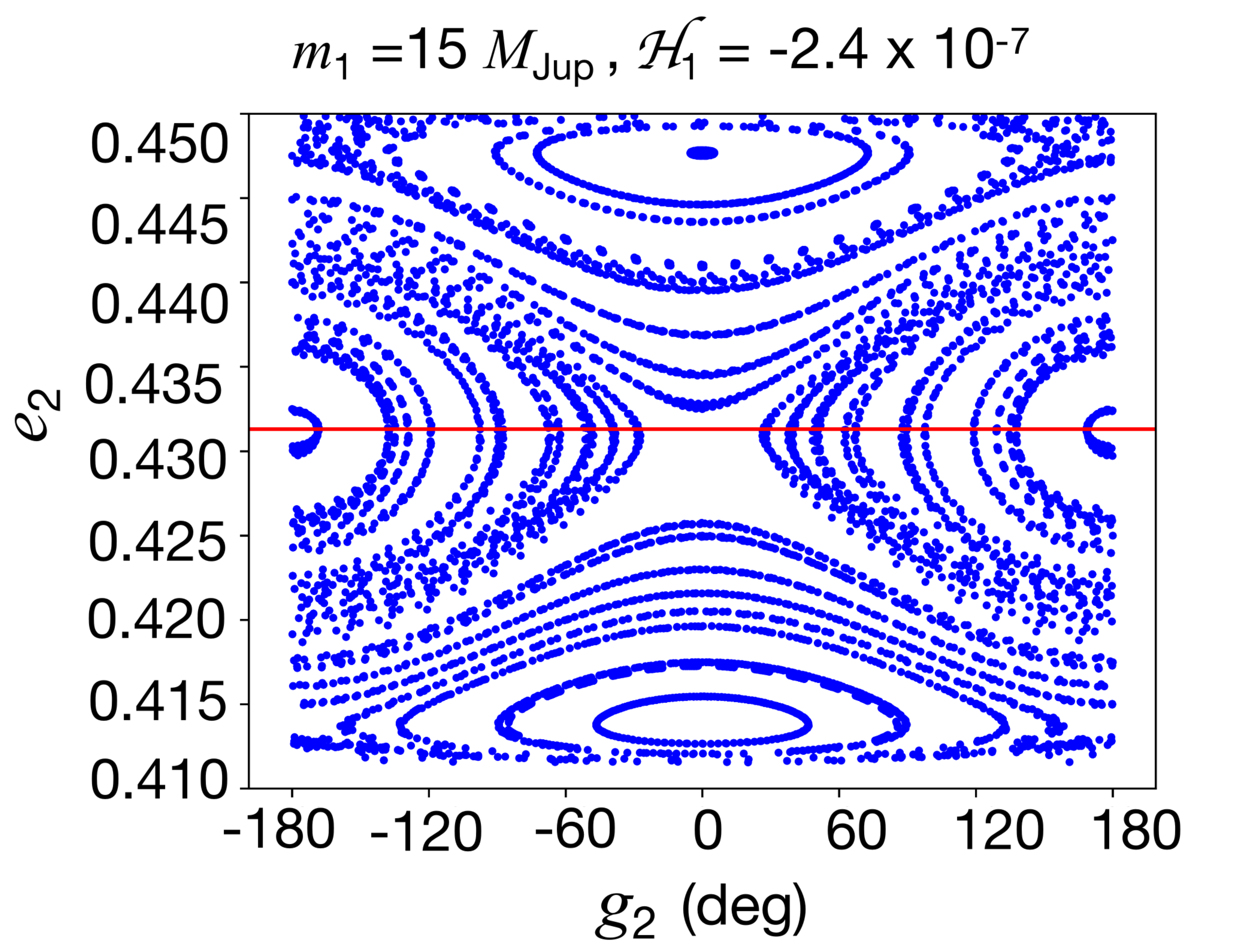}}\\
\subfigure[]{\includegraphics[width=0.4\textwidth]{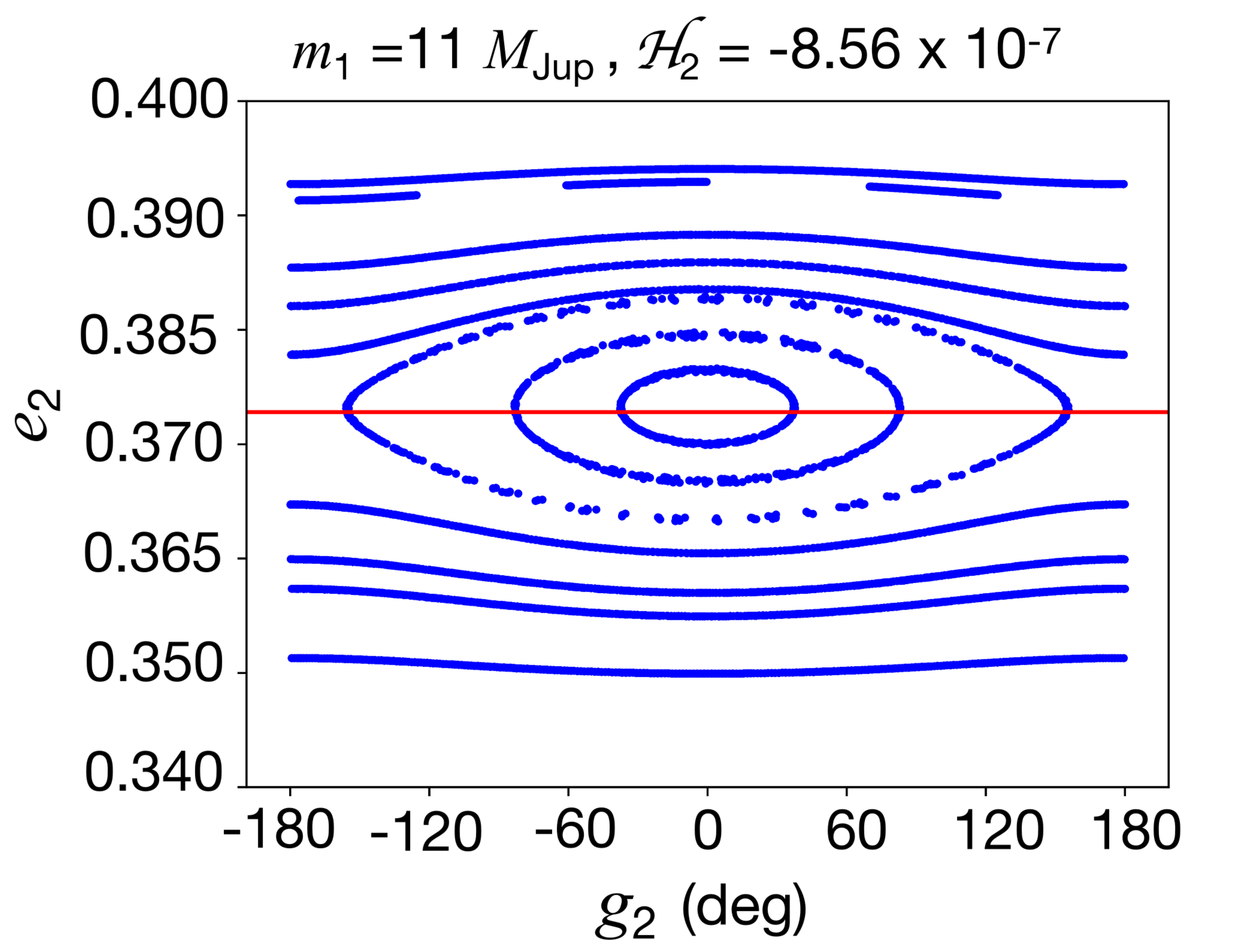}}
\subfigure[]{\includegraphics[width=0.4\textwidth]{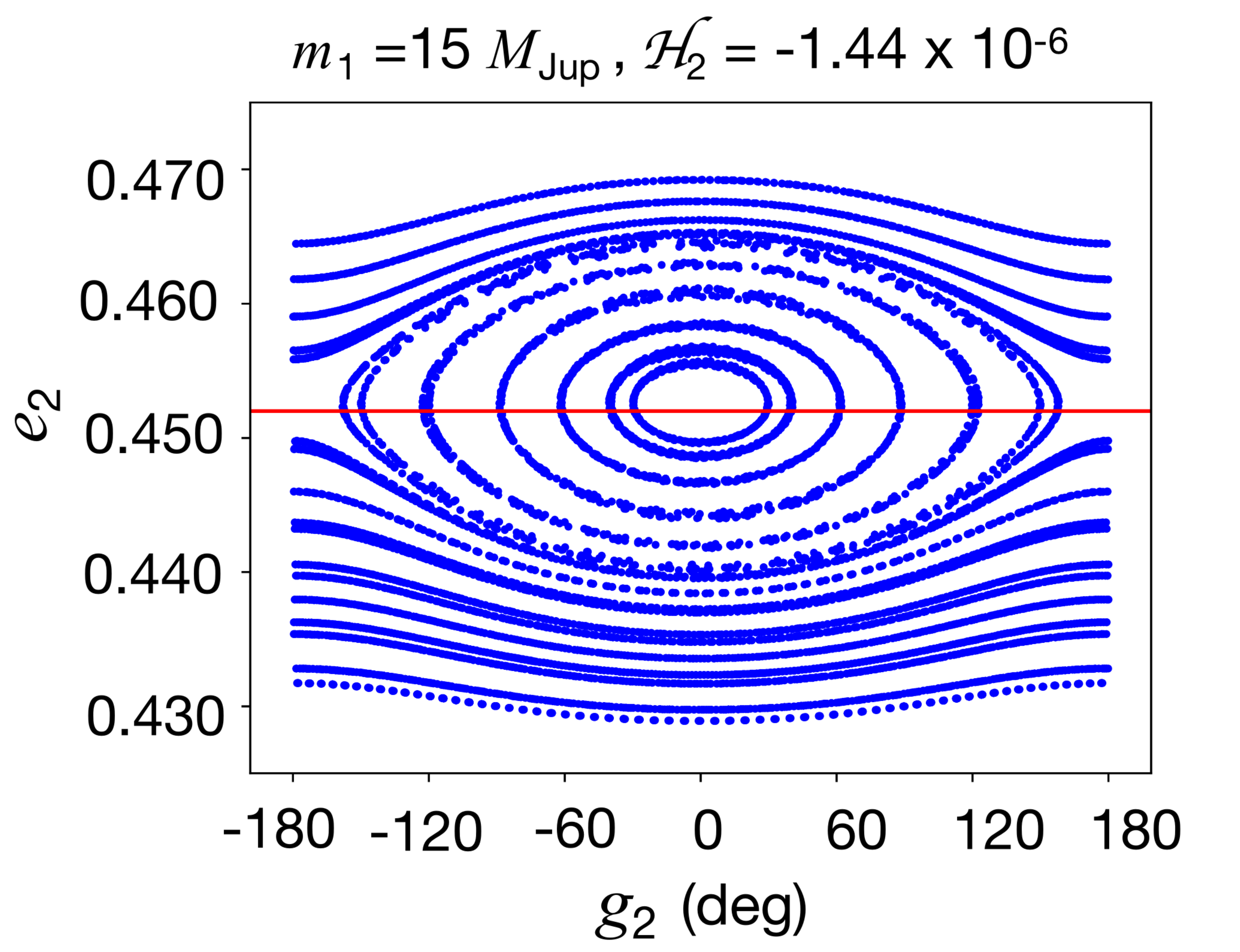}}\\
\subfigure[]{\includegraphics[width=0.4\textwidth]{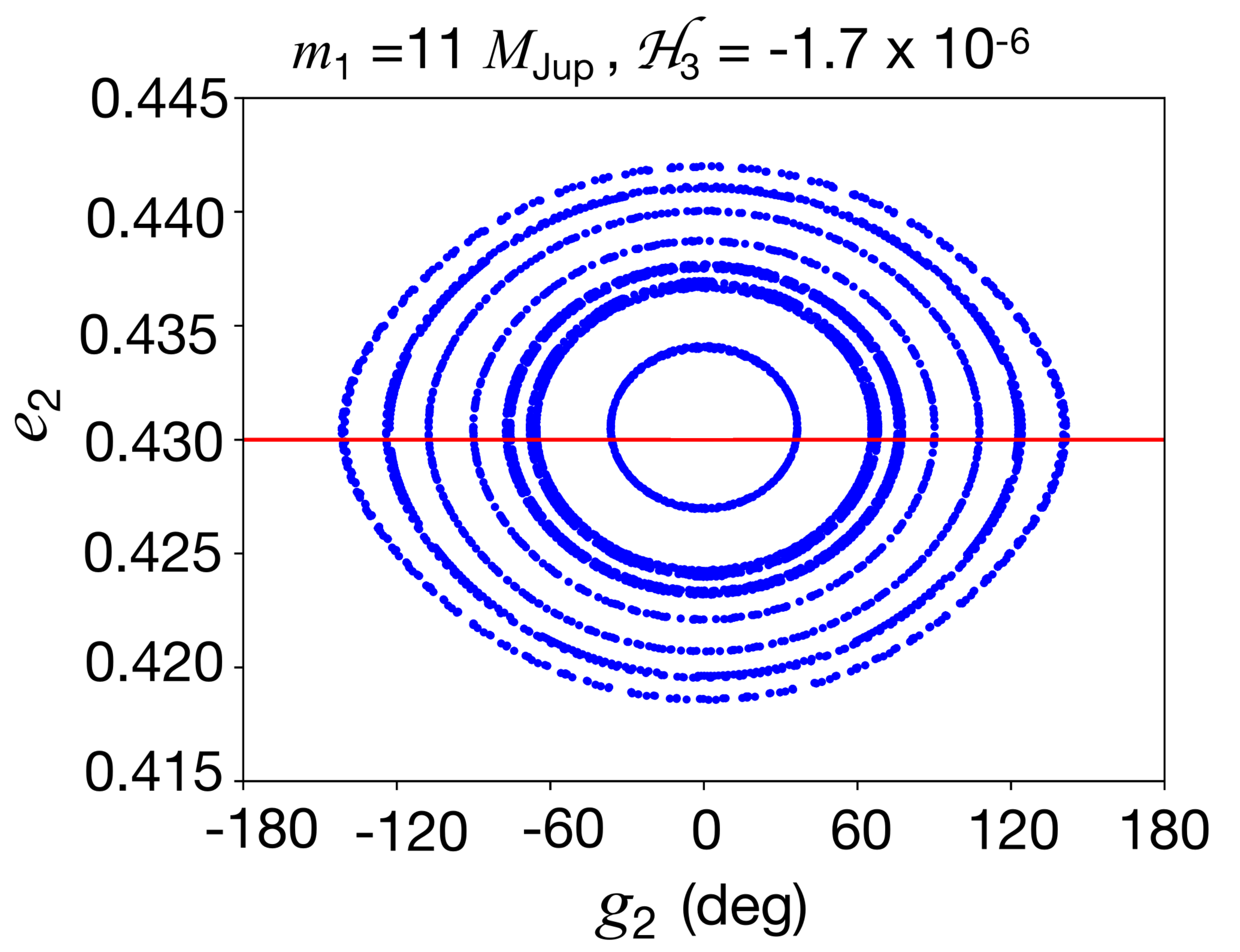}}
\subfigure[]{\includegraphics[width=0.4\textwidth]{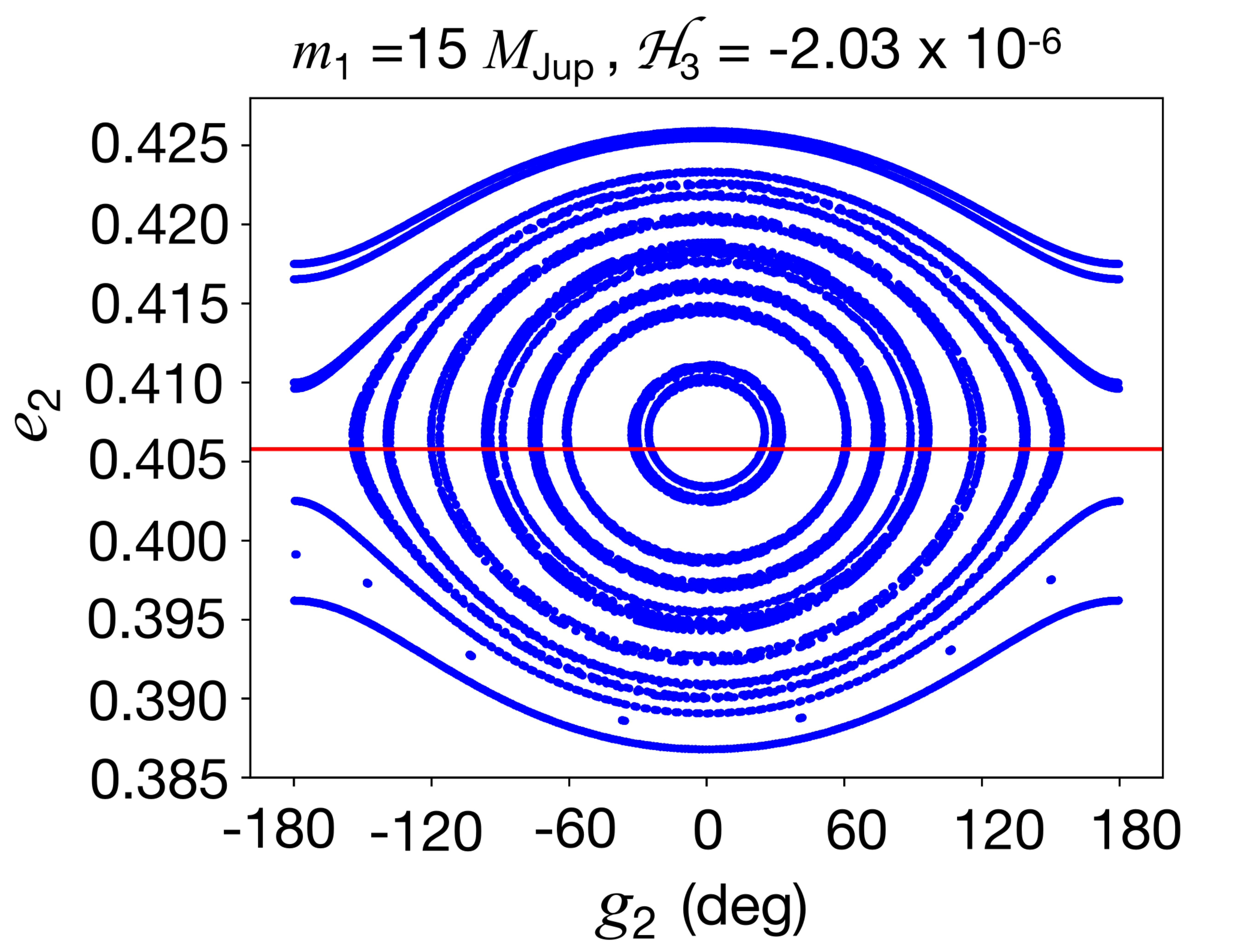}}
\caption{Surface of sections in the $g_1$ = 0 plane with different masses of $\gamma$ Cep Ab, by varying the perturbation Hamiltonian $\mathcal{H}(\Delta h \rightarrow \pi)$. Panels (a)(c)(e): dynamical maps for $m_1$ = 11 $M_{\mathrm{Jup}}$.  Panels (b)(d)(f): dynamical maps for $m_1$ = 15 $M_{\mathrm{Jup}}$. }
    \label{fig:sections}
\end{center}
\end{figure*}

For the secular evolution of $\gamma$ Cep Ab B system, one couple of free variables are $e_1$ and $g_1$ in the 2-DOF averaged problem. Thus we can first easily perform the qualitative analysis in the ($e_1$, $g_1$) plane \citep{Tan2020}. Given to the limit of $G_{\mathrm{tot}}$ in Section \ref{subsec:inclination1}, we select and fix the value of  $G_{\mathrm{tot}}$ to present the level curves of Hamiltonian in the parameter space of ($e_1$, $g_1$), as shown in Figure \ref{fig:Hcontour1}. The green stream lines for circulating orbits and blue circles for resonant orbits are easy to distinguish. With the change of perturbation Hamiltonian, the dynamical structure transitions from circulation to libration and the evolution interval of $e_1$ is varying. The smallest circles have the largest magnitude of $\mathcal{H}$, which indicates the secular resonance effect is the most intense, as shown in Figure \ref{fig:sections} for more details.

In addition to the representative plane of ($\sqrt{1-e_{1}^{2}}$, $\sqrt{1-e_{2}^{2}}$) and level curves in the ($e_1$, $g_1$) plane, we further investigate the stability of rolling-over orbits under the selected Hamiltonian with the Poincar\'{e} surface of section. For two-dimensional Hamiltonian system, the Poincar\'{e} surface of section is a commonly used method: under a given energy integral, the phase flow of the system is three-dimensional. By selecting a suitable section, the intersection point of the systematic phase flow and the section is projected in two dimension. In other words, the Poincar\'{e} surface of section reflects the geometry of the system dynamics. Given the integration of energy (i.e., the octupole perturbation Hamiltonian) and the selection of the section (i.e., the $g_1$ = 0 section), the surface of section can be generated by plotting intersection points in the $g_2-e_2$ frame for the plane of $g_1$ = 0.

The surface of section enables us to identify the order of orbital resonance and the dynamical stability of the system through a series of geometric structures. For a given, fixed value of the Hamiltonian, the orbital mode can be derived from only two of orbital parameters. Various orbital modes, including resonance, circulation and chaos, are distinguished by the shape that points distributed as in the section.

Here we present typical structures in surfaces of section of $\gamma$ Cep Ab B system in Figure \ref{fig:sections}. The surfaces of section are plotted in the $g_1$ = 0 plane for $m_1$ = $\{5, 9, 11, 15\}$ $M_{\mathrm{Jup}}$ with corresponding perturbation Hamiltonian. To set $e_2$ and $g_2$ initially in the uniform grid, then we solve other initial parameters based on the conservation of the system energy and the total angular momentum. There are two kinds of regions in these plots: closed circles are resonant orbits (quasi-periodic orbits) where $e_2$ and $g_2$ oscillate in the bounded regions. Streamlines denote circulative orbits when at least one of the parameters circulates.

Figure \ref{fig:sections} simply exhibits the dynamical structure of the surface of section for $m_1$ = 11 $M_{\mathrm{Jup}}$ and 15 $M_{\mathrm{Jup}}$. With the increasing of $\mathcal{H}$, the structure in surfaces of section appears to be diverse.  As shown in Figure \ref{fig:sections} (a), (c) and (e), when $m_1$ = 11 $M_{\mathrm{Jup}}$, the orbits with respect to $g_2$ = 180$^{\circ}$ disappear, whereas both orbits with fixed points of $g_2$ = 0$^{\circ}$ and 180$^{\circ}$ retain. These libration regions perfectly match orbits that would turn over. In Figure \ref{fig:sections} (b), (d) and (f), when $m_1$ = 15 $M_{\mathrm{Jup}}$, positions of the fixed angle of $g_2$ are the same as $m_1$ = 11 $M_{\mathrm{Jup}}$. Figure \ref{fig:sections} (a) shows that orbital flips of $\gamma$ Cep Ab are dominated by the octupole level resonance, and the oscillation amplitude of $e_2$ in these flipping cases is close to 0.01. In Figure \ref{fig:sections} (e) and (f), when $m_1$ = 11 $M_{\mathrm{Jup}}$, $\mathcal{H}=-1.7\times10^{-6}$, and $m_1$ = 15 $M_{\mathrm{Jup}}$, $\mathcal{H}=-2.03\times10^{-6}$ respectively, most orbits around initial values of $e_2$ could turn over in the secular evolution.
Additional simulations for $m_1$ = 5 and 9 $M_{\mathrm{Jup}}$ reveal the similar structure but with various oscillation intervals of $e_2$ when compared to above results.

The orbital flip criterion of $e_2$ can be evaluated with the conserved total angular momentum in Equation (\ref{equ:LGH}) when $i_2$ = 0$^{\circ}$ and $i_{\mathrm{mut}}$ = 90$^{\circ}$:
\begin{equation}
\label{equ:flipe2}
\begin{split}
e_{2,flip}=\sqrt{1-(\frac{G_{\mathrm{tot}}}{L_{2}})^2} .
\end{split}
\end{equation}
We mark out this flip criterion in Figure \ref{fig:sections} with red horizontal lines. Circles and streamlines crossing the red horizontal line represent those orbits inverting regularly. While ellipses above the red horizontal line in Figure \ref{fig:sections} (a-b) stand for orbits without inverting. {Here we find that the structures in the phase portrait become much clearer with a decrease of the perturbation Hamiltonian $\mathcal{H}$, where the chaotic orbits disappear, similar to those of \citet{Lei2022}}.

{Additionally, we compare the boundaries of regions of orbital flips in the ($e_{1,0}$, $i_{mut,0}$) space for non-restricted model with those of Figure 3 \citep{Lei2022}. By analyzing the results in the Poincar\'{e} surfaces of section, the inner orbit appears to be more stable in our hierarchical system when planetary eccentricity $0.5<e_{1,0}<0.6$, which is simply occupied by circulating and librating orbits.}

 In order to further confirm whether the orbital flip in  $\gamma$ Cep Ab is stable, we adopt the long-term stability criterion given by \citet{Mardling2001}:
\begin{equation}
\label{equ:stab1}
\frac{a_{2}}{a_{1}}>2.8\left(1+\frac{m_{2}}{m_{0}+m_{1}}\right)^{2 / 5} \frac{\left(1+e_{2}\right)^{2 / 5}}{\left(1-e_{2}\right)^{6 / 5}}\left(1-\frac{0.3 i_{\mathrm{mut}}}{180^{\circ}}\right),
\end{equation}
where $e_2$ and $i_{\mathrm{mut}}$ are time-varying, and the constants $a_1$, $a_2$, $m_0$, $m_1$ and $m_2$ can be moved to the left side of the expression. We then derive the new expression of this criterion by a new variable $S$:
\begin{equation}
\label{equ:stab2}
 S =  \frac{\left(1+e_{2}\right)^{2 / 5}}{\left(1-e_{2}\right)^{6 / 5}}\left(1-\frac{0.3  i_{\mathrm{mut}}}{180^{\circ}}\right).
 \end{equation}
For $m_1$ $\in$ [5, 15] $M_{\mathrm{Jup}}$, $a_1$ = 2.14 au, $a_2$ = 18.62 au, thus $S$ should meet $S < 2.807$ to maintain the stability of the system. The calculated maximum $S$ for rolling-over orbits in Figure \ref{fig:sections} (a)(c)(f) is 2.13, which is definitely within the stability criterion. In Figure \ref{fig:sections} (b)(d)(f), the calculated maximum $S$ is 2.64, providing stable cases from Equation \ref{equ:stab2}. Hence, the orbital flips in Figure \ref{fig:sections} are stable, without chaotic excitation of the binary's eccentricity and inclination.

Comparing with previous work on the stability of $\gamma$ Cep system, \citet{Haghighipour2004} implemented an extensive numerical study of the orbital stability of $\gamma$ Cep system and presented that the system can remain steady for $i_{\mathrm{mut}}$ $\in$ [0$^{\circ}$, 60$^{\circ}$] and $e_2 < 0.5$. This condition is also supported in this work since we constrain the eccentricity $e_2$ $\in$ [0.35, 0.45], which is well consistent with our simulation results. Set Figure \ref{fig:sections} (b) as an example, the critical value of the initial $i_{\mathrm{mut}}$ for orbital flips to occur is lower than 60$^{\circ}$, indicating a stable initial status according to \citet{Haghighipour2004}, regular oscillation of $e_1$ and $i_{\mathrm{mut}}$ make the system always back to the stable situation. \citet{Satyal2013} examined the stability and quasi-periodicity of $\gamma$ Cep, and explored the orbital stability for various inclinations and binary eccentricity $e_2$ through the reliability comparison of chaos indicator. They demonstrated the planet $\gamma$ Cep Ab can maintain stable for $e_2$ as high as 0.6 or for $i_1 \leq $ 25$^{\circ}$. For rolling-over orbits in our work, we demonstrate $\gamma$ Cep Ab can maintain stable as well when $e_2 \sim 0.4$ and $i_ {\mathrm{mut}}\sim $ 90$^{\circ}$. In this work, we further confirm that the flipping cases of $\gamma$ Cep Ab have a great possibility to be locked in Kozai resonance based on surfaces of section in Figure \ref{fig:sections} and the long-term stability criterion $S$.

From the simulations results, we also see that the planetary eccentricity could be stirred up to 0.9999 due to secular perturbation from the binary, as close approaches may eject the planet out of the system so that it would not be observed or its orbit could be shrunk owing to tides over the evolution. While this process takes too much time to be observed over the timescale of the EKL mechanisms, thus we employ $i_{\mathrm{mut}}$ to explore the stability of the system, as the evolution of $e_1$ and $i_{\mathrm{mut}}$ are coupled under this scenario. The extreme oscillations of inclination and eccentricity would enhance the rate of bringing the system to an unstable situation. \citet{Li2014a} showed there could be chaotic behaviour when the mutual inclination between the inner and the outer orbit remained high.

The instability of S-type planets may be induced by large eccentricity excitations. The planet may fall into the region of Roche lobe and merge into the primary, or run away from the inner binary system \citep{Eggleton1998, Kiseleva1998, Ford2000}.

\section{Maximum mutual inclinations of general S-type systems}\label{sec:secular_evolution2}
After exploring crucial issues on inclination excitation of $\gamma$ Cep Ab, we attempt to reveal the dynamical features of general S-type planets under the octupole level secular resonance. For secular evolution of general S-type planets, dynamical evolution should be extensively studied in a wider parameter space, thus more parameters are accounted for in this Section.

As the distribution of SMA of S-type planets has been introduced in Section \ref{sec:introduction}, $m_{1}$ refers to the mass of S-type planet while $m_{2}$ denotes that of the stellar companion. For general S-type planets in close-binary systems, the parameter space of $m_1$, $a_1/a_2$, $e_1$, $e_2$ should be enlarged. In this Section, the mass of the primary star is set to be 1 $M_{\odot}$, mass of the secondary companion is assumed to be 0.3 $M_{\odot}$ on the basis of the average stellar masses of detected star binary hosting S-type planets \citep{Raghavan2010, Moe2017}.

In this Section, we do not discuss the effect of the initial inclinations, we mainly focus on whether the S-type planets in the systems may have experienced orbital flips under the relative conditions concluded from previous Sections of this work, therefore we take the initial mutual inclination $i_{\mathrm{mut}}$ as 50$^{\circ}$, which is a bit larger than the critical inclination of 39$^{\circ}$ for orbital flip in the classical Kozai--Lidov theory. Moreover, we let $e_1$ and $e_2$ uniformly distribute in 0.0 $\sim$ 0.8, and yield a grid of $16\times16$. For each set of initial parameters, we integrate the system for 100 Myr. The argument of periastron is set as $g_1$ = 5$^{\circ}$ and $g_2$ = 0$^{\circ}$ with longitudes of node $h_1$ = 180$^{\circ}$ and $h_2$ = 0$^{\circ}$.

\begin{figure*}
\centering
\subfigure[]{\includegraphics[width=0.95\columnwidth,height=6.5cm]{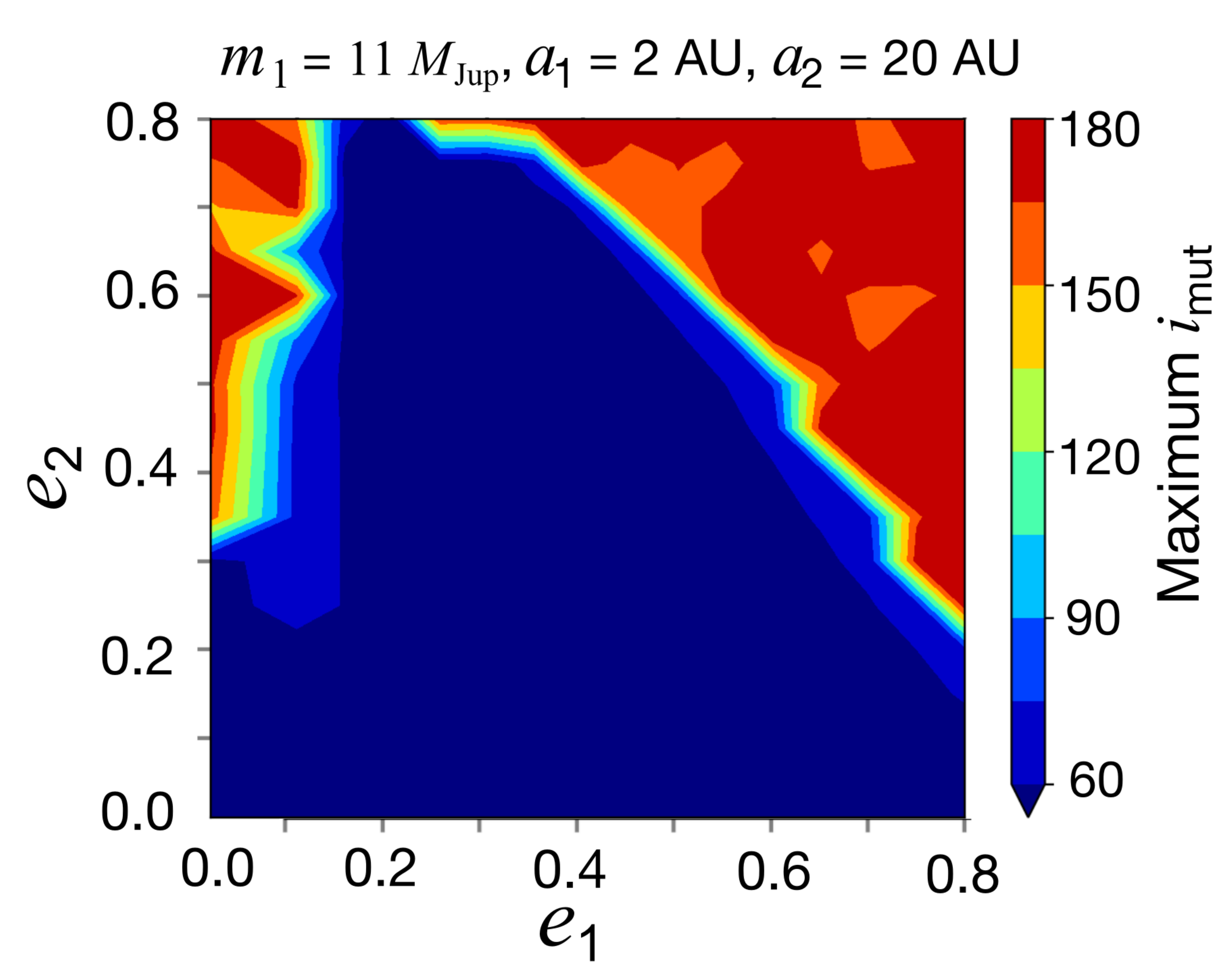}}
\subfigure[]{\includegraphics[width=0.95\columnwidth,height=6.5cm]{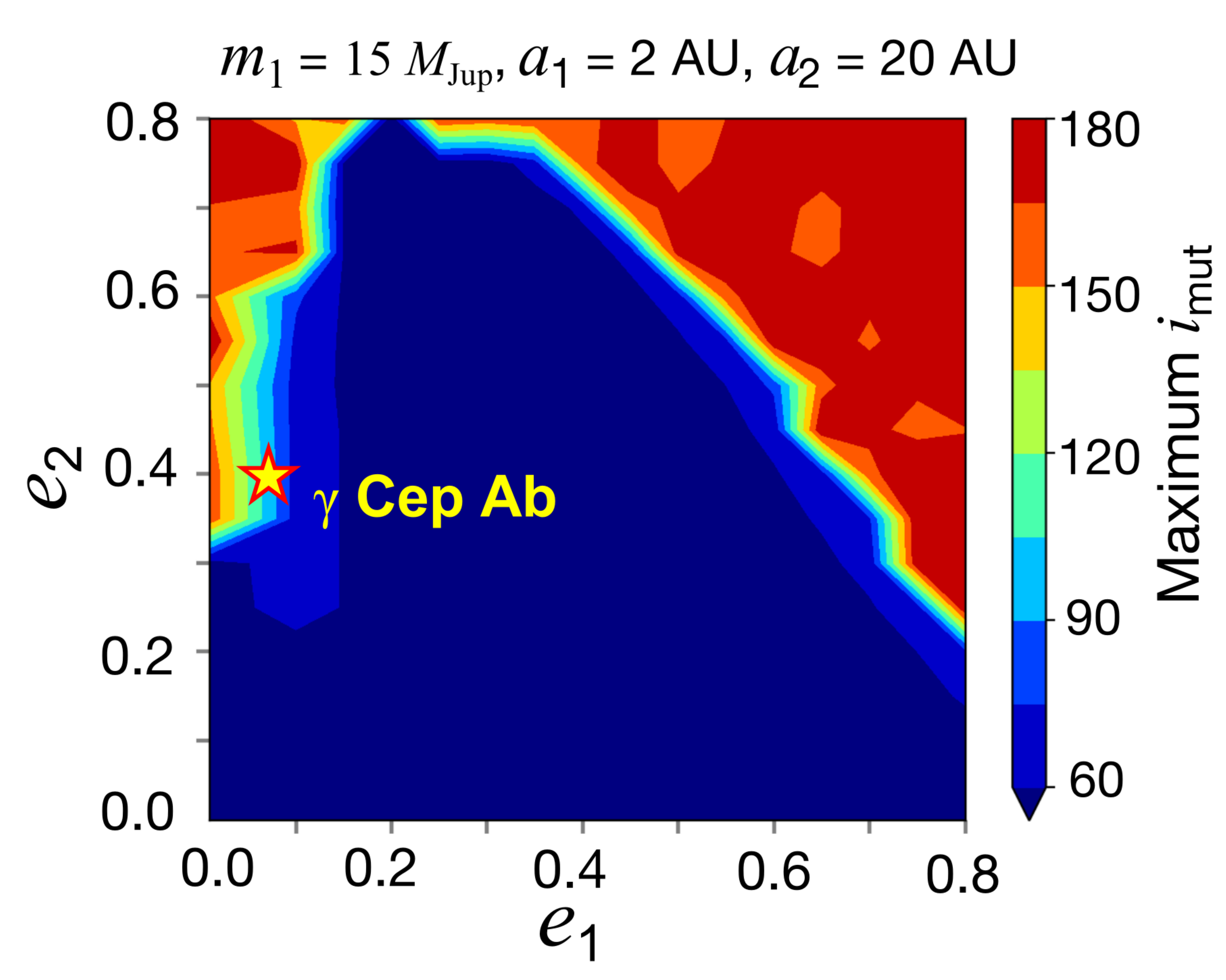}}
\caption{Distribution of the maximum $i_{\mathrm{mut}}$ for S-type planets with $a_1$ = 2 au and $a_2$ = 20 au. Panel (a): $m_1$ = 11 $M_{\mathrm{Jup}}$. Panel (b): $m_1$ = 15 $M_{\mathrm{Jup}}$. The colour bar on the right represents the maximum value of $i_{\mathrm{mut}}$ over 100 Myr. }
 \label{fig:imaxcontour2}
 \end{figure*}

In order to show the evolution of $\gamma$ Cep Ab from the current observed position and enlarge the parameter spaces of $e_1$, $e_2$ to present more general results, Figure \ref{fig:imaxcontour2} plots the distribution map of the maximum $i_{\mathrm{mut}}$ for $m_1$ = 11, 15 $M_{\mathrm{Jup}}$ for $a_1$ = 2 au, $a_2$ = 20 au.  Both panels show that orbital flips occur when $e_1$ and $e_2$ both larger than 0.2 or $e_1 < 0.2$ and $e_2 > 0.3$. We mark the orbital position of $\gamma$ Cep Ab by the yellow pentagram in Figure \ref{fig:imaxcontour2} (b). Flip constrains of $e_1$ and $e_2$ reveal that $\gamma$ Cep Ab still have a great possibility to maintain the flipping orbit. We use the new variable $p_{\mathrm{flip}}$ to describe the flip possibility of total 256 runs in each panel, in Figure \ref{fig:imaxcontour2} (a), $p_{\mathrm{flip}} = 0.301$, while in Figure \ref{fig:imaxcontour2} (b), $p_{\mathrm{flip}}$ is calculated to be 0.305, and the difference is mainly induced by the tiny changes of flip cases in the region of $e_1 < 0.3$.

As we mentioned in Section \ref{subsec:eklmechanism}, the SMA ratio $a_1/a_2$ is the specific element to evaluate the strength of the octupole level effect when $m_0$, $m_1$, and $e_2$ are fixed. We further set $a_1/a_2$ as a new independent variable, we choose $a_1$ = $\{1, 2, 5, 10\}$ au and $a_2$ = $\{10, 20, 50, 100\}$ au, which are selected from the "Target Area" in Figure \ref{fig:distributiona}, thus $a_1/a_2$ = $\{0.01, 0.02, 0.04, 0.05, 0.1\}$.  Additionally, we assume the S-type planetary mass to be 1 $M_{\mathrm{Jup}}$, according to the detected average of the minimum masses of S-type planets.

\begin{figure*}
\centering
	\includegraphics[width=2\columnwidth,height=15cm]{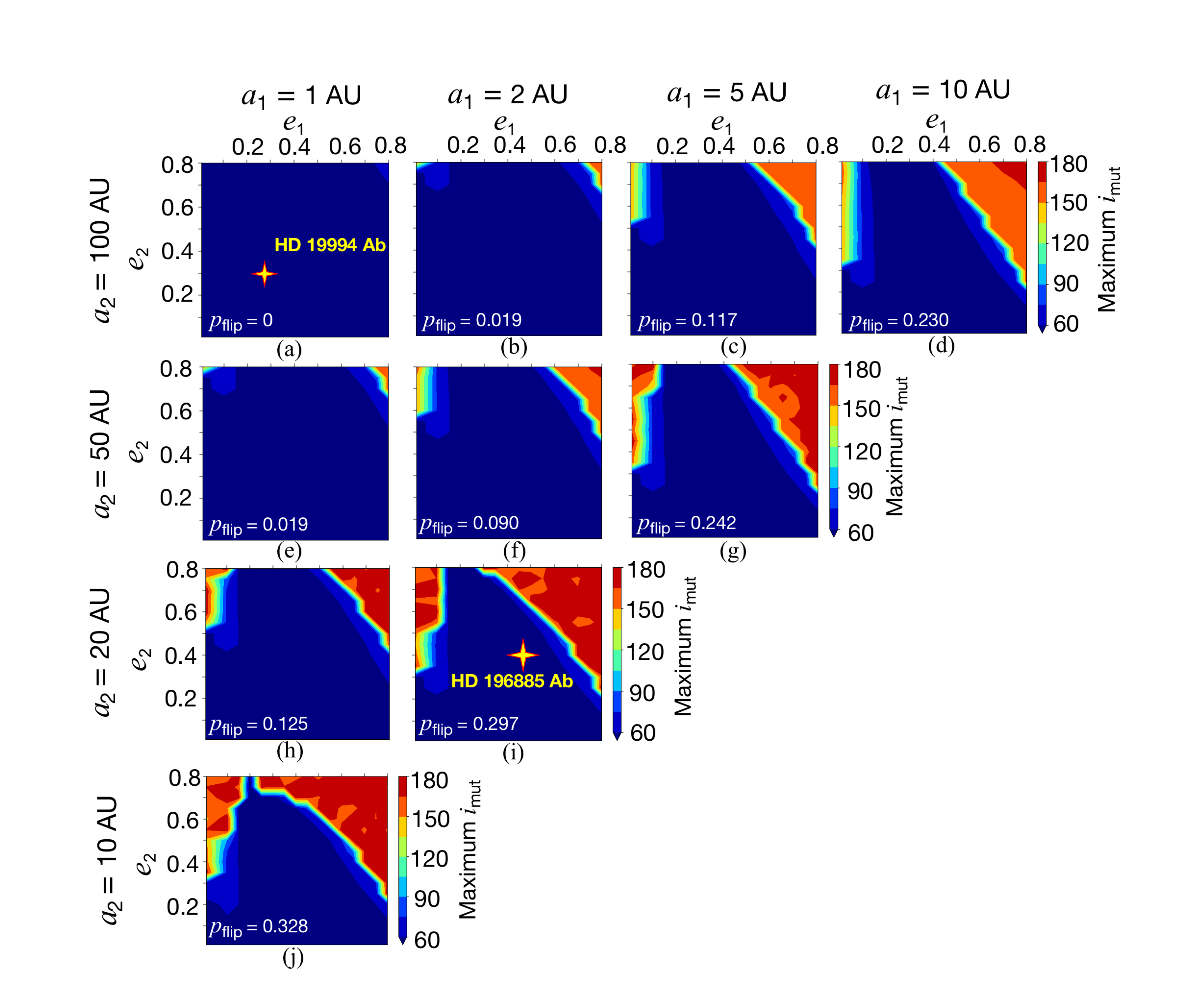}
\caption{Distribution of the maximum $i_{\mathrm{mut}}$ for S-type planetary systems with various initial SMA $a_1$ and eccentricity $e_1$, $e_2$. Each diagram has specific initial parameters $a_1$ and $a_2$. The colour bar on the right represents the range of the maximum value of $i_{\mathrm{mut}}$ over 100 Myr, regions in dark red represent orbits with $i_{\mathrm{mut}}$ excited up to 180$^{\circ}$.}
 \label{fig:imaxcontour}
 \end{figure*}

Figure \ref{fig:imaxcontour} shows the distribution of maximum orbital inclination in the parameter space for general S-type planets with mass of 1 $M_{\mathrm{Jup}}$. We attach the value of flip possibility $p_{\mathrm{flip}}$ in the lower left corner of each panel. For panels (d), (g), (i), (j) in Figure \ref{fig:imaxcontour} with $a_1/a_2$ = 0.1, the critical eccentricities for the orbital inverting are $e_2$ $\sim$ 0.3 when $e_1$ $\sim$ 0, and  $e_2$ $\sim$ 0.2 when $e_1 > 0.4$. For panels (c), (h) in Figure \ref{fig:imaxcontour} where $a_1/a_2$ = 0.05, the critical eccentricity for the orbital turning over are $e_2$ $\sim$ 0.5 when $e_1$ $\sim$ 0, and  $e_2$ $\sim$ 0.4 when $e_1 > 0.5$, $p_{\mathrm{flip}}$ in these two panels are both larger than 0.1. It should be noticed that in Figure \ref{fig:imaxcontour} (f) with $a_1/a_2$ = 0.04, the critical $e_1$ and $e_2$ for flips are both lifted by 0.1 than that in Figure \ref{fig:imaxcontour} (c) and (h), with $p_{\mathrm{flip}} < 0.1$. For the remaining plots with $a_1/a_2$ = 0.01 or 0.02, the flip regions in the parameter space gradually disappear.

We first conclude from Figure \ref{fig:imaxcontour} that the orbital flip occurs more easily with the increase of $a_1/a_2$, which scales the octupole strength, and the flip possibility arises simultaneously. The other major point in Figure \ref{fig:imaxcontour} is that when $a_1/a_2$ is fixed, the region for orbital flips in panels on the diagonal is also getting larger with the decrease of $a_2$, since the gravitational perturbation from the secondary star is getting stronger.

Comparing the distribution map of the maximum $i_{\mathrm{mut}}$ for $m_1$ = 11 and 15 $M_{\mathrm{Jup}}$ in Figure \ref{fig:imaxcontour2} and that of $m_1$ = 1 $M_{\mathrm{Jup}}$ in Figure \ref{fig:imaxcontour} (i). We notice that the flip possibilities in these three plots are respectively 0.301, 0.305 and 0.297, thus the number of flipping cases changes little with the planetary mass. \citet{Anderson2016} also investigated the ratio of all likely consequences of the inward migration of giants in the stellar binaries under the octupole perturbation and the tidal dissipation. They found that the fraction of systems that give rise to either hot-Jupiter formation or tidal disruption is constantly 11 -- 14 percent, having little variation with planetary mass, stellar type and tidal dissipation strength. Nevertheless, we still obtain some new findings related to the variation of planetary masses. As the planetary mass increases, original flips in the region of $e_1 \leq 0.1$ and $e_2$ $\sim$ 0.6 disappear, while there are new flips emerge in the region of $e_1$ $\in$ [0.2, 0.3] and $e_2 \sim 0.8$.

In \citet{Michtchenko2004}, the theoretical analysis of secular dynamics for three body systems were performed within the space ($e_1$, $e_2$), and the phase space structure depends upon the ratios of the planetary masses and their SMA. After presenting analysis over a wide range of the planetary mass and semi-major axis ratios, they showed that when both mass and semi-major axis ratios are far from unity, the domains of oscillation orbits decrease. By comparison, our results from Figures \ref{fig:imaxcontour2} and \ref{fig:imaxcontour} are consistent with those of \citet{Michtchenko2004}. In Figure \ref{fig:sections}, we find those reverting orbits are almost identical to oscillating cases, thus the magnitude of the flip possibility  $p_{\mathrm{flip}}$ (Figures \ref{fig:imaxcontour2} and \ref{fig:imaxcontour}) could  be expressed as the region within ($e_1$, $e_2$) plane dominated by oscillations. Thus we conclude that the flip possibility  $p_{\mathrm{flip}}$ goes down with the decreasing $a_1 / a_2$ and $m_1 / m_2$.

We further calculate the orbital flip maps of some specific potential inclined S-type planets, employing their real minimum planetary masses. Here we mark out possible orbital positions of S-type planets HD 19994 Ab \citep{Mayor2004} and HD 196885 Ab \citep{Chauvin2011} in Figure \ref{fig:imaxcontour}, since the emergence of flipping cases changes little with the planetary mass for $e_1 \geq 0.3$. The yellow diamond in panel (a) approximately give the observed parameters for the S-type planet HD 19994 Ab. HD 19994 Ab was discovered in 2003 by the radial velocity measurement with a minimum mass $m_1 \sin I_1  = 1.68$ $M_{\mathrm{Jup}}$, $a_1$ = $1.42\pm0.01$ au and $e_1$ = $0.3\pm0.04$. The binary HD 19994 AB consists of the primary of $m_0$ = 1.34 $M_{\odot}$ and the secondary of $m_2$ = 0.35 $M_{\odot}$, $a_2$ = 100 au, $e_2$ = 0.26. Figure \ref{fig:imaxcontour} (a) indicates that HD 19994 Ab has a high probability of retaining the mutual inclination below 60$^{\circ}$ for the minimum mass.

Aside from HD 19994 Ab, we also discuss the flip possibility of HD 196885 Ab with $m_1 \sin I_1  = 2.98$ $M_{\mathrm{Jup}}$, $a_1$ = $2.0$ au and $e_1$ = $0.48$. HD 196885 AB binary consists of the primary star with $m_0$ = 1.33 $M_{\odot}$ and the secondary with $m_2$ = 0.55 $M_{\odot}$, $a_2$ = 23 au. The problem is that we do not know the eccentricity of the secondary star. Here we make some assumptions of $e_2$, if $e_2 \geq 0.65$, then the orbit of HD 196885 Ab will roll over for $m_1$ = 1 $M_{\mathrm{Jup}}$, otherwise, it will not flip. Since we have found that the planetary mass will not change the orbital flip possibility for $e_1\geq 0.3$, this assumption is still valid as $m_1$ increases.

\section{Conclusions and discussion}\label{sec:conclusions}
In this work, we employ the non-restricted EKL mechanism to shed light on the secular evolution of the inclined S-type planet $\gamma$ Cep Ab. With a wide range of parameters of SMA, eccentricity, and planetary masses, we perform numerical simulations in relation to the octupole level effects to extensively investigate the orbital flip possibility of the potential inclined S-type planets in general systems. Here we summarize major results as follows:

\begin{enumerate}[1)]
\item
We first derive the posterior distributions of orbital parameters of $\gamma$ Cep Ab and the star companion in $\gamma$ Cep system with more accurate estimation uncertainties with N-body model, using the MCMC ensemble sampler and the initial parameters independently. The minimum planetary mass is further estimated.

\item
Then we employ the EKL mechanism to explain the origin of high inclination of $\gamma$ Cep Ab. With initial conditions selected from the ($e_1$, $i_{\mathrm{mut}}$) plane, we show that when $m_1$ = 15 $M_{\mathrm{Jup}}$, it is easier for $\gamma$ Cep Ab to reach the target $i_{\mathrm{mut}}$ over 113$^{\circ}$ while the initial $i_{\mathrm{mut}} < 60^{\circ}$ and $e_1 < 0.7$.  Our investigation further indicates that relatively small values of $\mathcal{H}$ and low initial $i_{\mathrm{mut}}$ may lead to a bit longer timescale for the first orbital flip. The libration and circulation regions in the ($e_1$, $g_1$) plane and Poincar\'{e} surfaces of section, as well as the  secular stability criterion, confirm that the flipping orbits of $\gamma$ Cep Ab have a great possibility to retain stable.

\item
This work further extends the application of the EKL mechanism to general S-type planets. We take $m_1$, $a_1$, $e_2$ as independent variables, and the orbital flips can be observed in the selected space of $a_1$ and  $a_2$. The most intense orbital inclination excitation occurs when $a_1/a_2=0.1$ and $e_2$ $\sim$ 0.8. As the planetary mass increases, the original flips at $e_1 \leq 0.1$ and $e_2 \sim 0.6$ are suppressed, whereas the new flips emerge for $e_1$ $\in$ [0.2, 0.3] and $e_2 \sim 0.8$, with little effect on the total flip possibility.
\end{enumerate}

In this study, we mainly focus on the extremely high mutual inclination or transformation between the prograde and retrograde orbits in binary systems over the timescale of secular resonance. Note that the EKL mechanism can trigger high eccentricities of the planet, and the final orbit of the planet is closely related to $e_1$ and tidal dissipation. The star companion can drive the planet approach the central star through eccentricity excitation in the secular chaos, and the tidal dissipation will eventually circularize them into hot Jupiters \citep{Wu2003,Wu2011}. 	However, there still remain several open questions for forthcoming investigation, e.g., in what kinds of mechanisms the excited eccentricity of the planet is reduced to be approximately zero.

Actually, the comprehensive effect of the EKL mechanism and the tidal theory will be explored in our future study. We will construct a double-averaged model that involves the octupole level secular resonance and the equilibrium tides of spinning planets. The extreme evolution of the periastron distance induced by the secular resonance could significantly vary the tidal dissipation timescale. Our simulations for hot Jupiters in the binary system indicate that the eccentricity of the planet can decrease from 0.8 to 0 over less than $10^4$ years. Recent studies also show that the combination of high-eccentricity migration (HEM) and the Kozai--Lidov scenario can be applicable to the formation of diverse exoplanets \citep{Petrovich2019, Christopher2021}.

 \citet{Anderson2016} explored inward migration for giant planets in stellar binaries via EKL and showed that the gas-giants can be circularized as a hot Jupiter. For S-type planets falling into the region dominated by the tide, the secular evolution in combination with tidal dissipation will give a more comprehensive picture of dynamical evolution. Alternative studies suggest that the mean motion resonance in mixture with secular resonance as well as other scenarios can play a vital role in the evolution of giant planets in binary systems or hosted by single stars \citep{Haghighipour2006, Bazso2017, Liu2020}. Hence, this requires extensive investigations with additional high-precision observations of space-based/ground-based telescopes to better understand the complicated dynamics and formation of S-type planets.

\section*{Acknowledgements}
We thank the anonymous referee for constructive comments
and suggestions to significantly improve the original manuscript. This work is financially supported by the National Natural Science Foundation of China (Grant Nos. 12033010, 11773081), the B-type Strategic Priority Program of the Chinese Academy of Sciences (Grant No. XDB41000000), Foundation of Minor Planets of the Purple Mountain Observatory.
%%%%%%%%%%%%%%%%%%%% REFERENCES %%%%%%%%%%%%%%%%%%

% The best way to enter references is to use BibTeX:
%%\bibliographystyle{AASJournal}

\bibliography{refs}{}

\begin{thebibliography}{}
\expandafter\ifx\csname natexlab\endcsname\relax\def\natexlab#1{#1}\fi
\providecommand{\url}[1]{\href{#1}{#1}}
\providecommand{\dodoi}[1]{doi:~\href{http://doi.org/#1}{\nolinkurl{#1}}}
\providecommand{\doeprint}[1]{\href{http://ascl.net/#1}{\nolinkurl{http://ascl.net/#1}}}
\providecommand{\doarXiv}[1]{\href{https://arxiv.org/abs/#1}{\nolinkurl{https://arxiv.org/abs/#1}}}

\bibitem[{{Anderson} {et~al.}(2016){Anderson}, {Storch}, \&
  {Lai}}]{Anderson2016}
{Anderson}, K.~R., {Storch}, N.~I., \& {Lai}, D. 2016, \mnras, 456, 3671,
  \dodoi{10.1093/mnras/stv2906}

\bibitem[{{Andrade-Ines} {et~al.}(2016){Andrade-Ines}, {Beaug{\'e}},
  {Michtchenko}, \& {Robutel}}]{Andrade-Ines2016}
{Andrade-Ines}, E., {Beaug{\'e}}, C., {Michtchenko}, T., \& {Robutel}, P. 2016,
  Celestial Mechanics and Dynamical Astronomy, 124, 405,
  \dodoi{10.1007/s10569-015-9669-5}

\bibitem[{{Antognini}(2015)}]{Antognini2015}
{Antognini}, J.~M.~O. 2015, \mnras, 452, 3610, \dodoi{10.1093/mnras/stv1552}

\bibitem[{{Armstrong} {et~al.}(2014){Armstrong}, {Osborn}, {Brown}, {Faedi},
  {G{\'o}mez Maqueo Chew}, {Martin}, {Pollacco}, \& {Udry}}]{Armstrong2014}
{Armstrong}, D.~J., {Osborn}, H.~P., {Brown}, D.~J.~A., {et~al.} 2014, \mnras,
  444, 1873, \dodoi{10.1093/mnras/stu1570}

\bibitem[{{Artymowicz} \& {Lubow}(1994)}]{Artymowicz1994}
{Artymowicz}, P., \& {Lubow}, S.~H. 1994, \apj, 421, 651,
  \dodoi{10.1086/173679}

\bibitem[{{Bazs{\'o}} {et~al.}(2017){Bazs{\'o}}, {Pilat-Lohinger}, {Eggl},
  {Funk}, {Bancelin}, \& {Rau}}]{Bazso2017}
{Bazs{\'o}}, {\'A}., {Pilat-Lohinger}, E., {Eggl}, S., {et~al.} 2017, \mnras,
  466, 1555, \dodoi{10.1093/mnras/stw3095}

\bibitem[{{Butler} {et~al.}(2006){Butler}, {Wright}, {Marcy}, {Fischer},
  {Vogt}, {Tinney}, {Jones}, {Carter}, {Johnson}, {McCarthy}, \&
  {Penny}}]{Butler2006}
{Butler}, R.~P., {Wright}, J.~T., {Marcy}, G.~W., {et~al.} 2006, \apj, 646,
  505, \dodoi{10.1086/504701}

\bibitem[{{Campbell} {et~al.}(1988){Campbell}, {Walker}, \&
  {Yang}}]{Campbell1988}
{Campbell}, B., {Walker}, G.~A.~H., \& {Yang}, S. 1988, \apj, 331, 902,
  \dodoi{10.1086/166608}

\bibitem[{{Chauvin} {et~al.}(2011){Chauvin}, {Beust}, {Lagrange}, \&
  {Eggenberger}}]{Chauvin2011}
{Chauvin}, G., {Beust}, H., {Lagrange}, A.~M., \& {Eggenberger}, A. 2011, \aap,
  528, A8, \dodoi{10.1051/0004-6361/201015433}

\bibitem[{{Eggleton} {et~al.}(1998){Eggleton}, {Kiseleva}, \&
  {Hut}}]{Eggleton1998}
{Eggleton}, P.~P., {Kiseleva}, L.~G., \& {Hut}, P. 1998, \apj, 499, 853,
  \dodoi{10.1086/305670}

\bibitem[{{Eggleton} \& {Kiseleva-Eggleton}(2001)}]{Eggleton2001}
{Eggleton}, P.~P., \& {Kiseleva-Eggleton}, L. 2001, \apj, 562, 1012,
  \dodoi{10.1086/323843}

\bibitem[{{Fabrycky} \& {Tremaine}(2007)}]{Fabrycky2007}
{Fabrycky}, D., \& {Tremaine}, S. 2007, \apj, 669, 1298, \dodoi{10.1086/521702}

\bibitem[{{Ford}(2006)}]{Ford2006}
{Ford}, E.~B. 2006, \apj, 642, 505, \dodoi{10.1086/500802}

\bibitem[{{Ford} {et~al.}(2000){Ford}, {Kozinsky}, \& {Rasio}}]{Ford2000}
{Ford}, E.~B., {Kozinsky}, B., \& {Rasio}, F.~A. 2000, \apj, 535, 385,
  \dodoi{10.1086/308815}

\bibitem[{{Foreman-Mackey} {et~al.}(2013){Foreman-Mackey}, {Hogg}, {Lang}, \&
  {Goodman}}]{Foreman-Mackey2013}
{Foreman-Mackey}, D., {Hogg}, D.~W., {Lang}, D., \& {Goodman}, J. 2013, \pasp,
  125, 306, \dodoi{10.1086/670067}

\bibitem[{{Gaia Collaboration} {et~al.}(2018){Gaia Collaboration}, {Brown},
  {Vallenari}, {Prusti}, {de Bruijne}, {Babusiaux}, {Bailer-Jones}, {Biermann},
  {Evans}, {Eyer}, {Jansen}, {Jordi}, {Klioner}, {Lammers}, {Lindegren},
  {Luri}, {Mignard}, {Panem}, {Pourbaix}, {Randich}, {Sartoretti}, {Siddiqui},
  {Soubiran}, {van Leeuwen}, {Walton}, {Arenou}, {Bastian}, {Cropper},
  {Drimmel}, {Katz}, {Lattanzi}, {Bakker}, {Cacciari}, {Casta{\~n}eda},
  {Chaoul}, {Cheek}, {De Angeli}, {Fabricius}, {Guerra}, {Holl}, {Masana},
  {Messineo}, {Mowlavi}, {Nienartowicz}, {Panuzzo}, {Portell}, {Riello},
  {Seabroke}, {Tanga}, {Th{\'e}venin}, {Gracia-Abril}, {Comoretto},
  {Garcia-Reinaldos}, {Teyssier}, {Altmann}, {Andrae}, {Audard},
  {Bellas-Velidis}, {Benson}, {Berthier}, {Blomme}, {Burgess}, {Busso},
  {Carry}, {Cellino}, {Clementini}, {Clotet}, {Creevey}, {Davidson}, {De
  Ridder}, {Delchambre}, {Dell'Oro}, {Ducourant},
  {Fern{\'a}ndez-Hern{\'a}ndez}, {Fouesneau}, {Fr{\'e}mat}, {Galluccio},
  {Garc{\'\i}a-Torres}, {Gonz{\'a}lez-N{\'u}{\~n}ez}, {Gonz{\'a}lez-Vidal},
  {Gosset}, {Guy}, {Halbwachs}, {Hambly}, {Harrison}, {Hern{\'a}ndez},
  {Hestroffer}, {Hodgkin}, {Hutton}, {Jasniewicz}, {Jean-Antoine-Piccolo},
  {Jordan}, {Korn}, {Krone-Martins}, {Lanzafame}, {Lebzelter}, {L{\"o}ffler},
  {Manteiga}, {Marrese}, {Mart{\'\i}n-Fleitas}, {Moitinho}, {Mora}, {Muinonen},
  {Osinde}, {Pancino}, {Pauwels}, {Petit}, {Recio-Blanco}, {Richards},
  {Rimoldini}, {Robin}, {Sarro}, {Siopis}, {Smith}, {Sozzetti}, {S{\"u}veges},
  {Torra}, {van Reeven}, {Abbas}, {Abreu Aramburu}, {Accart}, {Aerts},
  {Altavilla}, {{\'A}lvarez}, {Alvarez}, {Alves}, {Anderson}, {Andrei},
  {Anglada Varela}, {Antiche}, {Antoja}, {Arcay}, {Astraatmadja}, {Bach},
  {Baker}, {Balaguer-N{\'u}{\~n}ez}, {Balm}, {Barache}, {Barata}, {Barbato},
  {Barblan}, {Barklem}, {Barrado}, {Barros}, {Barstow}, {Bartholom{\'e}
  Mu{\~n}oz}, {Bassilana}, {Becciani}, {Bellazzini}, {Berihuete}, {Bertone},
  {Bianchi}, {Bienaym{\'e}}, {Blanco-Cuaresma}, {Boch}, {Boeche}, {Bombrun},
  {Borrachero}, {Bossini}, {Bouquillon}, {Bourda}, {Bragaglia}, {Bramante},
  {Breddels}, {Bressan}, {Brouillet}, {Br{\"u}semeister}, {Brugaletta},
  {Bucciarelli}, {Burlacu}, {Busonero}, {Butkevich}, {Buzzi}, {Caffau},
  {Cancelliere}, {Cannizzaro}, {Cantat-Gaudin}, {Carballo}, {Carlucci},
  {Carrasco}, {Casamiquela}, {Castellani}, {Castro-Ginard}, {Charlot},
  {Chemin}, {Chiavassa}, {Cocozza}, {Costigan}, {Cowell}, {Crifo}, {Crosta},
  {Crowley}, {Cuypers}, {Dafonte}, {Damerdji}, {Dapergolas}, {David}, {David},
  {de Laverny}, {De Luise}, {De March}, {de Martino}, {de Souza}, {de Torres},
  {Debosscher}, {del Pozo}, {Delbo}, {Delgado}, {Delgado}, {Di Matteo},
  {Diakite}, {Diener}, {Distefano}, {Dolding}, {Drazinos}, {Dur{\'a}n},
  {Edvardsson}, {Enke}, {Eriksson}, {Esquej}, {Eynard Bontemps}, {Fabre},
  {Fabrizio}, {Faigler}, {Falc{\~a}o}, {Farr{\`a}s Casas}, {Federici},
  {Fedorets}, {Fernique}, {Figueras}, {Filippi}, {Findeisen}, {Fonti},
  {Fraile}, {Fraser}, {Fr{\'e}zouls}, {Gai}, {Galleti}, {Garabato},
  {Garc{\'\i}a-Sedano}, {Garofalo}, {Garralda}, {Gavel}, {Gavras}, {Gerssen},
  {Geyer}, {Giacobbe}, {Gilmore}, {Girona}, {Giuffrida}, {Glass}, {Gomes},
  {Granvik}, {Gueguen}, {Guerrier}, {Guiraud}, {Guti{\'e}rrez-S{\'a}nchez},
  {Haigron}, {Hatzidimitriou}, {Hauser}, {Haywood}, {Heiter}, {Helmi}, {Heu},
  {Hilger}, {Hobbs}, {Hofmann}, {Holland}, {Huckle}, {Hypki}, {Icardi},
  {Jan{\ss}en}, {Jevardat de Fombelle}, {Jonker}, {Juh{\'a}sz}, {Julbe},
  {Karampelas}, {Kewley}, {Klar}, {Kochoska}, {Kohley}, {Kolenberg},
  {Kontizas}, {Kontizas}, {Koposov}, {Kordopatis}, {Kostrzewa-Rutkowska},
  {Koubsky}, {Lambert}, {Lanza}, {Lasne}, {Lavigne}, {Le Fustec}, {Le
  Poncin-Lafitte}, {Lebreton}, {Leccia}, {Leclerc}, {Lecoeur-Taibi},
  {Lenhardt}, {Leroux}, {Liao}, {Licata}, {Lindstr{\o}m}, {Lister}, {Livanou},
  {Lobel}, {L{\'o}pez}, {Managau}, {Mann}, {Mantelet}, {Marchal}, {Marchant},
  {Marconi}, {Marinoni}, {Marschalk{\'o}}, {Marshall}, {Martino}, {Marton},
  {Mary}, {Massari}, {Matijevi{\v{c}}}, {Mazeh}, {McMillan}, {Messina},
  {Michalik}, {Millar}, {Molina}, {Molinaro}, {Moln{\'a}r}, {Montegriffo},
  {Mor}, {Morbidelli}, {Morel}, {Morris}, {Mulone}, {Muraveva}, {Musella},
  {Nelemans}, {Nicastro}, {Noval}, {O'Mullane}, {Ord{\'e}novic},
  {Ord{\'o}{\~n}ez-Blanco}, {Osborne}, {Pagani}, {Pagano}, {Pailler},
  {Palacin}, {Palaversa}, {Panahi}, {Pawlak}, {Piersimoni}, {Pineau}, {Plachy},
  {Plum}, {Poggio}, {Poujoulet}, {Pr{\v{s}}a}, {Pulone}, {Racero}, {Ragaini},
  {Rambaux}, {Ramos-Lerate}, {Regibo}, {Reyl{\'e}}, {Riclet}, {Ripepi}, {Riva},
  {Rivard}, {Rixon}, {Roegiers}, {Roelens}, {Romero-G{\'o}mez}, {Rowell},
  {Royer}, {Ruiz-Dern}, {Sadowski}, {Sagrist{\`a} Sell{\'e}s}, {Sahlmann},
  {Salgado}, {Salguero}, {Sanna}, {Santana-Ros}, {Sarasso}, {Savietto},
  {Schultheis}, {Sciacca}, {Segol}, {Segovia}, {S{\'e}gransan}, {Shih},
  {Siltala}, {Silva}, {Smart}, {Smith}, {Solano}, {Solitro}, {Sordo}, {Soria
  Nieto}, {Souchay}, {Spagna}, {Spoto}, {Stampa}, {Steele},
  {Steidelm{\"u}ller}, {Stephenson}, {Stoev}, {Suess}, {Surdej}, {Szabados},
  {Szegedi-Elek}, {Tapiador}, {Taris}, {Tauran}, {Taylor}, {Teixeira},
  {Terrett}, {Teyssandier}, {Thuillot}, {Titarenko}, {Torra Clotet}, {Turon},
  {Ulla}, {Utrilla}, {Uzzi}, {Vaillant}, {Valentini}, {Valette}, {van Elteren},
  {Van Hemelryck}, {van Leeuwen}, {Vaschetto}, {Vecchiato}, {Veljanoski},
  {Viala}, {Vicente}, {Vogt}, {von Essen}, {Voss}, {Votruba}, {Voutsinas},
  {Walmsley}, {Weiler}, {Wertz}, {Wevers}, {Wyrzykowski}, {Yoldas},
  {{\v{Z}}erjal}, {Ziaeepour}, {Zorec}, {Zschocke}, {Zucker}, {Zurbach}, \&
  {Zwitter}}]{Gaia2018}
{Gaia Collaboration}, {Brown}, A.~G.~A., {Vallenari}, A., {et~al.} 2018, \aap,
  616, A1, \dodoi{10.1051/0004-6361/201833051}

\bibitem[{{Gaia Collaboration} {et~al.}(2021){Gaia Collaboration}, {Brown},
  {Vallenari}, {Prusti}, {de Bruijne}, {Babusiaux}, {Biermann}, {Creevey},
  {Evans}, {Eyer}, {Hutton}, {Jansen}, {Jordi}, {Klioner}, {Lammers},
  {Lindegren}, {Luri}, {Mignard}, {Panem}, {Pourbaix}, {Randich}, {Sartoretti},
  {Soubiran}, {Walton}, {Arenou}, {Bailer-Jones}, {Bastian}, {Cropper},
  {Drimmel}, {Katz}, {Lattanzi}, {van Leeuwen}, {Bakker}, {Cacciari},
  {Casta{\~n}eda}, {De Angeli}, {Ducourant}, {Fabricius}, {Fouesneau},
  {Fr{\'e}mat}, {Guerra}, {Guerrier}, {Guiraud}, {Jean-Antoine Piccolo},
  {Masana}, {Messineo}, {Mowlavi}, {Nicolas}, {Nienartowicz}, {Pailler},
  {Panuzzo}, {Riclet}, {Roux}, {Seabroke}, {Sordo}, {Tanga}, {Th{\'e}venin},
  {Gracia-Abril}, {Portell}, {Teyssier}, {Altmann}, {Andrae}, {Bellas-Velidis},
  {Benson}, {Berthier}, {Blomme}, {Brugaletta}, {Burgess}, {Busso}, {Carry},
  {Cellino}, {Cheek}, {Clementini}, {Damerdji}, {Davidson}, {Delchambre},
  {Dell'Oro}, {Fern{\'a}ndez-Hern{\'a}ndez}, {Galluccio}, {Garc{\'\i}a-Lario},
  {Garcia-Reinaldos}, {Gonz{\'a}lez-N{\'u}{\~n}ez}, {Gosset}, {Haigron},
  {Halbwachs}, {Hambly}, {Harrison}, {Hatzidimitriou}, {Heiter},
  {Hern{\'a}ndez}, {Hestroffer}, {Hodgkin}, {Holl}, {Jan{\ss}en}, {Jevardat de
  Fombelle}, {Jordan}, {Krone-Martins}, {Lanzafame}, {L{\"o}ffler}, {Lorca},
  {Manteiga}, {Marchal}, {Marrese}, {Moitinho}, {Mora}, {Muinonen}, {Osborne},
  {Pancino}, {Pauwels}, {Petit}, {Recio-Blanco}, {Richards}, {Riello},
  {Rimoldini}, {Robin}, {Roegiers}, {Rybizki}, {Sarro}, {Siopis}, {Smith},
  {Sozzetti}, {Ulla}, {Utrilla}, {van Leeuwen}, {van Reeven}, {Abbas}, {Abreu
  Aramburu}, {Accart}, {Aerts}, {Aguado}, {Ajaj}, {Altavilla}, {{\'A}lvarez},
  {{\'A}lvarez Cid-Fuentes}, {Alves}, {Anderson}, {Anglada Varela}, {Antoja},
  {Audard}, {Baines}, {Baker}, {Balaguer-N{\'u}{\~n}ez}, {Balbinot}, {Balog},
  {Barache}, {Barbato}, {Barros}, {Barstow}, {Bartolom{\'e}}, {Bassilana},
  {Bauchet}, {Baudesson-Stella}, {Becciani}, {Bellazzini}, {Bernet}, {Bertone},
  {Bianchi}, {Blanco-Cuaresma}, {Boch}, {Bombrun}, {Bossini}, {Bouquillon},
  {Bragaglia}, {Bramante}, {Breedt}, {Bressan}, {Brouillet}, {Bucciarelli},
  {Burlacu}, {Busonero}, {Butkevich}, {Buzzi}, {Caffau}, {Cancelliere},
  {C{\'a}novas}, {Cantat-Gaudin}, {Carballo}, {Carlucci}, {Carnerero},
  {Carrasco}, {Casamiquela}, {Castellani}, {Castro-Ginard}, {Castro Sampol},
  {Chaoul}, {Charlot}, {Chemin}, {Chiavassa}, {Cioni}, {Comoretto}, {Cooper},
  {Cornez}, {Cowell}, {Crifo}, {Crosta}, {Crowley}, {Dafonte}, {Dapergolas},
  {David}, {David}, {de Laverny}, {De Luise}, {De March}, {De Ridder}, {de
  Souza}, {de Teodoro}, {de Torres}, {del Peloso}, {del Pozo}, {Delbo},
  {Delgado}, {Delgado}, {Delisle}, {Di Matteo}, {Diakite}, {Diener},
  {Distefano}, {Dolding}, {Eappachen}, {Edvardsson}, {Enke}, {Esquej}, {Fabre},
  {Fabrizio}, {Faigler}, {Fedorets}, {Fernique}, {Fienga}, {Figueras},
  {Fouron}, {Fragkoudi}, {Fraile}, {Franke}, {Gai}, {Garabato},
  {Garcia-Gutierrez}, {Garc{\'\i}a-Torres}, {Garofalo}, {Gavras}, {Gerlach},
  {Geyer}, {Giacobbe}, {Gilmore}, {Girona}, {Giuffrida}, {Gomel}, {Gomez},
  {Gonzalez-Santamaria}, {Gonz{\'a}lez-Vidal}, {Granvik},
  {Guti{\'e}rrez-S{\'a}nchez}, {Guy}, {Hauser}, {Haywood}, {Helmi}, {Hidalgo},
  {Hilger}, {H{\l}adczuk}, {Hobbs}, {Holland}, {Huckle}, {Jasniewicz},
  {Jonker}, {Juaristi Campillo}, {Julbe}, {Karbevska}, {Kervella}, {Khanna},
  {Kochoska}, {Kontizas}, {Kordopatis}, {Korn}, {Kostrzewa-Rutkowska},
  {Kruszy{\'n}ska}, {Lambert}, {Lanza}, {Lasne}, {Le Campion}, {Le Fustec},
  {Lebreton}, {Lebzelter}, {Leccia}, {Leclerc}, {Lecoeur-Taibi}, {Liao},
  {Licata}, {Lindstr{\o}m}, {Lister}, {Livanou}, {Lobel}, {Madrero Pardo},
  {Managau}, {Mann}, {Marchant}, {Marconi}, {Marcos Santos}, {Marinoni},
  {Marocco}, {Marshall}, {Martin Polo}, {Mart{\'\i}n-Fleitas}, {Masip},
  {Massari}, {Mastrobuono-Battisti}, {Mazeh}, {McMillan}, {Messina},
  {Michalik}, {Millar}, {Mints}, {Molina}, {Molinaro}, {Moln{\'a}r},
  {Montegriffo}, {Mor}, {Morbidelli}, {Morel}, {Morris}, {Mulone}, {Munoz},
  {Muraveva}, {Murphy}, {Musella}, {Noval}, {Ord{\'e}novic}, {Orr{\`u}},
  {Osinde}, {Pagani}, {Pagano}, {Palaversa}, {Palicio}, {Panahi}, {Pawlak},
  {Pe{\~n}alosa Esteller}, {Penttil{\"a}}, {Piersimoni}, {Pineau}, {Plachy},
  {Plum}, {Poggio}, {Poretti}, {Poujoulet}, {Pr{\v{s}}a}, {Pulone}, {Racero},
  {Ragaini}, {Rainer}, {Raiteri}, {Rambaux}, {Ramos}, {Ramos-Lerate}, {Re
  Fiorentin}, {Regibo}, {Reyl{\'e}}, {Ripepi}, {Riva}, {Rixon}, {Robichon},
  {Robin}, {Roelens}, {Rohrbasser}, {Romero-G{\'o}mez}, {Rowell}, {Royer},
  {Rybicki}, {Sadowski}, {Sagrist{\`a} Sell{\'e}s}, {Sahlmann}, {Salgado},
  {Salguero}, {Samaras}, {Sanchez Gimenez}, {Sanna}, {Santove{\~n}a},
  {Sarasso}, {Schultheis}, {Sciacca}, {Segol}, {Segovia}, {S{\'e}gransan},
  {Semeux}, {Shahaf}, {Siddiqui}, {Siebert}, {Siltala}, {Slezak}, {Smart},
  {Solano}, {Solitro}, {Souami}, {Souchay}, {Spagna}, {Spoto}, {Steele},
  {Steidelm{\"u}ller}, {Stephenson}, {S{\"u}veges}, {Szabados}, {Szegedi-Elek},
  {Taris}, {Tauran}, {Taylor}, {Teixeira}, {Thuillot}, {Tonello}, {Torra},
  {Torra}, {Turon}, {Unger}, {Vaillant}, {van Dillen}, {Vanel}, {Vecchiato},
  {Viala}, {Vicente}, {Voutsinas}, {Weiler}, {Wevers}, {Wyrzykowski}, {Yoldas},
  {Yvard}, {Zhao}, {Zorec}, {Zucker}, {Zurbach}, \& {Zwitter}}]{Gaia2021}
---. 2021, \aap, 649, A1, \dodoi{10.1051/0004-6361/202039657}

\bibitem[{{Gellert} {et~al.}(1977){Gellert}, {Hellwich}, {K{\"a}Stner},
  {Hirsch}, \& {Reichardt}}]{Gellert1977}
{Gellert}, W., {Hellwich}, M., {K{\"a}Stner}, H., {Hirsch}, K.~A., \&
  {Reichardt}, H. 1977, {The VNR Concise Encyclopedia of Mathematics} No.
  Chapter 3, 320--668

\bibitem[{{Giuppone} {et~al.}(2011){Giuppone}, {Leiva}, {Correa-Otto}, \&
  {Beaug{\'e}}}]{Giuppone2011}
{Giuppone}, C.~A., {Leiva}, A.~M., {Correa-Otto}, J., \& {Beaug{\'e}}, C. 2011,
  \aap, 530, A103, \dodoi{10.1051/0004-6361/201016375}

\bibitem[{{Gong} \& {Ji}(2018)}]{Gong2018}
{Gong}, Y.-X., \& {Ji}, J. 2018, \mnras, 478, 4565,
  \dodoi{10.1093/mnras/sty1300}

\bibitem[{{Haghighipour}(2004)}]{Haghighipour2004}
{Haghighipour}, N. 2004, in American Institute of Physics Conference Series,
  Vol. 713, The Search for Other Worlds, ed. S.~S. {Holt} \& D.~{Deming},
  269--272, \dodoi{10.1063/1.1774536}

\bibitem[{{Haghighipour}(2006)}]{Haghighipour2006}
{Haghighipour}, N. 2006, \apj, 644, 543, \dodoi{10.1086/503351}

\bibitem[{{Harrington}(1968)}]{Harrington1968}
{Harrington}, R.~S. 1968, \aj, 73, 190, \dodoi{10.1086/110614}

\bibitem[{{Hatzes} {et~al.}(2003){Hatzes}, {Cochran}, {Endl}, {McArthur},
  {Paulson}, {Walker}, {Campbell}, \& {Yang}}]{hatzes2003}
{Hatzes}, A.~P., {Cochran}, W.~D., {Endl}, M., {et~al.} 2003, \apj, 599, 1383,
  \dodoi{10.1086/379281}

\bibitem[{{Henrard} \& {Libert}(2008)}]{Henrard2008}
{Henrard}, J., \& {Libert}, A.-S. 2008, Celestial Mechanics and Dynamical
  Astronomy, 102, 177, \dodoi{10.1007/s10569-007-9111-8}

\bibitem[{{Jang-Condell} {et~al.}(2008){Jang-Condell}, {Mugrauer}, \&
  {Schmidt}}]{Jang-Condell2008}
{Jang-Condell}, H., {Mugrauer}, M., \& {Schmidt}, T. 2008, \apjl, 683, L191,
  \dodoi{10.1086/591791}

\bibitem[{{Ji} {et~al.}(2022){Ji}, {Li}, {Zhang}, {Fang}, {Li}, {Wang}, {Cao},
  {Deng}, {Li}, {Xian}, {Gao}, {Zhang}, {Li}, {Liu}, {Qi}, {Jin}, {Liu},
  {Chen}, {Li}, {Dong}, {Zhu}, \& {CHES Consortium}}]{Ji2022}
{Ji}, J.-H., {Li}, H.-T., {Zhang}, J.-B., {et~al.} 2022, Research in Astronomy
  and Astrophysics, 22, 072003, \dodoi{10.1088/1674-4527/ac77e4}

\bibitem[{{Katz} {et~al.}(2011){Katz}, {Dong}, \& {Malhotra}}]{Katz2011}
{Katz}, B., {Dong}, S., \& {Malhotra}, R. 2011, \prl, 107, 181101,
  \dodoi{10.1103/PhysRevLett.107.181101}

\bibitem[{{Kiseleva} {et~al.}(1998){Kiseleva}, {Eggleton}, \&
  {Mikkola}}]{Kiseleva1998}
{Kiseleva}, L.~G., {Eggleton}, P.~P., \& {Mikkola}, S. 1998, \mnras, 300, 292,
  \dodoi{10.1046/j.1365-8711.1998.01903.x}

\bibitem[{{Kostov} {et~al.}(2014){Kostov}, {McCullough}, {Carter}, {Deleuil},
  {D{\'\i}az}, {Fabrycky}, {H{\'e}brard}, {Hinse}, {Mazeh}, {Orosz},
  {Tsvetanov}, \& {Welsh}}]{Kostov2014}
{Kostov}, V.~B., {McCullough}, P.~R., {Carter}, J.~A., {et~al.} 2014, \apj,
  787, 93, \dodoi{10.1088/0004-637X/787/1/93}

\bibitem[{{Kozai}(1962)}]{Kozai1962}
{Kozai}, Y. 1962, \aj, 67, 591, \dodoi{10.1086/108790}

\bibitem[{{Lee} \& {Peale}(2003)}]{Lee2003}
{Lee}, M.~H., \& {Peale}, S.~J. 2003, \apj, 592, 1201, \dodoi{10.1086/375857}

\bibitem[{{Lei}(2019)}]{Lei2019}
{Lei}, H. 2019, \mnras, 490, 4756, \dodoi{10.1093/mnras/stz2917}

\bibitem[{{Lei}(2021)}]{Lei2021}
---. 2021, \mnras, 506, 1879, \dodoi{10.1093/mnras/stab1789}

\bibitem[{{Lei}(2022)}]{Lei2022}
---. 2022, \aj, 163, 214, \dodoi{10.3847/1538-3881/ac5fa8}

\bibitem[{{Lei} {et~al.}(2018){Lei}, {Circi}, \& {Ortore}}]{Lei2018}
{Lei}, H., {Circi}, C., \& {Ortore}, E. 2018, \mnras, 481, 4602,
  \dodoi{10.1093/mnras/sty2619}

\bibitem[{{Li} {et~al.}(2014{\natexlab{a}}){Li}, {Naoz}, {Holman}, \&
  {Loeb}}]{Li2014a}
{Li}, G., {Naoz}, S., {Holman}, M., \& {Loeb}, A. 2014{\natexlab{a}}, \apj,
  791, 86, \dodoi{10.1088/0004-637X/791/2/86}

\bibitem[{{Li} {et~al.}(2014{\natexlab{b}}){Li}, {Naoz}, {Kocsis}, \&
  {Loeb}}]{Li2014b}
{Li}, G., {Naoz}, S., {Kocsis}, B., \& {Loeb}, A. 2014{\natexlab{b}}, \apj,
  785, 116, \dodoi{10.1088/0004-637X/785/2/116}

\bibitem[{{Libert} \& {Henrard}(2006)}]{Libert2006}
{Libert}, A.-S., \& {Henrard}, J. 2006, \icarus, 183, 186,
  \dodoi{10.1016/j.icarus.2006.02.007}

\bibitem[{{Libert} \& {Henrard}(2007)}]{Libert2007}
{Libert}, A.~S., \& {Henrard}, J. 2007, \aap, 461, 759,
  \dodoi{10.1051/0004-6361:20065767}

\bibitem[{{Libert} \& {Henrard}(2008)}]{Libert2008}
{Libert}, A.-S., \& {Henrard}, J. 2008, Celestial Mechanics and Dynamical
  Astronomy, 100, 209, \dodoi{10.1007/s10569-007-9113-6}

\bibitem[{{Lidov}(1962)}]{Lidov1962}
{Lidov}, M.~L. 1962, \planss, 9, 719, \dodoi{10.1016/0032-0633(62)90129-0}

\bibitem[{{Lithwick} \& {Naoz}(2011)}]{Lithwick2011}
{Lithwick}, Y., \& {Naoz}, S. 2011, \apj, 742, 94,
  \dodoi{10.1088/0004-637X/742/2/94}

\bibitem[{{Liu} \& {Ji}(2020)}]{Liu2020}
{Liu}, B., \& {Ji}, J. 2020, Research in Astronomy and Astrophysics, 20, 164,
  \dodoi{10.1088/1674-4527/20/10/164}

\bibitem[{{Mardling} \& {Aarseth}(2001)}]{Mardling2001}
{Mardling}, R.~A., \& {Aarseth}, S.~J. 2001, \mnras, 321, 398,
  \dodoi{10.1046/j.1365-8711.2001.03974.x}

\bibitem[{{Mart{\'\i}} \& {Beaug{\'e}}(2012)}]{Marti2012}
{Mart{\'\i}}, J.~G., \& {Beaug{\'e}}, C. 2012, \aap, 544, A97,
  \dodoi{10.1051/0004-6361/201219403}

\bibitem[{{Mayor} {et~al.}(2004){Mayor}, {Udry}, {Naef}, {Pepe}, {Queloz},
  {Santos}, \& {Burnet}}]{Mayor2004}
{Mayor}, M., {Udry}, S., {Naef}, D., {et~al.} 2004, \aap, 415, 391,
  \dodoi{10.1051/0004-6361:20034250}

\bibitem[{{Michtchenko} \& {Malhotra}(2004)}]{Michtchenko2004}
{Michtchenko}, T.~A., \& {Malhotra}, R. 2004, \icarus, 168, 237,
  \dodoi{10.1016/j.icarus.2003.12.010}

\bibitem[{{Moe} \& {Di Stefano}(2017)}]{Moe2017}
{Moe}, M., \& {Di Stefano}, R. 2017, \apjs, 230, 15,
  \dodoi{10.3847/1538-4365/aa6fb6}

\bibitem[{{Naoz}(2016)}]{Naoz2016}
{Naoz}, S. 2016, \araa, 54, 441, \dodoi{10.1146/annurev-astro-081915-023315}

\bibitem[{{Naoz} \& {Fabrycky}(2014)}]{Naoz2014}
{Naoz}, S., \& {Fabrycky}, D.~C. 2014, \apj, 793, 137,
  \dodoi{10.1088/0004-637X/793/2/137}

\bibitem[{{Naoz} {et~al.}(2013){Naoz}, {Farr}, {Lithwick}, {Rasio}, \&
  {Teyssandier}}]{Naoz2013}
{Naoz}, S., {Farr}, W.~M., {Lithwick}, Y., {Rasio}, F.~A., \& {Teyssandier}, J.
  2013, \mnras, 431, 2155, \dodoi{10.1093/mnras/stt302}

\bibitem[{{Nelson} {et~al.}(2016){Nelson}, {Robertson}, {Payne}, {Pritchard},
  {Deck}, {Ford}, {Wright}, \& {Isaacson}}]{Nelson2016}
{Nelson}, B.~E., {Robertson}, P.~M., {Payne}, M.~J., {et~al.} 2016, \mnras,
  455, 2484, \dodoi{10.1093/mnras/stv2367}

\bibitem[{{Neuh{\"a}user} {et~al.}(2007){Neuh{\"a}user}, {Mugrauer},
  {Fukagawa}, {Torres}, \& {Schmidt}}]{neuhauser2007}
{Neuh{\"a}user}, R., {Mugrauer}, M., {Fukagawa}, M., {Torres}, G., \&
  {Schmidt}, T. 2007, \aap, 462, 777, \dodoi{10.1051/0004-6361:20066581}

\bibitem[{{O'Connor} {et~al.}(2021){O'Connor}, {Liu}, \&
  {Lai}}]{Christopher2021}
{O'Connor}, C.~E., {Liu}, B., \& {Lai}, D. 2021, \mnras, 501, 507,
  \dodoi{10.1093/mnras/staa3723}

\bibitem[{{Perets} \& {Fabrycky}(2009)}]{Perets2009}
{Perets}, H.~B., \& {Fabrycky}, D.~C. 2009, \apj, 697, 1048,
  \dodoi{10.1088/0004-637X/697/2/1048}

\bibitem[{{Petrovich} {et~al.}(2019){Petrovich}, {Deibert}, \&
  {Wu}}]{Petrovich2019}
{Petrovich}, C., {Deibert}, E., \& {Wu}, Y. 2019, \aj, 157, 180,
  \dodoi{10.3847/1538-3881/ab0e0a}

\bibitem[{{Raghavan} {et~al.}(2010){Raghavan}, {McAlister}, {Henry}, {Latham},
  {Marcy}, {Mason}, {Gies}, {White}, \& {ten Brummelaar}}]{Raghavan2010}
{Raghavan}, D., {McAlister}, H.~A., {Henry}, T.~J., {et~al.} 2010, \apjs, 190,
  1, \dodoi{10.1088/0067-0049/190/1/1}

\bibitem[{{Reffert} \& {Quirrenbach}(2011)}]{reffert2011}
{Reffert}, S., \& {Quirrenbach}, A. 2011, \aap, 527, A140,
  \dodoi{10.1051/0004-6361/201015861}

\bibitem[{{Rein} \& {Spiegel}(2015)}]{Rein2015}
{Rein}, H., \& {Spiegel}, D.~S. 2015, \mnras, 446, 1424,
  \dodoi{10.1093/mnras/stu2164}

\bibitem[{{Satyal} {et~al.}(2013){Satyal}, {Quarles}, \& {Hinse}}]{Satyal2013}
{Satyal}, S., {Quarles}, B., \& {Hinse}, T.~C. 2013, \mnras, 433, 2215,
  \dodoi{10.1093/mnras/stt888}

\bibitem[{{Schwarz} {et~al.}(2016){Schwarz}, {Funk}, {Zechner}, \&
  {Bazs{\'o}}}]{Schwarz2016}
{Schwarz}, R., {Funk}, B., {Zechner}, R., \& {Bazs{\'o}}, {\'A}. 2016, \mnras,
  460, 3598, \dodoi{10.1093/mnras/stw1218}

\bibitem[{{Shappee} \& {Thompson}(2013)}]{Shappee2013}
{Shappee}, B.~J., \& {Thompson}, T.~A. 2013, \apj, 766, 64,
  \dodoi{10.1088/0004-637X/766/1/64}

\bibitem[{{Sidorenko}(2018)}]{Sidorenko2018}
{Sidorenko}, V.~V. 2018, Celestial Mechanics and Dynamical Astronomy, 130, 4,
  \dodoi{10.1007/s10569-017-9799-z}

\bibitem[{{Tan} {et~al.}(2020){Tan}, {Hou}, {Liao}, {Wang}, \&
  {Tang}}]{Tan2020}
{Tan}, P., {Hou}, X., {Liao}, X., {Wang}, W., \& {Tang}, J. 2020, \aj, 160,
  139, \dodoi{10.3847/1538-3881/aba89c}

\bibitem[{{Teyssandier} {et~al.}(2013){Teyssandier}, {Naoz}, {Lizarraga}, \&
  {Rasio}}]{Teyssandier2013}
{Teyssandier}, J., {Naoz}, S., {Lizarraga}, I., \& {Rasio}, F.~A. 2013, \apj,
  779, 166, \dodoi{10.1088/0004-637X/779/2/166}

\bibitem[{{Thompson}(2011)}]{Thompson2011}
{Thompson}, T.~A. 2011, \apj, 741, 82, \dodoi{10.1088/0004-637X/741/2/82}

\bibitem[{{Tokovinin}(1997)}]{Tokovinin1997}
{Tokovinin}, A.~A. 1997, Astronomy Letters, 23, 727

\bibitem[{{Torres}(2007)}]{torres2007}
{Torres}, G. 2007, \apj, 654, 1095, \dodoi{10.1086/509715}

\bibitem[{{Valtonen} \& {Karttunen}(2006)}]{Valtonen2006}
{Valtonen}, M., \& {Karttunen}, H. 2006, {The Three-Body Problem}

\bibitem[{{von Zeipel}(1910)}]{vonZeipel1910}
{von Zeipel}, H. 1910, Astronomische Nachrichten, 183, 345,
  \dodoi{10.1002/asna.19091832202}

\bibitem[{{Walker} {et~al.}(1992){Walker}, {Bohlender}, {Walker}, {Irwin},
  {Yang}, \& {Larson}}]{Walker1992}
{Walker}, G. A.~H., {Bohlender}, D.~A., {Walker}, A.~R., {et~al.} 1992, \apjl,
  396, L91, \dodoi{10.1086/186524}

\bibitem[{{Wu} \& {Lithwick}(2011)}]{Wu2011}
{Wu}, Y., \& {Lithwick}, Y. 2011, \apj, 735, 109,
  \dodoi{10.1088/0004-637X/735/2/109}

\bibitem[{{Wu} \& {Murray}(2003)}]{Wu2003}
{Wu}, Y., \& {Murray}, N. 2003, \apj, 589, 605, \dodoi{10.1086/374598}

\bibitem[{{Xie} {et~al.}(2010){Xie}, {Zhou}, \& {Ge}}]{Xie2010}
{Xie}, J.-W., {Zhou}, J.-L., \& {Ge}, J. 2010, \apj, 708, 1566,
  \dodoi{10.1088/0004-637X/708/2/1566}

\end{thebibliography}
\bibliographystyle{aasjournal}
%%%%%%%%%%%%%%%%%%%%%%%%%%%%%%%%%%%%%%%%%%%%%%%%%%

\end{CJK*}
\end{document}